\documentstyle[eqsecnum,aps,graphicx,times]{revtex}
\begin{document}

\draft \title{Direct Photon Production in 158~{\it A}~GeV
   $^{208}$Pb\/+\/$^{208}$Pb Collisions}

\author{ M.M.~Aggarwal,$^{1}$ A.~Agnihotri,$^{2}$ Z.~Ahammed,$^{3}$
   A.L.S.~Angelis,$^{4}$ V.~Antonenko,$^{5}$ V.~Arefiev,$^{6}$
   V.~Astakhov,$^{6}$ V.~Avdeitchikov,$^{6}$ T.C.~Awes,$^{7}$
   P.V.K.S.~Baba,$^{8}$
% ...
}

\author{(WA98 Collaboration)}

\address{$^{1}$~University of Panjab, Chandigarh 160014, India}
\address{$^{2}$~University of Rajasthan, Jaipur 302004, Rajasthan,
   India} \address{$^{3}$~Variable Energy Cyclotron Centre, Calcutta
   700 064, India} \address{$^{4}$~University of Geneva, CH-1211 Geneva
   4,Switzerland} \address{$^{5}$~RRC ``Kurchatov Institute'', 
   RU-123182 Moscow,
   Russia} \address{$^{6}$~Joint Institute for Nuclear Research,
   RU-141980 Dubna, Russia} \address{$^{7}$~Oak Ridge National
   Laboratory, Oak Ridge, Tennessee 37831-6372, USA}
\address{$^{8}$~University of Jammu, Jammu 180001, India}
% ...

\date{\today} \maketitle
\begin{abstract}
   A measurement of direct photon production in
   $^{208}$Pb\/+\/$^{208}$Pb collisions at 158~{\it A}~GeV has been
   carried out in the CERN WA98 experiment.  The invariant yield or
   upper limit of direct photons as a function of transverse momentum
   in the interval $0.5 < p_T < 4$ GeV/c is presented.  A significant
direct photon excess is observed at $p_T > 1.5$ GeV/c in central
   collisions. The results are
   compared to proton-induced results and to theoretical predictions.
   Implications for the dynamics of high-energy heavy-ion collisions
   are discussed.
\end{abstract}

% body of paper here
\section{\label{sec:intro} INTRODUCTION}

A major current goal of the field of nuclear physics is
the experimental confirmation of the existence of a new phase
of strongly interacting matter, the quark gluon plasma (QGP)~\cite{QM99},
which is predicted to exist according to lattice calculations
of quantum chromodynamics. 
Enhanced production of strange hadrons, photons, and dileptons, 
and suppression of $J/\psi$ mesons
are some of the proposed consequences of QGP formation.
Both $J/\psi$ suppression~\cite{Abr97} and strangeness 
enhancement~\cite{And99} have been observed in 
relativistic heavy-ion collisions with strongly enhanced nuclear
effects. While these observations naturally lead to the 
conclusion that the initial phase of 
the collision consisted of a hot and dense system 
with strong rescattering, which may be
explained by the assumption of QGP formation, the direct experimental 
detection of QGP through observation of direct emission of real or virtual 
photons from the quark matter remains to be attained. 

Historically, photons and lepton-pairs were the probes first suggested
to use to search for evidence of quark-gluon plasma formation in
ultrarelativistic heavy ion collisions. During the collision, real
photons are produced mainly by scattering amongst the electrically
charged objects while virtual (i.e. massive) photons, which later
decay into pairs of oppositely charged leptons, or dileptons, are
produced mainly by particle-antiparticle annihilations.  Once
produced, the real and virtual photons will interact with the
surrounding hot dense matter through the electromagnetic interaction
only. The resulting small interaction cross section implies a long
mean free path in the dense matter with the consequence that the
photons are likely to escape unscathed once produced. As a result,
real and virtual photons carry information about the conditions of the
matter from which they were produced throughout the entire history of
the heavy ion collision, including especially the initial hot dense
phase. Therefore, if the initial phase includes a quark-gluon plasma
which radiates real and virtual photons differently than would dense
hadronic matter this difference may be apparent in the photon and
dilepton spectra observed by the experimentalist.  This is in contrast
to hadrons which, due to their extremely short mean free path in the
hot dense matter, are unlikely to escape until the system has cooled
and expanded to the low temperature and low density freezeout stage.
As a result, quark-gluon plasma formation during the initial stage of
the collision will make its presence evident via hadronic probes only
if it alters the macroscopic features of the system, such as its
strangeness content or collective flow, in a way which is different
from dense hadronic matter {\em and} if these altered features are
preserved until the time of freezeout. Thus the electromagnetic and
hadronic probes provide complementary information.  Since the real and
virtual photon emission rate is greatest in hot dense matter the
electromagnetic probes should carry information mostly about the
dynamics (or thermodynamics) of the initial phase of the collision,
while hadronic probes carry information dominantly about the late
stage of the collision.

Originally, Feinberg~\cite{Fei76} and Shuryak~\cite{Shu78} suggested
that thermal emission might be an important process in hadron-induced
and even lepton-induced reactions when a large multiplicity of
particles are produced in the final state. In particular they pointed
out that rescattering amongst the produced particles in local thermal
equilibrium during the later stages of the interaction would give rise
to real and virtual photon emission. (Bjorken and Weisberg made
similar suggestions at that time about the possible importance of
rescattering~\cite{Bjo76}).  Such a mechanism could explain the, at
that time, puzzling excess dilepton yield observed at intermediate
dilepton masses, $M\leq 5$ GeV/c$^2$~\cite{Shu78}. 
Feinberg~\cite{Fei76} speculated upon the nature of the hot prematter remaining
after the interaction and suggested that it may even be gluonic matter
with embedded quarks. Shuryak~\cite{Shu78} went on to assume formation
of such a quark-gluon plasma in order to calculate the emission rates
by perturbative QCD methods.

While it remains unknown whether quark-gluon plasma may be produced in
hadron-induced reactions, it was suggested shortly afterwards that
relativistic collisions of heavy ions provide conditions likely to
result in the production of quark-gluon plasma. Initially it was
suggested that such a plasma might occur at incident laboratory
energies as low as a few GeV per nucleon~\cite{Chi78} due to
compression of the colliding nuclei and the resulting high baryon
density. Another estimate based on extrapolations of known properties
of NN and NA collisions at energies of $E_{cm} \ge 30$ GeV suggested
that the fragmentation regions of AA collisions were likely to result
in quark-gluon plasma formation~\cite{Ani80}. Later calculations
solving the relativistic hydrodynamic equations indicated that the
highest energy densities would instead occur in the mid-rapidity
region with energy densities considered sufficient for QGP formation
at incident energies of around 400 GeV per nucleon~\cite{Kaj83}.

Almost concurrent with the suggestions to use relativistic heavy ions
as a means to produce the QGP in the laboratory were suggestions to
use dilepton or photon measurements to diagnose whether QGP has been
formed. First estimates considering only the lowest order elementary
processes and thermal parton distributions~\cite{Dom81,Kaj81}
concluded that the thermal dilepton emission rate from the QGP should
exceed that from a hadronic gas in the mass region below the $\rho$
\cite{Dom81}, and that real and virtual photons should provide
accurate information on the temperature of the plasma~\cite{Kaj81}.  A
simple counting estimate indicated an expected photon enhancement
relative to the number of pions in the case of QGP formation
\cite{Hal82}.  First calculations which performed the space-time
integration of the lowest order production rates by solving the
relativistic hydrodynamic equations confirmed~\cite{Chi82} that the
dilepton yield in the mass region below the $\rho$ was sensitive to
the initial temperature of the QGP and to the critical temperature.
Alternatively, it was suggested that the ratio of the simultaneously
observed photon and dilepton pair yield might provide a signal which
was sensitive to QGP formation while being insensitive to the details
of the collision dynamics~\cite{Sin83}.

While these initial estimates indicated that photons and dileptons
should be useful probes to diagnose the presence of QGP, the rate
estimates themselves were suspect since lowest order perturbative
calculations had been applied at energies, or temperatures, similar to
$\Lambda_{QCD}$, the QCD scale factor. The dilepton and photon rate
estimates were put on firmer ground when McLerran and Toimela
\cite{McL85}, following the suggestion by Feinberg~\cite{Fei76} that
the photon and dilepton rates could be determined from the expectation
value of the electromagnetic current correlation function,
demonstrated that for each order the emission rates had an invariant
form with thermal structure functions entering in a manner exactly
analogous to the usual structure functions for finding a quark or
gluon in a hadron. Also, it was observed that terms which contribute
to the dilepton or photon emission which are problematic at the basic
diagram level, are regularized in the QGP. For example, dilepton
emission from quark-antiquark annihilation is infrared divergent in
the limit of zero mass gluons while in the plasma gluons propagate as
plasma oscillations with a plasmon mass which provides a cutoff to
eliminate the divergences for small gluon momenta~\cite{McL85}.

Later, using the resummation techniques of Braaten and 
Pisarski~\cite{Pis89,Bra90}, Kapusta et al.~\cite{Kap91} demonstrated that the
photon emission rates of quark gluon matter and hadronic matter
were very similar. As a consequence, it could be concluded that 
while photon emission was not {\it per se} a signature of quark
gluon matter, detection of the emitted photons could provide
a good measurement of the temperature of the hot and dense matter.

Recently, the situation has changed again with the demonstration
by Aurenche et al.~\cite{prd:aur98} 
that the contribution to the photon emission rate 
from two-loop diagrams are significantly larger than the lowest
order contributions of 
the Compton ($q(\overline{q})g \rightarrow q(\overline{q})\gamma $)
and annihilation ($q\overline{q} \rightarrow g\gamma $) processes
which were previously thought to dominate the photon emission rate
from the quark matter. The two-loop diagrams were
shown to give a large bremsstrahlung
($qq(g) \rightarrow qq(g)\gamma $) contribution and a contribution
from a previously
neglected process of $q\overline{q}$ annihilation accompanied by
q(g) rescattering. This annihilation with rescattering process is 
found to dominate the photon emission rate of the quark matter at
large transverse momenta. Inclusion of these rates in hydrodynamic
model calculations of heavy-ion collisions has recently shown that 
photon yield from the quark matter may be significantly larger
than the photon yield from the hadronic matter~\cite{epj:sri99}. 
The direct photons may therefore dominantly carry information about
the quark gluon plasma.

A large body of data on prompt photon emission exists for
proton-induced reactions on targets of protons, anti-protons, and
light nuclei
\cite{prl:mcl83,zpc:bad86,prd:dem87,npb:ang86,plb:ada95,zpc:ann82,zpc:bon88,zpc:bon88pi,npb:ang89,sjn:aak90,prd:alv93,prl:apa98,prd:apa99,plb:bal98,prl:abe94}. 
The prompt photon 
measurements have provided important input on gluon
structure functions~\cite{rmp:owe87}. It is now
possible to perform complete and fully consistent next-to-leading
order (NLO) QCD calculations of the prompt photon cross 
sections.  In general, the prompt photon data can be
well described from fixed target energies up to Tevatron
energies~\cite{jpg:vog97} which provides an important
foundation for the intrepretation of direct photon production in
nucleus-nucleus collisions. 
In the past,  discrepancies with calculation
have sometimes been attributed to effects of intrinsic 
$k_T$ smearing arising
from higher order contributions such as soft-gluon 
emissions~\cite{prd:apa99,rmp:owe87,prd:hus95}. While the evidence for 
intrinsic $k_T$ effects remains under debate~\cite{epj:aur99},
the observed trend of an underestimated prompt photon yield at low
transverse momentum and low incident energy is suggestive of an
intrinsic $k_T$ effect~\cite{jpg:vog97,epj:aur99}. A thorough
understanding of the source of this discrepancy will be important
in the search for thermal direct photons at low transverse momentum
in nucleus-nucleus collisions at the presently available low 
incident energies.

First attempts to observe direct photon production 
in ultrarelativistic heavy-ion collisions with oxygen and sulphur
beams found no significant excess~\cite{Ake90,Alb91,Bau96,prl:alb96}.
The WA80 collaboration \cite{prl:alb96} provided the most interesting
result with a $p_{T}$ dependent upper limit on the direct
photon production in S+Au collisions at 200$A$GeV. 
This result was subsequently used by several authors to rule out a
simple version of the hadron gas scenario 
~\cite{prl:sri94,prc:neu95,prc:dum95,plb:arb95} and to establish 
an upper limit on the initial temperature of $T_{i} = 250 
\,\mathrm{MeV}$ \cite{prc:sol97}. In this paper, the
first observation of direct photons from ultrarelativistic
heavy-ion collisions is reported for central 158~{\it A}~GeV
 $^{208}$Pb\/+\/$^{208}$Pb collisons.
The implications of the result are discussed.

The organization of the paper is as follows: A description of the
WA98 experimental setup including the event selection and photon
spectrometer are presented in the next section. A general description
of the WA98 direct photon analysis method is given in Sec.~III.
The details of the data analysis including a presentation 
of the various corrections and their associated errors for extraction
of the inclusive photon, $\pi^0$, and $\eta$ yields is given in 
Sec.~IV.  The final inclusive photon, $\pi^0$, and
$\eta$  distributions are presented in Sec.~V.
The direct photon result is also presented in Sec.~V
and the results are compared to calculation and discussed. 
A summary and conclusion is given in Sec.~VI.

\section{\label{sec:wa98} WA98 EXPERIMENTAL SETUP}

\begin{figure}[h]
\begin{center}
   \includegraphics[scale=0.9]{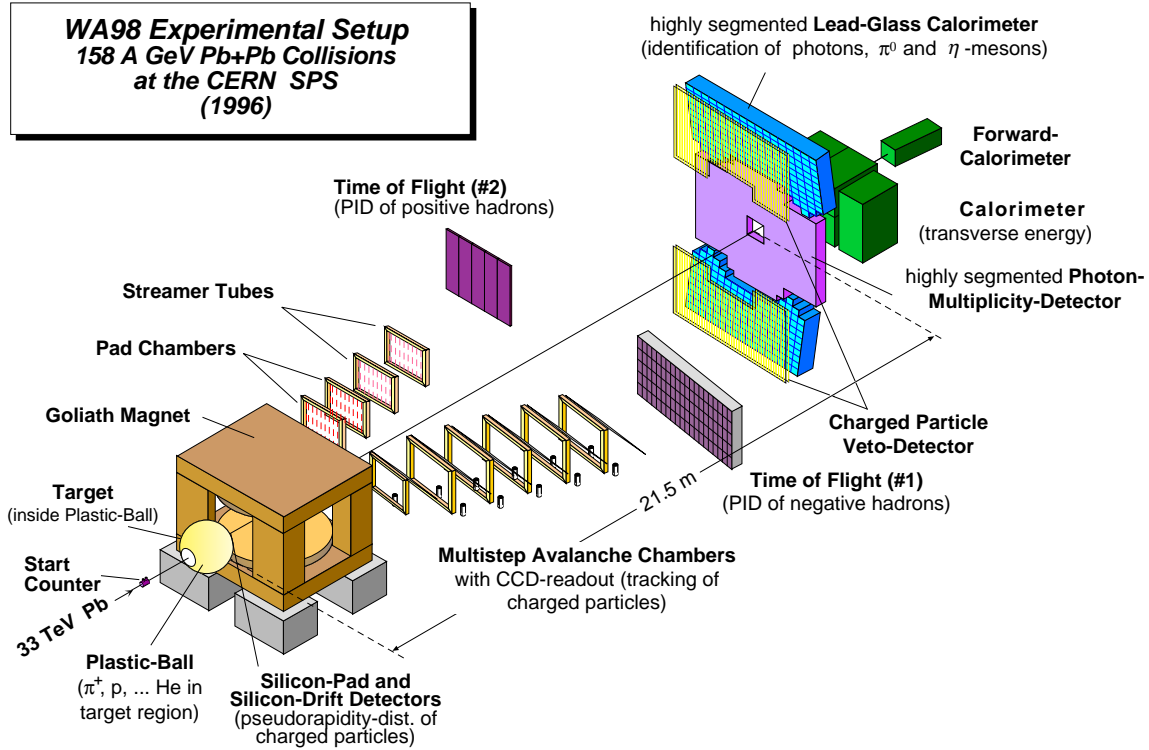}
\caption{The WA98 experimental setup.}
\label{fig:wa98}
\end{center}
\end{figure}

The CERN experiment WA98 is a general-purpose apparatus which consists
of large acceptance photon and hadron spectrometers together with
several other large acceptance devices which allow to measure various
global variables on an event-by-event basis. The experiment took data
with the 158~{\it A}~GeV $^{208}$Pb beams from the SPS in 1994, 1995,
and 1996. The results presented here were obtained from analysis of
the 1995 and 1996 data sets.  The layout of the WA98 experiment as it
existed during the final WA98 run period in 1996 is shown in
Fig.~\ref{fig:wa98}.

%%%fig1%%%

\subsection{\label{sec:detectors} Detector Subsytems}

Each 158~{\it A}~GeV $^{208}$Pb beam particle is qualified in a series
of trigger counters located upstream of the target. The $^{208}$Pb
target is mounted in a thin target wheel at the center of a 20 cm
diameter spherical thin-walled aluminum vacuum chamber located within
the Plastic Ball detector.  The target wheel has 5 target positions,
one of which was empty for non-target background measurements.

The Plastic Ball detector consists of 655 modules which provide energy
measurement and particle identification of charged pions and light
particles by $\Delta E-E$ measurement~\cite{nim:bad82}.  The Plastic
Ball detector provides particle measurement over the interval $-1.7 <
\eta < 1.3$.  Each module comprises a slow 4~mm thick CaF$_2$ $\Delta
E$ scintillator followed by a fast plastic scintillator readout by a
common photomultiplier.  The signals from the photomultipliers of the
two forward-most rings of Plastic Ball modules, subtending the angular
region from $30^\circ$ to $50^\circ$, are split and a portion of the
signals are analog summed to provide an online energy signal for
trigger purposes. This signal is used to suppress interactions
downstream of the target.

The target vacuum chamber is extended downstream in a $30^\circ$
conical vacuum chamber which contains the Silicon Drift Detector (SDD)
and the Silicon Pad Multiplicity Detector (SPMD)~\cite{nim:lin97},
each consisting of 300 ${\mu}$m thick silicon wafers. The SDD and SPMD
are located 12.5~cm and 30~cm downstream from the target,
respectively.  These detectors provide charged particle multiplicity
measurement over the intervals of $2.5 < \eta < 3.75$ and $2.3 < \eta
< 3.75$, respectively.

Charged particle momentum measurement and particle identification is
accomplished using two tracking spectrometer arms. The momentum
measurement is accomplished by magnetic analysis in a large (1.6~m)
aperture dipole magnet called GOLIATH which provides 1.6~Tm bending
power. Both tracking spectrometers use straight-line tracking outside
the magnetic field. Particle identification is obtained using
time-of-flight measured with scintillator slat detectors in each
tracking arm.  In the normal magnetic field configuration the
negative tracks are deflected to the right, looking downstream, into
the first tracking arm. The first tracking arm consists of six
planes of multi-step avalanche chambers~\cite{nim:rub95}. The active
area of the first tracking chamber is $1.2 \times 0.8$~m$^2$ while
that of the other five chambers is $1.6 \times 1.2$~m$^2$. The
chambers produce UV photons by means of a photoemissive vapor 
which are then converted to visible light via wavelength shifter
plates. On exiting the chamber, the visible light is reflected
$45^\circ$ by thin-foil mirrors to CCD cameras equipped with
two-stages of image intensifiers. Each CCD pixel viewed a chamber area
of about $3.1 \times 3.1$~mm$^2$.

The second tracking arm measures positive-charged tracks in the normal
field condition. It consists of four chambers of $1.6 \times
1.2$~m$^2$~\cite{nim:car99}. The first two chambers are multi-step
avalanche chambers similar to those of the first tracking arm, but
with the avalanche signal collected directly on an anode plane with
pad readout. In total about 35000 pads per chamber are read out.  The
last two tracking chambers consist of streamer tubes read out with
6000 pads each. The second tracking arm was installed and operated for
the 1996 run period only.

The Photon Multiplicity Detector (PMD)~\cite{nim:agg96} is located at
a distance of 21.5~m downstream from the target. The PMD is a large
(21~m$^2$) preshower detector consisting of 3.3 radiation lengths of
lead used to convert and count photons. The lead converter is backed
by over 50000 scintillator tiles individually wrapped and readout via
wavelength shifter optical fibers coupled in groups to a set of CCD
cameras with image intensifiers. The PMD provides a photon
multiplicity measurement over the interval $2.8 < \eta < 4.4$.

The WA98 photon spectrometer comprises the LEad-glass photon Detector
Array (LEDA), and a charged particle veto detector. The photon
spectrometer is divided into two halves placed above and below the
beam plane to benefit from the charge-sweeping effect of the GOLIATH
magnet, and is located at about the same distance as the PMD. It
provides photon energy measurement over the interval $2.4 < \eta <
3.0$.  The photon spectrometer is described in more detail below.

Further downstream, at a distance of 24.7~m from the target, the total
transverse energy is measured in the MIRAC calorimeter
\cite{nim:awe89}.  The MIRAC is a sampling calorimeter with 180
calorimeter towers readout on two sides with wavelength shifter plates
coupled to photomultipliers.  Each tower is segmented longitudinally
to provide separate measurement of the electromagnetic and hadronic
energy deposit.  A portion of the signal from each photomultiplier of
MIRAC is split off and the analog signal is 
summed with appropriate weight to form a
total transverse energy signal for trigger purposes.  The MIRAC is
deployed in a rectangular wall 3.3~m wide by 2.4~m high centered on
the beam axis with a central aperture 61~cm wide by 23~cm high through
which the beam passes.  A portion of the MIRAC coverage overlaps
with the PMD preshower detector.  The MIRAC provides total transverse
energy measurement with varying azimuthal coverage over the interval
$3.2 < \eta < 5.4$.

Finally, the total energy of the uninteracting beam, or of the
residual beam fragments and produced particles emitted near to zero
degrees, is measured in the Zero Degree Calorimeter (ZDC). The ZDC
consists of 35 lead/scintillator sampling calorimeter modules of $15
\times 15$~cm$^2$ cross sectional area each.  For each module the
scintillator is read out from the side with a fast wavelength shifter
plate coupled to a photomultiplier with an active base. This allowed
stable operation with intensities up to 1 MHz of the full 33 TeV
$^{208}$Pb beam.  The ZDC modules were stacked in an array 7 modules
wide by 5 modules high. Since it serves as
the WA98 beam stop the ZDC is located in a
shielded cave (not shown in Fig.~\ref{fig:wa98}) for radiological
protection reasons. The cave is located just behind MIRAC with an
entrance aperture the same size as the aperture through MIRAC.

In order to minimize backgrounds, WA98 has been designed with
attention to minimize the amount of material in the beamline and in
the flight paths of the detected particles.  Thus, except for trigger
detectors in the beamline and a small air gap in the GOLIATH magnet,
the beam is transported in an evacuated beampipe  
(not shown in Fig.~\ref{fig:wa98}) from the
point of extraction from the SPS through the entire experiment until
just before being stopped in the ZDC.  A trapezoidal chamber extends
the vacuum beyond the silicon detectors to the entrance of the GOLIATH
magnet. It ends with a $1.4 \times 1.$~m$^2$ exit window of 125
$\mu$m thick mylar suspended by a kevlar mesh of 240 $\mu$m average
thickness. For the 1995 run period a 2.5 mm thick aluminum ring of 15
mm diameter with a 11 mm diameter hole was attached to the exit window
at the location of the beam exit.  The thick mylar and kevlar mesh was
removed from the ring aperature and replaced by a thin mylar foil.
While the purpose of the thin foil had been to reduce downstream
interactions, the ring caused significant interactions from the beam
halo and so the exit window was replaced with a homogeneous mesh and
mylar foil for the 1996 run period.  After a 75~cm air gap, the vacuum
continues with a series of 0.5~mm thick carbon fiber beam tubes.  The
first beam tube is of 5~cm diameter and 1.44~m length followed by a
second tube of 10~cm diameter and 4.5~m length.  A third beam tube
continues through the experiment with a 20~cm diameter to the front of
the MIRAC calorimeter where it attaches to a rectangular vacuum pipe
which defines the aperture in MIRAC and terminates the vacuum at the
rear of MIRAC just before the entrance to the ZDC.

The PMD, SDD, and tracking spectrometers are not used in the
present analysis and will not be discussed further. Additional details
on event selection and on the photon spectrometer are given next.

\subsection{\label{sec:trigger} Event Selection}

The WA98 trigger detectors comprise a nitrogen gas \v{C}erenkov
counter~\cite{nim:chu96} to provide a fast start signal ($\leq~30$~ps
time resolution), beam-halo veto counters, and the MIRAC calorimeter.
A clean beam trigger is defined as a signal in the start counter,
located 3.5 m upstream of the target, with no coincident signal in the
veto scintillator counter (which had a 3 mm diameter circular hole and
was located 2.7 m upstream of the target), or in beam halo
scintillator counters which covered the region from the veto counter
to 25 cm transverse to the beam axis.  Beam fragments from upstream
interactions are rejected by use of a high threshold on the start
counter signal, set just below the $^{208}$Pb signal. Short timescale
pileup events are vetoed by an anti-coincidence requirement with a
higher threshold start signal, set just above the $^{208}$Pb signal.
Additional background event rejection is performed offline using the
amplitude and timing information from the trigger detectors. For
purposes of background rejection each of the trigger logic signals is
copied multiple times and recorded on TDCs with various delayed starts
or delayed stops which allow to inspect the time period immediately
preceeding or following the trigger event. This set of TDCs allows to
reject pileup beam particles or interactions over preceeding or
following time ranges of 100~ns, 500~ns, or 10~$\mu$s in the offline
analysis.

The MIRAC calorimeter provides an analog total transverse energy sum
for centrality selection for online trigger purposes.  The WA98
minimum bias trigger requires a clean beam trigger with a MIRAC
transverse energy signal which exceeds a low threshold.  Two
additional trigger signals are derived from the MIRAC transverse
energy signal using thresholds set somewhat above and far above the
minimum bias threshold. These three MIRAC thresholds define three
non-overlapping event classes which are used to define the WA98
physics triggers.  The thresholds were adjusted such that the
so-called peripheral event class, between the lowest and
next-to-lowest thresholds, corresponded to about 20\% of the minimum
bias event rate and the central event class, above the highest
threshold, corresponded to about 10\% of the minimum bias event rate.
The remaining $\approx 70\%$ of the minimum bias cross section between
central and peripheral event classes is referred to as the
not-so-central event class. Taken together the three event classes
were equivalent to the minimum bias event class. In normal run
operation the central event class triggers were taken without prescale
factor while the peripheral event triggers were typically downscaled
by a factor of two and the not-so-central triggers were usually
prescaled by a factor of 32 (after deadtime suppression) for the 1995
run period. For the 1996 run period the peripheral and not-so-central
event classes were typically downscaled by a factor of 4 and 16,
respectively.  Downscaled beam triggers and in-spill pedestal triggers
were also taken at a low rate as well as various out-of-spill
calibration triggers for monitoring and calibration purposes for the
various detectors.  In order to obtain absolute cross section
information, all trigger logic signals were counted with scalers
before and after deadtime suppression, and after application of
downscale factors. The scalers were recorded between spills.

In order to obtain the maximum data rate for the direct photon
measurement, WA98 was operated with three different event types.  The
event types were distinguished by different groups of detectors with
different readout deadtimes, varying up to about $5$ms, $10$~ms,
or $15$~ms for event types one, two, or three, respectively. Event
type one included the trigger detectors, MIRAC, ZDC, Plastic Ball, and
the photon spectrometer.  Event type two also included the PMD, SPMD,
and SDD. Event type three further included the tracking spectrometers.
Zero-suppressed data volumes of about 50 kbyte/event were produced for
central collisions. The experiment operated with a typical beam
intensity of $\approx 0.5$ MHz $^{208}$Pb delivered to target over an
effective SPS spill of about 2.5 s during the 14.4 s machine cycle.
About 250 events were recorded per spill with a typical deadtime of
about 80\%.

\subsection{\label{sec:ledaspec} Photon Spectrometer}

The WA98 photon spectrometer consists of a large area lead-glass
detector array, LEDA, supplemented with a charged particle veto (CPV)
detector placed immediately in front of it (see Fig.~\ref{fig:wa98}).
The spectrometer has an unobstructed view of the target through the
vacuum chamber exit window at the entrance to the GOLIATH magnet.  The
photon spectrometer is separated into two nearly symmetric halves
above and below the beam plane in the two regions of reduced charged
particle density which result from the sweeping action of the GOLIATH
magnet. 
%%tca The lower half of the detector is slightly smaller due to
%%tca space limitations imposed by the floor of the experimental hall. 
The two detector halves are inclined by an angle of $8^\circ$ such that
photons near the center of the detector impinge with normal incidence.
The maximum deviation from normal incidence to the detector surface is
less than $9^\circ$ at the detector corners.  The perpendicular
distance to the front surface of the lead-glass is 22.1~m. This
distance was chosen to allow the photon measurement near mid-rapidity
while maintaining a maximum local particle hit occupancy below 3$\%$,
which is necessary to insure that overlapping shower effects remain
manageable.

\begin{figure}[hbt]
\begin{center}
   \includegraphics[scale=0.4]{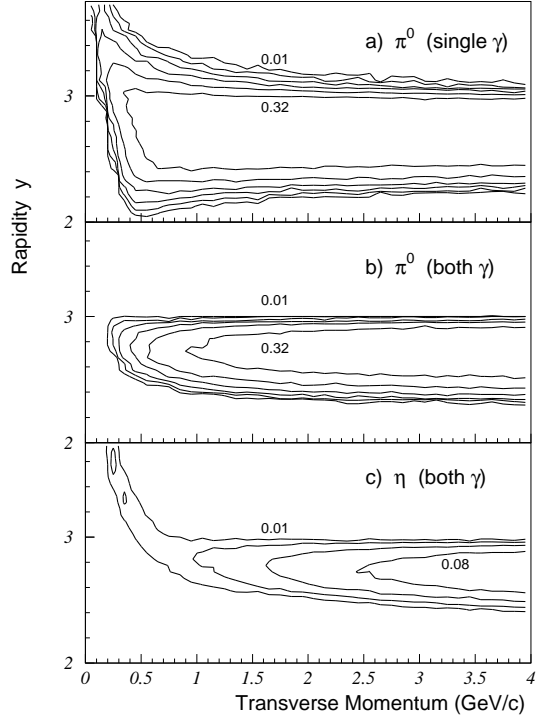}
\caption{The acceptance of the WA98 photon spectrometer in parent
   particle rapidity versus transverse momentum for a) single photons
   from $\pi^{0}$ decay and for photon-pairs from b) $\pi^{0}$ decay or
   c) $\eta$ decay with a 750 MeV photon energy threshold. The contours
   show the fraction of full phi acceptance in steps of factors of two.
   }
\label{fig:acceptance}
\end{center}
\end{figure}

The acceptance of the photon spectrometer for $\pi^{0}$ and $\eta$
detection, in rapidity and transverse momentum of the $\pi^{0}$ or
$\eta$, is shown in Fig.~\ref{fig:acceptance}. The acceptance is
calculated for a 750~MeV photon energy threshold. The acceptance in
part a) is shown for detection of a single photon from the decaying
$\pi^{0}$.  It indicates the phase space region over which $\pi^{0}$'s
contribute photons into the acceptance of the spectrometer.  The
acceptance for simultaneous detection of both photons from the
$\pi^{0}$ and $\eta$ two-photon decay branch is shown in parts b) and
c), respectively. The acceptance covers the region $2.4 < y < 3.0$,
near mid-rapidity ($y_{cm}=2.9$).

%%fig2%%
\vskip-0.4in

\subsubsection{\label{sec:leda} Lead-Glass Detector}

The lead-glass detector comprises 10,080 individual lead-glass
modules. Each module is a $4 \times 4 \times 40$~cm$^3$ (14.3
radiation lengths) TF1 lead-glass block with photomultiplier readout.
The sides of each block are wrapped in an aluminized mylar reflective
foil and sealed in a PVC plastic shrink tube of 0.15~mm wall
thickness.  Twenty-four lead-glass modules are epoxied together in an
array 6 modules wide by 4 modules high to form a super-module. Each
super-module has its own calibration and gain monitoring system based
on a set of 3 LEDs mounted inside a sealed reflecting front cover dome
\cite{nim:pei96}. Each lead-glass module views the reflected LED light
through an aperture on the front surface, while the LED light is
simultaneously monitored by a PIN-photodiode.

All 10,080 lead-glass modules were calibrated with 10 GeV electrons in
the X1 beamline in the west area of the CERN SPS during the period of
fall 1993 to spring 1994. The calibration beam was used to determine
the GeV equivalent of the photodiode-normalized LED light viewed by
each lead-glass module. The LED system allowed the calibration to be
maintained after the lead-glass was installed in the WA98 experimental
area with a new readout system. The energy and position resolution,
and the non-linearity of the lead-glass detector were measured in the
same test beam using electrons of incident energies from 3 GeV to 20
GeV. The measured energy resolution could be parameterized as
\cite{nim:pei96}
\begin{equation}
\sigma/E = (5.5 \pm 0.6)\% / \sqrt{E} + (0.8 \pm 0.2) \%
\label{e_res}
\end{equation}
and the measured position resolution could be parameterized as
\begin{equation}
\sigma_x = (8.35 \pm 0.25) {\rm ~mm} / \sqrt{E} + (0.15 \pm 0.07)
{\rm ~mm}
\label{x_res}
\end{equation}
with energy measured in GeV.

Each lead-glass module is read out by an FEU-84 photomultiplier with
individually controlled high voltage. The high voltage is generated
on-base with custom developed \cite{nim:neu95} Cockcroft-Walton
voltage-multiplier type bases. The bases are controlled using a VME
based processor and controllers.  
The photomultiplier signals are digitized with a
custom-built ADC system \cite{nim:win94} which was installed in the
fall of 1994. The system features a fast shaping amplifier with dual
gain ranges separated by a factor of 8 in gain.  Each gain range is
digitized with 10-bits resolution for 13-bits of effective dynamic
range. The ADC system includes an analog memory in which the
integrated signal is sampled and stored at 20 MHz in a ring buffer 16
cells deep. The analog memory provides the latency needed ($\approx
400$~ns) for the WA98 trigger decision without the need for cable
delay of the photomultiplier signals.  The readout system also
includes a constant fraction discriminator with TAC for time-of-flight
measurement, and overlapping module current sums for possible trigger
purposes, neither of which are used for the present analysis.

\begin{figure}[hbt]
\begin{center}
   \includegraphics[scale=0.5]{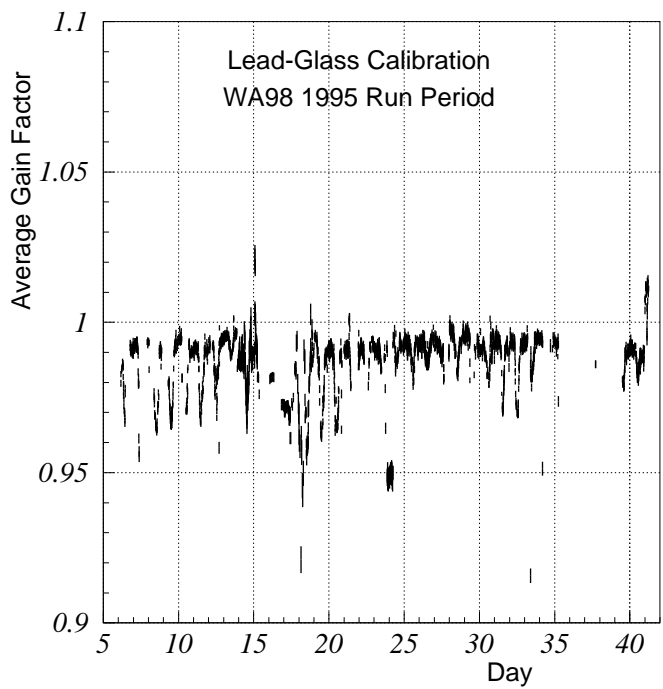}
\caption{The stability of the lead-glass calibration as determined
   with the LED monitoring system. The time-dependent gain factor,
   averaged over all modules, is shown as a function of time for the
   entire 1995 run period. The vertical size of the points indicates
   the rms of the distribution of time-dependent gains of all modules.}
\label{fig:stability}
\end{center}
\end{figure}

After calibration and installation in WA98 with the new readout
system, the high voltage of each module was adjusted to set the
full-scale ADC value at 40 GeV, based on the GeV-equivalent of the
calibrated LED light. During the period of datataking, the LED system
was pulsed and all lead-glass modules and photodiodes were read out
and recorded between spills at a frequency of a few Hz.  These
calibration events were used offline to provide time-dependent gain
correction factors for each module. The gain correction factors were
stored in a database and applied on a run-by-run basis in the offline
analysis. The overall stability of the lead-glass system is indicated
in Fig.~\ref{fig:stability} where the time-dependent gain correction,
averaged over all lead-glass modules, is plotted as a function of time
during the 1995 run period.  The rms of the distribution of module
gains at a given time is indicated by the vertical bars. The smallness
of the rms values throughout the run period demonstrates 
the stability of the high
voltage and readout systems while the diurnal variation of the average
gain factors suggests a sensitivity of the photomultiplier gains to
the temperature in the experimental hall. Fig.~\ref{fig:stability}
gives a good indication of the magnitude and importance of the
time-dependent gain corrections which have been applied.

%%fig3%%

\subsubsection{\label{sec:ledacpv} Charged Particle Veto Detector}

In order to tag and thereby directly deduce the fraction of 
showers observed in the lead-glass originating from charged hadron, the photon
spectrometer is supplemented by a Charged Particle Veto (CPV) detector
which covers the lead-glass region of acceptance~\cite{the:rey99}.  
The two sections of
the CPV each consist of 86 Iarocci-type plastic streamer tubes in a
single layer.  Each tube is tilted by $30^\circ$ to avoid normal
particle incidence which would result in a 7\% geometrical
inefficiency due to the streamer tube walls.  The streamer tubes are
operated with a gas mixture of 10\% argon, 30\% isobutane, and 60\%
carbon dioxide at atmospheric pressure.  A streamer discharge induces
a charge signal on externally mounted pads which have a size of 42~mm
$\times$ 7~mm.  Groups of 16 pads are connected to a charge sensitive
chip which converts a charge signal into a 6 bit ADC value.  In total
49120 pads and 3070 chips are necessary to read out the 19~m$^2$
active area of the CPV.  The CPV detector is nearly transparent to high
energy photons with only 2.0\% of incident photons converting and 
producing detectable signals inside the streamer tubes.  The CPV was under
construction at the time of the 1995 $^{208}$Pb run and was fully
operational only for the 1996 run period.  As a result, the CPV has
been used in the analysis of the 1996 data set only.

By employing magnetic field off data, with straight-line trajectories
from the target, the silicon pad multiplicity detector can be used
together with the lead-glass detector to determine the average CPV
efficiency {\it in situ}. The SPMD consists of four quadrants each
divided into 1012 pads with 46 azimuthal divisions and 22 radial
divisions. Each pad has roughly equal size in $\Delta{\eta} \times
\Delta{\phi}$ of about $0.065 \times 2^\circ$. The efficiency for
detecting a charged particle in the SPMD was measured in a test beam
to be better than $99\%$.
%By employing the SPMD and lead-glass as reference detectors
%to find charged particle
%tracks which have traversed the CPV, its efficiency has been determined
%in the low multiplicity environment of $p\/+\/Pb$ reactions to be 96~\%.

\section{\label{sec:method} DATA ANALYSIS METHOD}

In this section we discuss the WA98 direct photon analysis method. An
overview of the method is presented followed by a discussion of
the details of the photon identification criteria and the photon and
$\pi^0$ efficiency determination. These are the main sources of
systematical error in the present direct photon analysis. Next, the method to
determine the charged particle contamination in the photon yield is
discussed.  Then the calculation of the photon background expected
from radiative decays of long-lived resonances is described. Finally,
the extraction of the direct photon excess is discussed.

\subsection{\label{sec:overview} Direct Photon Analysis Overview}

Due to the high photon multiplicity in central Pb+Pb collisions, and
the limited photon spectrometer acceptance, it is not feasible to
identify isolated single direct photons on an event-by-event basis.
Instead, in the direct photon analysis presented here, the transverse
momentum distribution of direct photons is determined on a statistical
basis. In brief, the direct photon excess is extracted from the
difference between the measured inclusive photon yield and the photon
yield predicted from a calculation of the radiative decays of
long-lived resonances. Among such decay photons, the $\pi^0$ and
$\eta$ comprise the largest source, contributing roughly 97\% of the
photon yield according to the expected relative abundances of produced
particles (see Fig.~\ref{fig:photon_pt_bkg}).  Therefore, in order to
maximize the sensitivity of the measurement to a direct photon excess,
it is imperative to accurately determine the $\pi^0$ and $\eta$ yield.

In the WA98 measurement, the $\pi^0$ and $\eta$ yield are determined
via their two-photon decay branch for exactly the same event sample
for which the inclusive photon yield is measured. This eliminates all
systematical error sources related to absolute cross section
normalization or centrality selection. In fact, this analysis method
allows to determine the decay background correctly even if data
sets with very different centralities or run conditions were combined
arbitrarily since the averaged decay photon
distribution would follow directly from the averaged $\pi^0$ and
$\eta$ distributions. 
Thus, for example, contributions from background sources such
as a secondary target will not produce an apparent photon excess in
this analysis, as long as their contribution to the $\pi^0$ and $\eta$
yield can be extracted from the two-photon invariant mass peaks. Such
background sources would distort the extracted $\pi^0$ and $\eta$
transverse momentum distributions, but this distortion would also be
reflected in the inclusive photon distribution. Similarly, a
distortion of the photon momentum distribution due to a calibration
error or non-linearity of the detector response would be reflected in
the momentum (and mass) distribution of the reconstructed $\pi^0$'s
and $\eta$'s. This means that the sensitivity of the direct photon
search to detector calibration or non-linearity errors is reduced in
this analysis. Furthermore, the momentum dependence of the $\pi^0$
invariant mass peak provides an {\it in situ} means to verify and
quantify the accuracy of the detector calibration.

A major source of systematical error in the present analysis is the
determination of the photon detection efficiency.  Roughly speaking,
since the photon detection efficiency enters quadratically in the
efficiency correction of the $\pi^0$ yield extracted via its
two-photon decay, but only linearly in the photon yield correction, an
error in the photon detection efficiency directly modifies the
apparent photon excess.  
%%tca For example, if the photon detection
%%tca efficiency is overestimated, the corrected inclusive photon yield will
%%tca be underestimated. At the same time the corrected $\pi^0$ yield would
%%tca then be even further underestimated due to its quadratic dependence on
%%tca the photon efficiency. This would result in a greater underestimate of
%%tca the decay photon yield and hence an apparent direct photon excess.
Thus a major emphasis of the present analysis is to demonstrate an
accurate determination of the identification efficiencies and
associated systematical errors. This is accomplished by applying
different photon identification criteria having very different efficiencies
and sensitivities to backgrounds, and verifying that the
final corrected results are consistent in all cases.

Another source of error may be due to mis-identified non-photon
backgrounds.  Since the number of charged hadrons exceeds the number
of photons by about a factor of three at large transverse momenta,
they pose a large potential background of apparent excess photons if
mis-identified as photons. Fortunately, high energy hadrons deposit
only a small fraction of their incident energy in the lead-glass
detector (its 40~cm length is about one interaction length). As a
result, showers with large energy deposit, or large apparent
transverse momentum, are predominantly photons with a hadron
contamination of only about $10\%$. Since hadronic showers typically
have large transverse dimension the hadron shower contamination can
be further reduced by a factor of 2-3 by excluding showers with large
width. Since the magnetic field alters the distribution and apparent
transverse momenta of the charged hadrons while leaving the neutral
particle distributions unchanged, a comparison of the extracted
neutral shower result for magnetic field on and field off provides a
consistency check of the charged hadron rejection.  The Charged
Particle Veto detector is used to determine the charged hadron
contribution to the photon spectrum.

Another source of apparent photons is neutrons and anti-neutrons. This
contribution is estimated by simulation only. Its contribution can
similarly be reduced by excluding showers with large width.
Consistency in the final result with different shower identification
criteria and run conditions provides confirmation that the background
contributions are properly eliminated.

\subsection{\label{sec:pid} Particle Identification and Yield Determination}

The most critical requirement of the direct photon search is an
accurate determination of the inclusive photon and $\pi^0$ yields. In
general, the accuracy of the yield determination is verified by using
different identification criteria with large differences in efficiency
and background sensitivity, and demonstrating consistent final
results.  The $\pi^0$ yield is largely insensitive to background
particles since the $\pi^0$'s are self-identified by their mass peak
in the two-photon invariant mass spectrum. The effect of background
particles is mainly to increase the combinatorial background in the
invariant mass spectrum which makes the problem of extraction of the
peak content more difficult. On the other hand, charged hadrons and
neutrons are significant backgrounds to the photon yield
determination.  In the present analysis, the Charged Particle Veto
detector is used to identify charged hadron showers and remove the
charged hadron contribution from the photon spectrum. At the same
time, care must be taken not to remove converted photons from the
photon or $\pi^0$ data.

The photon yield extraction involves the following steps:
\begin{itemize}
\item[$\Gamma$-1.] The photon identification criteria are applied to
   the reconstructed showers and a photon candidate $p_T$ spectrum is
   accumulated.
\item[$\Gamma$-2.] The normalized target-out background photon
   candidate $p_T$ spectrum is subtracted, if necessary.
\item[$\Gamma$-3.] The CPV detector is used to determine the charged
   hadron contamination included in the photon candidate spectrum. The
   charged shower contribution is subtracted from the photon candidate
   distribution to produce the uncorrected neutral shower $p_T$
   spectrum.
\item[$\Gamma$-4.] The neutral shower spectrum is corrected for photon
   conversions and for the reconstruction efficiency.
\item[$\Gamma$-5.] The neutron and anti-neutron contamination, based
   on simulation, is removed to produce the raw photon $p_T$ spectrum,
   within the lead-glass detector acceptance.
\item[$\Gamma$-6.] The raw photon spectrum is corrected for the
   geometrical acceptance to produce the final photon $p_T$ spectrum.
\end{itemize}

The $\pi^0$ (or $\eta$) yield extraction involves the following steps:
\begin{itemize}
\item[$\Pi$-1.] The photon identification criteria are applied to the
   reconstructed showers to produce a list of photon candidates for
   each event.
\item[$\Pi$-2.] The invariant mass of each photon pair within an event
   is calculated and sorted into invariant mass histograms according to
   the $p_T$ of the photon pair. An invariant mass histogram is
   accumulated for each $p_T$ bin to be used in the final $\pi^0$ $p_T$
   spectrum.
\item[$\Pi$-3.] The photons are simultaneously used to construct 
   artificial mixed
   events of similar multiplicity for each centrality class. The mixed
   events are analyzed in exactly the same manner as the real events to
   produce background invariant mass spectra as a function of $p_T$.
\item[$\Pi$-4.] The final mixed event invariant mass spectra are
   normalized and subtracted from the the final real event invariant
   mass spectra to remove the combinatorial background from the real
   event spectra.
\item[$\Pi$-5.] The normalized target-out final invariant mass spectra
   are subtracted from the final invariant mass spectra, if necessary.
\item[$\Pi$-6.] The final invariant mass spectra are analyzed to
   extract the content in the $\pi^0$ (or $\eta$) peak at each $p_T$.
   The result is the uncorrected $\pi^0$ $p_T$ spectrum.
\item[$\Pi$-7.] The $\pi^0$ $p_T$ spectrum is corrected for the
   $\pi^0$ reconstruction efficiency and losses due to photon
   conversions to produce the raw $\pi^0$ $p_T$ spectrum within the
   lead-glass detector acceptance.
\item[$\Pi$-8.] The raw $\pi^0$ spectrum is corrected for the
   geometrical acceptance and photon energy threshold to produce the
   final $\pi^0$ $p_T$ spectrum.
\end{itemize}

For both the photon and $\pi^0$ analysis, the shower reconstruction
itself involves the following steps \cite{nim:ber92}:  First, all
lead-glass detector modules with energy deposit are analyzed and
contiguous modules are associated together as a cluster. The list of
clusters is then analyzed to determine the number of local maxima in
each cluster.  Clusters with a single maximum are treated as single
showers. Clusters with multiple maxima are assumed to result from
overlapping showers, with one shower per maximum. The energy deposit in
each module in the cluster is partitioned to the overlapping showers
according to the distance of the module from the shower maxima,
assuming all showers to have electromagnetic radial shower profiles.
The individual showers are then analyzed to calculate the total shower
energy, position, and spatial dispersion (width) \cite{nim:ber92}.  The
shower positions are calculated with a logarithmic weighting of the
energy deposit \cite{nim:awe92} and are projected to the front surface
of the lead-glass detector, correcting for the shower depth and the
non-projective geometry. The distance from the shower position to the
nearest hit in the CPV is also extracted.  All of this information is
recorded on Data Summary Tapes (DSTs) as an intermediate analysis
step.

After application of the minimum energy threshold of 750 MeV used in
the analysis and acceptance calculations, the showers are subject to
various sets of further identification criteria.  The different
criteria result in varying non-photon background contaminations and
photon ($\pi^0$) identification efficiencies, which must then be
determined. In order to avoid shower distortions near the detector
edges, or around ``dead'' detector modules, an edge cut is applied to
require that the reconstructed shower position lies beyond a specified
distance from the detector edges or dead modules. A distance cut of
two module widths from the detector edge and 1.5 modules widths from
the center of a dead module was used.

In the present analysis, the photon selection has been made with the
following shower identification criteria:
\begin{itemize}
\item[S1.] Use all reconstructed showers.
\item[S2.] Use only narrow showers which have a dispersion (width) which
   is less than a specified value.
\item[S3.] Use only showers which have no associated CPV hit.
\item[S4.] Use narrow showers satisfying the dispersion cut with no
   CPV hit.
\end{itemize}
The first condition will have the highest photon identification
efficiency but largest non-photon background contribution, while the
last condition will have the lowest efficiency but lowest background
contributions.  For the extraction of the photon yield, the criteria
S1 and S2 are not entirely independent from the criteria S3 and S4
since the CPV detector is used in the first case also to determine the
charged hadron contamination. The distinction is mainly in the manner in
which the data is processed and in how the corrections are applied.
In particular, with criteria S3 and S4 the neutral shower distribution
is acquired directly, but must be corrected for photons which were
rejected due to random associated hits in the CPV, or due to photon
conversions, while for criteria S1 and S2 the charged shower
distributions are extracted using the CPV and corrected for random
associated hits and conversions and then the corrected charged shower
contribution is removed from the total shower distribution to obtain
the neutral shower distribution.  For the $\pi^0$ yield extraction the
CPV is not used at all when criteria S1 or S2 are used with the result
that no corrections for random CPV hits are needed and smaller
conversion corrections are required.  Criteria S2 and S4 make use of
the fact that the transverse size of hadronic showers, with large
energy deposit, is significantly greater than that of electromagnetic
showers in the lead-glass calorimeter. The cut on the shower
dispersion is chosen to accept more than $99\%$ of the
isolated photon showers while rejecting hadron showers by a factor of
2-3.  However, the shower dispersion cut is more likely to lose
electromagnetic showers in the case of shower overlap.

For additional consistency checks, the data has also been analyzed
with the shower energy threshold increased from 750 MeV to 1.5 GeV,
with the outer edge module cut increased to three module widths, and
with a photon energy asymmetry cut applied in the $\pi^0$ analysis.

\subsection{\label{sec:pideff} Particle Reconstruction Efficiency}

The large particle multiplicities in central Pb\/+\/Pb collisions 
result in
module occupancies in the WA98 lead-glass detector of up to $20\%$, 
which poses a special problem for the direct photon search.  These
large occupancies result in overlapping showers in which photons may
be lost, mis-identified, or significantly altered in position or
energy. This results in a significant dependence of the photon and
$\pi^0$ identification efficiency on the centrality of the collision.
Furthermore, the position and energy resolution, and even the energy
scale, will be centrality dependent due to the effect of shower
overlap. For an accurate direct photon search it is imperative to
accurately determine and account for these effects. For the present
analysis this has been accomplished by the method of randomly
inserting test showers into real events and studying how they are
altered and the efficiency with which they are recovered. This
procedure has been used to determine the $\gamma$, $\pi^0$, and $\eta$
reconstruction efficiency.

\begin{figure}[h]
\begin{center}
   \includegraphics[scale=0.4]{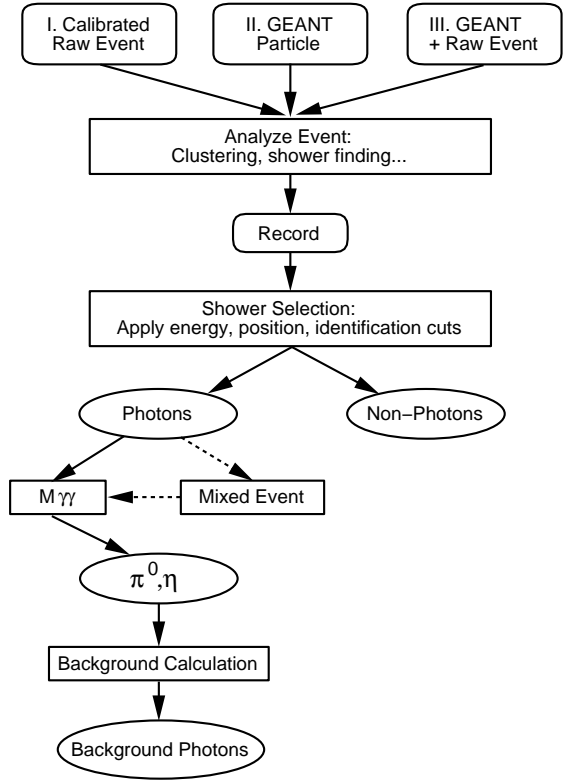}\vspace{0.5in}
\caption{An outline of the analysis procedure using the
   method of GEANT shower overlap used for the photon and $\pi^{0}$
   efficiency determination. The intermediate results are recorded to
   DSTs after the event analysis step. The shower selection and further
   analysis is performed on the DSTs.}
\label{fig:geant}
\end{center}
\end{figure}

For this reconstruction procedure the WA98 experiment geometry was
implemented in GEANT \cite{crn:geant} with the GEANT tracking
parameters for LEDA adjusted to reproduce test beam measurements of
the LEDA response to electrons.  The generation and transport
of \v{C}erenkov photons in GEANT was parameterized and this
parameterized \v{C}erenkov response was used in the full WA98 GEANT
simulation due to the prohibitive CPU-time consumption of the full
\v{C}erenkov tracking in GEANT.  Single $\pi^0$'s ($\eta$'s) were
simulated with a uniform distribution in transverse momentum and
pseudo-rapidity over the LEDA acceptance and the decay photons were
tracked through GEANT. The simulated LEDA response was recorded to
create a library of photon test showers in the form of digitized LEDA
signals. Only simulated $\pi^0$'s with both decay photons in the
nominal LEDA acceptance were recorded.  The reconstruction efficiency
was then extracted with the procedure illustrated in
Fig.~\ref{fig:geant}.  First the raw data were calibrated and the
event was characterized by the trigger detectors and other detectors
of WA98.  Then three passes were made through the LEDA event analysis
software. In the first pass, the calibrated LEDA data was analyzed to
perform the clustering and shower characterization as described above.
The position, energy, shower dispersion, and distance to nearest hit
in the CPV were saved for all identified showers together with the
calibrated information from the other WA98 detectors. Next, a $\pi^0$
event was read from the shower library and inserted into an empty LEDA
raw event. Simulated signals in dead modules were eliminated and the
event was then analyzed as a real event. The shower information
reconstructed from the simulated showers in the empty LEDA was
recorded together with the primary $\pi^0$ and photon information
prior to the GEANT response. Finally, the real LEDA event was
overlaid with the simulated LEDA event and the signals were summed.
Then the superimposed event was analyzed. In this step the position,
energy, shower dispersion, and distance to nearest hit in the CPV were
saved only for all new showers which were not in the list of showers
found in the original raw event. Also, all showers of the original
event found to be missing from the reanalyzed overlap event were
marked as lost. All of this information was stored for each event in a
single pass through the WA98 raw data and recoded on the Data Summary
Tapes.  Thereafter the data analysis was performed from the DSTs since
they could easily be analyzed multiple times, or simultaneously, with
different shower selection criteria.

In order to determine the photon reconstruction efficiency it is
necessary to determine which shower in the overlap event corresponds
to the simulated photon incident on the LEDA.  In the first step, the
empty LEDA GEANT shower information is analyzed to determine the
reconstructed position of the highest energy shower in the vicinity of
the incident GEANT photon. The overlap event is then analyzed to find
the shower nearest to that position.  If that shower has less than
twice the energy\footnote{The factor of two change in energy criterion
   determines who ``eats'' whom when showers overlap and is necessary
   to avoid double counting.} of the shower in the empty LEDA event it
is taken as the reconstructed photon. Otherwise the photon is
considered to be lost. After the associated shower is identified, it
is tested to determine whether it passes the photon energy threshold
requirement and whether it passes the detector edge cut. Also, if its
position falls on the location of a so-called ``bad'' module with
questionable gain, as described below, it is eliminated.  Finally, it
is tested against the various shower identification criteria S1-S4
described above.

The efficiency corrections are made as a function of the measured
transverse momenta. The photon reconstruction efficiency can be
constructed as a two-dimensional response matrix which 
transforms the transverse
momentum of the incident photon into the transverse momentum of the
reconstructed test shower, if it passes all identification criteria
(the reconstructed transverse momentum is set to zero if the criteria
are not satisfied). This two-dimensional efficiency matrix must then
be inverted and applied to the measured transverse momentum
distribution to obtain the final efficiency corrected result~\cite{the:buc99}.
Alternatively, the efficiency can be applied as an iterative
one-dimensional correction. In this case, one-dimensional histograms
are accumulated of the incident transverse momentum and reconstructed
transverse momentum with each entry weighted according to the final
transverse momentum spectrum and rapidity distribution.  The
reconstruction efficiency is given as the ratio of reconstructed to
input distributions. This one-dimensional efficiency determination
must be iterated until the weighted input distribution matches the
final measured distribution.

%%fig4%%

The $\pi^0$ reconstruction efficiency is similarly obtained by
requiring that the showers from both of the decay photons
simultaneously pass the photon identification criteria. The $\pi^0$
mass and transverse momentum is then calculated from the momenta of
the two reconstructed photon showers. It is then required
that the reconstructed mass fall within the $\pi^0$ peak mass
integration region, and it may be additionally required that the
photons pass an asymmetry ($\alpha=|E_1-E_2|/|E_1+E_2|$) cut on their
reconstructed energies. The $\pi^0$ reconstruction efficiency is
applied as a one-dimensional function of the transverse momentum. As
in the photon case, the correction is determined from the ratio of
reconstructed to input transverse momentum distributions. The input
$\pi^0$ transverse momentum and rapidity weights must be adjusted by
iteration until the assumed input transverse momentum distribution is
the same as the final corrected distribution.

It should be noted that while these correction factors have been
referred to as efficiency corrections, they might more properly be
termed response corrections. They include essentially all detector
effects other than the nominal detector acceptance.  Specifically,
they include the effects of detector edge cuts, dead and bad modules,
energy resolution, and distortions or loss due to shower overlap, as
well as the efficiency to satisfy the specified identification
criteria. In particular, the correction for smearing due to overlap
and energy resolution can result in efficiency correction factors 
which exceed unity.

\subsection{\label{sec:simulation} Background Calculation
   and Direct Photon Excess}

As described above, the direct photon excess is obtained from the
difference between the measured inclusive photon yield and the
background photon yield expected from radiative decays of long-lived
final state hadrons.  The background photon yield in the WA98 LEDA
acceptance is calculated by a Monte Carlo simulation of radiative
decays of hadrons. The most important input to this calculation is the
measured WA98 $\pi^0$ yield, which is extracted from the same data
sample used to obtain the inclusive photon yield (see
Fig.~\ref{fig:geant}).  Photons from $\pi^0$ decay account for about
$80-90\%$ of the total expected background from radiative decays.  It
is important to note that the background photon yield attributed to
$\pi^0$ decay includes both directly produced $\pi^0$'s, as well as
those from hadronic decays with $\pi^0$'s in their final
state.  Thus, photons resulting from the $3\pi^0$ decay branch of the
$\eta$ are taken into account via the measured $\pi^0$ yield. On the
other hand, the lifetime of the $K^0_L$ is sufficiently long that few
of the $3\pi^0$ weak decays occur in front of the LEDA detector and
therefore there is little contribution to the $\pi^0$ yield, and hence
little contribution to the background photon yield. However, the weak
decay of the $K^0_S$ to $2\pi^0$ is a special case.  The $K^0_S$
lifetime is such that a substantial fraction of the decays are
distributed over the distance between the target and the LEDA
detector. As with the $K^0_L$ contribution, those decays which occur
beyond the LEDA distance do not contribute to the $\pi^0$ or
background photon yield.  On the other hand, photons from a $\pi^0$
produced in a $K^0_S$ decay will have correctly
measured energies but will be assumed to be produced at the target
location and therefore will have an incorrect opening angle. This will
result in a reconstructed $\pi^0$ invariant mass which is incorrect.
While $K^0_S$ decays which occur close to the target will have a
reconstructed $\pi^0$ mass which falls into the $\pi^0$ identification
window, and so its decay photon contribution will be included via the
measured $\pi^0$ yield, some fraction of the $K^0_S$ decays will occur
sufficiently far from the target that their $\pi^0$ decays will not be
properly identified. Only this portion of the $K^0_S$ decay photon
contribution must be included in the calculated photon background.

The $2\gamma$ decay of the $\eta$ is the second most important
contribution to the photon decay background after the $\pi^0$
contribution. Together the $\pi^0$ and $\eta$ photon decays constitute
approximately $97\%$ of the expected radiative decay background (see
Fig.~\ref{fig:photon_pt_bkg}).  Compared to the $\pi^0$ yield
measurement the $\eta$ measurement is more difficult due to the
smaller production rate, the smaller $2\gamma$ decay branching ratio,
and the resulting smaller signal to combinatorial background ratio in
the $2\gamma$ invariant mass distribution.  In the present analysis,
the $\eta$ yield is measured with modest statistical accuracy over a
limited transverse momentum range due to these difficulties. In order
to extrapolate the measured $\eta$ transverse momentum distribution
into unmeasured regions it is assumed that the $\eta$ yield obeys
$m_T$-scaling. This is the phenomenological observation
%\cite{prc:fei72,npb:bou76}
\cite{npb:bou76} that the differential invariant cross sections,
plotted as a function of the transverse mass $m_T=\sqrt{m_0^2 +
   p_T^2}$, for the various hadrons, $h$, have the same form, $f(m_T)$,
with a normalization factor, $C_h$, which can vary but is found to 
be the same for many species:
  \begin{equation}
E\frac{d^3\sigma_h}{dp^3} =  C_h \cdot f(m_T).
\label{eq:mt_scaling}
\end{equation}
Quite different theoretical explanations
\cite{prd:ban74,plb:shu80,rnc:hag83} can account for this observation.
For various proton and pion-induced reactions at similar incident
energies
\cite{npb:bus76,npb:bad77,prl:don78,plb:kou79,prl:pov83,zpc:agu91,zpc:aps92,epj:aga98}
it is observed that the $\eta$ yield obeys $m_T$-scaling to good
accuracy \cite{plb:ake86}. A scaling factor relative to $\pi^0$
production of $R_{\eta/\pi^0} = C_{\eta}/C_{\pi^0} = 0.55$ 
is obtained for the case of
proton-induced reactions~\cite{plb:ake86,plb:alb95}. Similarly, the
$\eta$ yield is found to be consistent with $m_T$-scaling in minimum
bias sulphur-induced reactions~\cite{plb:alb95} at 200~{\it A}~GeV
incident energy.

Collective transverse flow will affect the spectrum of produced
particles according to their mass with the result that $m_T$-scaling
might be violated in collisions of very heavy ions. Evidence for
collective transverse flow has been observed at the SPS for central
$^{208}$Pb\/+\/$^{208}$Pb collisions with estimated average transverse
flow velocities as large as $\beta_T \approx 0.5$
\cite{prl:bea97,epj:app98,prl:agg99}.  For a particle of mass $m$ and
transverse momentum $p_T$ the effective $m_T$ inverse slope, or
temperature $T_{eff}$, will be modified as~\cite{prc:nix98}
\begin{equation}
T_{eff}=\frac{\sqrt{1-\beta_T^2}}{1-\beta_T\sqrt{1+m^2/p_T^2}} T
\label{t_flow}
\end{equation}
where $\beta_T$ is the average transverse flow velocity and $T$ is the
thermal temperature. While the modification of the transverse mass
spectrum is seen to decrease with increasing $p_T$,  
the effect can be significant. As an example, a transverse flow
of $\beta_T = 0.5$ would increase $T_{eff}$ for the $\eta$ by about
$4\%$ at $p_T=2$ GeV/c which would result in an increase of about
$50\%$ in the $\eta$ yield at $p_T=2$ GeV/c.

It has been suggested that if chiral symmetry restoration occurs in
the hot dense system formed in relativistic heavy ion collisions, then
the masses of the $\eta$ and $\eta'$ mesons might decrease with an
associated increase in their production rates
\cite{prd:kap96,prd:hua96}. These initial estimates suggested that the
$\eta$ and $\eta'$ yields might be increased by as much as a factor of
3 and 10, respectively. Once produced, the $\eta$ and $\eta'$ are
expected to interact relatively little in the dense matter
with the result that they would
survive to the final state to decay with their vacuum masses and
contribute significantly to the decay background to produce excess
photons and dileptons~\cite{prd:kap96}. On the other hand, more recent
calculations within the context of the non-linear sigma model suggest
that the temperature dependence of the $\eta$ and $\eta'$ masses and
mixing are negligible~\cite{prd:jal98}.  In view of these significant
uncertainties in the extrapolation of the $\eta$ yield from
proton-induced reactions to central $^{208}$Pb\/+\/$^{208}$Pb
collisions it is important to measure the $\eta$ yield directly for
central collisions to provide experimental constraints on its possible
contribution to the background photon yield.

Besides the $\pi^0$ and $\eta$, other hadrons with radiative decays
which may contribute to the background photon yield are listed in
Table~\ref{table1}~\cite{epj:pdb98}. The production rates of these
other hadrons are not measured in this experiment.  As for the $\eta$,
their production has been assumed to follow $m_T$-scaling with the
same $m_T$ spectrum as the measured $\pi^0$ spectrum and with relative
normalizations $R_{X/\pi^0}$ (equivalent to the asymptotic ratio as
$p_T \rightarrow \infty$) given in Table~\ref{table1}. Within
experimental errors the ratio $R_{X/\pi^0}(p_T \rightarrow \infty)
\approx 1$ independent of incident energy for the $\rho$ and $\omega$
\cite{npb:bou76,zpc:agu91,epj:aga98,plb:dia80} and for the $\eta'$
\cite{plb:dia80}. For the $K^0_S$ a ratio of $R_{K^0_S/\pi^0} \approx
0.4$ \cite{prl:dao73,zpc:der91,zpc:alb94} is observed for
proton-induced reactions with indications for an increased ratio for
nucleus-induced reactions \cite{zpc:alb94,plb:aba96}.

Of the other radiative decays listed in Table~\ref{table1} only the
$\eta'$ and $\omega$ are expected to contribute more than one percent
of the background photons (see Fig.~\ref{fig:photon_pt_bkg}). The
$\eta'$ is notable in that it might be significantly enhanced due to
the mechanism discussed above. While the $\eta'$ production rate is
not determined in the present measurement, it can be constrained by
the $\eta$ measurement due to its $65.5\%$ branching ratio to
$\pi\pi\eta$.

To summarize, the background photon yield in the acceptance of the 
WA98 lead-glass
detector is calculated by a Monte Carlo simulation of
radiative decays of all hadrons listed in Table~\ref{table1}. The
various hadrons are assumed to have the same transverse mass spectrum
as the measured WA98 $\pi^0$ transverse mass spectrum for each event
class.  The yields of the other hadrons relative to the $\pi^0$ yield
are given by the $m_T$-scaling factors listed in Table~\ref{table1}.
The one exception is the $\eta$ scaling factor for central
$^{208}$Pb\/+\/$^{208}$Pb collisions where the measured $\eta$
$m_T$-scaling factor is used. The Monte Carlo program uses the JETSET
7.3 routines \cite{cern:jet89} to implement the hadron decays with
proper branching ratios and decay distributions. The hadrons are
assumed to have a Gaussian rapidity distribution centered on
mid-rapidity $y=2.9$ with a width of $\sigma_y=1.3$ according to
measurements for $^{208}$Pb\/+\/$^{208}$Pb collisions
\cite{plb:agg99}.

%%insert table
\begin{table}[h]
\caption[table1]{Dominant radiative decays which contribute to the 
inclusive photon
background. $R_{X/\pi^0}(p_T \rightarrow \infty)$ is the assumed
asymptotic ratio of yield of hadron X relative to $\pi^0$'s, or
equivalently the $m_T$-scaling factor.  $\sigma(X)/\sigma(\pi^0)$
is the ratio of integrated yields which would be obtained for an
exponential spectrum in $m_T$ with a slope of 200 MeV/c$^2$ and the
assumed $R_{X/\pi^0}$ values. It is only an indication of the
ratio of integrated yields since the actual value will depend
on the measured $\pi^0$ $m_T$ spectrum.
The listed decays and branching ratios are taken
from Ref.~\cite{epj:pdb98}.
}
\label{table1}
\begin{tabular}{|cddd|cd|}
State & Mass & $R_{X/\pi^0}(p_T \rightarrow \infty)$ &
  $\sigma(X)/\sigma(\pi^0)$ &
  Decay Branch & Branching Ratio  \\
\tableline
$\pi^0$    & 134.98 &      & & $\gamma\gamma$      &  98.798\% \\
            &        &      & & $e^+e^-\gamma$      &   1.198\% \\
\tableline
$\eta$     & 547.3  & 0.55 (0.486) & 0.08 & $\gamma\gamma$      &  39.21\% \\
            &        &      & & $\pi^+\pi^-\gamma$  &   4.77\% \\
            &        &      & & $e^+e^-\gamma$      &  4.9$\cdot$10$^{-3}$ \\
            &        &      & & $\pi^0\gamma\gamma$ &  7.1$\cdot$10$^{-4}$ \\
            &        &      & & $\mu^+\mu^-\gamma$  &  3.1$\cdot$10$^{-4}$ \\
\tableline
$\rho^0$   & 770.0  & 1.0 & 0.05 & $\pi^+\pi^-\gamma$  &  9.9$\cdot$10$^{-3}$\\
            &        &      & & $\pi^0\gamma$       &  7.9$\cdot$10$^{-4}$\\
\tableline
$\omega$   & 781.9  & 1.0  & 0.05 & $\pi^0\gamma$       &   8.5\% \\
            &        &      & & $\eta\gamma$        &  6.5$\cdot$10$^{-4}$\\
\tableline
$\eta'$    & 957.8  & 1.0  & 0.02& $\rho\gamma$        &  30.2\% \\
            &        &      & & $\omega\gamma$      &   3.01\% \\
            &        &      & & $\gamma\gamma$      &   2.11\% \\
\tableline
$K^0_S$    & 497.7  & 0.4  & 0.07 & $(\pi^0\pi^0)$      & (31.39\%) \\
\tableline
$\Sigma^0$ & 1192.6 & 1.0  & 0.007 & $\Lambda\gamma$     & 100.\% \\
\end{tabular}
\end{table}

\section{\label{sec:details} DATA ANALYSIS DETAILS}

In this section, detailed results are presented for the extraction of
the $\gamma$, $\pi^0$, and $\eta$ yield and their error estimates, as
required for the direct photon analysis.  First, a description of the
data sample selection is presented.  This is followed by a discussion
of the analysis involving the Charged Particle Veto detector. The CPV
is used to determine the charged particle contamination in the photon
shower sample. Next follows a description of the details of the
extraction of the inclusive photon transverse momentum distributions,
including a discussion of the background contributions from charged
particles and neutrons, and losses due to photon conversions. The
photon identification efficiency for the various methods is discussed
together with a summary of the estimated systematical error on the
inclusive photon measurement. Next, the $\pi^0$ yield extraction is
described.  This includes a discussion of the yield extraction method,
efficiencies, backgrounds, and systematical error. Finally, the $\eta$
yield extraction is described. The final results and the extraction of
the direct photon excess are described in the following section.

\subsection{\label{sec:selection} Data Selection}

The present analysis has been performed using the event samples
summarized in Table~\ref{table2}. The data have been taken over six
week run periods in 1995 and in 1996 with the 158~{\it A}~GeV
$^{208}$Pb beam of the SPS on $^{208}$Pb targets of 495 and 239
mg/cm$^2$, respectively.  During both run periods most data were taken
with the GOLIATH magnet on, as required for the WA98 tracking
spectrometer measurements.  The 
minimum bias cross sections for the various data sets, after
subtraction of the target out backgrounds, are given in
Table~\ref{table2}.  Because of the change in the apparent transverse
momenta of the charged particles due to the deflection in the magnetic
field, the apparent transverse energy measured in MIRAC is increased
with magnet on compared to the actual transverse energy. With the
fixed low transverse energy trigger threshold, this resulted in
larger minimum bias cross sections for the magnet on
data sets. During the 1995 datataking period the vacuum exit window at
the entrance to the GOLIATH magnet produced a significant background
of downstream interactions which satisfied the minimum bias
transverse energy threshold. These downstream interactions were
eliminated by requiring an interaction at the target location by the
requirement of a hit in the Plastic Ball in the angular region from
$30^\circ$ to $50^\circ$ using the Plastic Ball trigger (described in
Sec.~\ref{sec:detectors}) in coincidence with the minimum bias
trigger. For the 1996 run period, the vacuum exit window was changed
resulting in fewer downstream interactions.  As a result, the Plastic
Ball trigger was not required in the online trigger which resulted in
less biased minimum bias cross sections.  The rms variations of the
minimum bias cross sections determined on a run-by-run basis are also
given in Table~\ref{table2}. The measured variation gives an
indication of the uncertainty in the measured absolute cross sections
due to normalization and background corrections. The background
corrections were obtained from special empty target runs with no
target in the target location. Due to the lower event rates, and
resulting low deadtime and lack of need to be downscaled, the empty
target data was taken with similar number of integrated beam
triggers as obtained for the Pb data. Specifically, the 1995 and 1996
empty target data corresponded to a factor of 2.3 and 1.4 fewer beam
triggers than the 1995 and 1996 Pb target data, respectively.

%%insert table
\begin{table}[hbt]
\caption[table2]{Summary of the event selection used for the present
   analysis. The table shows the WA98 158~{\it A}~GeV
  $^{208}$Pb\/+\/$^{208}$Pb  minimum bias cross sections for
the 1995 and 1996 run periods under conditions of magnetic field on
  and field off. The rms of the minimum bias cross section extracted
run by run is indicated in parenthesis. The central event class is
defined by those events falling above a high cut on the
measured transverse energy where the cut is chosen
to give a cross section of 635 mb above the cut. Similarly,
the peripheral event class is defined by those events falling
below a low transverse energy cut where the cut is chosen to give
a total cross section of 4910 mb above the cut.
The number of peripheral and central events used
under each condition for the present analysis are listed.}
\label{table2}
\begin{tabular}{lccdd}
Run Period & $^{208}$Pb Target & Minimum Bias &
  Peripheral Collisions & Central\ Collisions  \\
&Thickness (mg/cm$^2$) & $\sigma_{min bias}$ (mb) &
   N$_{Events}$ &  N$_{Events}$ \\
\tableline
1995 Field On & 495. & 6192. (56.) & 2694528. &  2879652.\\
1995 Field Off & 495. &  5971. (30.) & 159291. & 213170.\\
1996 Field On & 239. &  6451. (53.) & 1203407. & 2937565.\\
1996 Field Off & 239. &  6202. (58.) & 222917. & 650521.\\
\end{tabular}
\end{table}

\begin{figure}[hbt]
\begin{center}
   \includegraphics[scale=0.6]{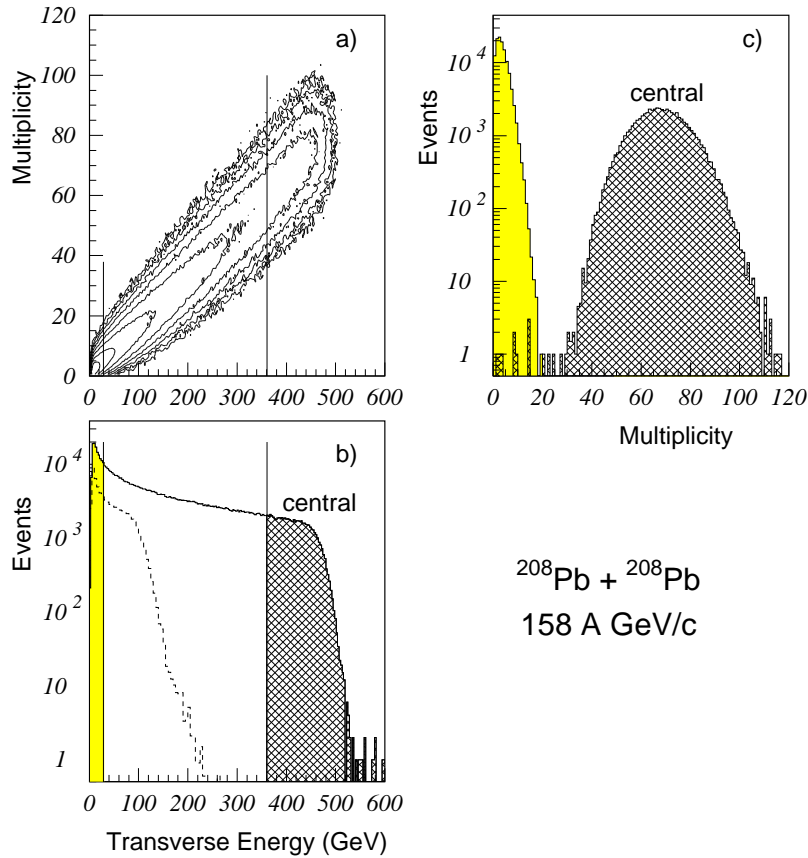}
\caption{The correlation between the number of showers
   observed in the lead-glass detector and the measured transverse
   energy for 158~{\it A}~GeV $^{208}$Pb\/+\/$^{208}$Pb collisions for
   a sample of the 1995 magnet on data.  Part b) shows the projection
   onto the transverse energy axis with the shaded regions indicating
   the peripheral and central event class selections used for the
   present analysis. The distribution is uncorrected for the empty
   target contribution which is shown by the dashed histogram with
   arbitrary relative normalization.  Part c) shows the projection onto
   the lead-glass multiplicity axis for the selected events.}
\label{fig:events}
\end{center}
\end{figure}

The direct photon analysis has been performed for event selections
corresponding approximately to the $20\%$ most peripheral and $ 10\%$
most central portions of the minimum bias cross sections.  These event
classes are defined by cuts on the total transverse energy, measured
in MIRAC, as calculated in the offline analysis.  The selections
correspond closely to the online trigger event classes described in
Sec.~\ref{sec:trigger}. More precisely, the transverse energy cut
which defines the central event sample was chosen to correspond to a
most central cross section of 635~mb, or impact parameters less than
about 4.5~fm, for all data sets.  Similarly, the transverse energy cut
which defines the peripheral event sample was chosen to correspond to
a cross section of 4910~mb above the transverse energy cut, or to a
peripheral event sample with impact parameters greater than about
12.5~fm.  Due to the variation of the minimum bias cross section for
the different data sets analyzed (see Table~\ref{table2}), the meaning
of the peripheral event class (for example, as reflected in the
particle multiplicity) depended on the data sample.  These event class
definitions are shown in Fig.~\ref{fig:events} for the 1995 magnet on
data set where the multiplicity of showers in the lead-glass fiducial
region with energy above 750 MeV is plotted versus the total
transverse energy. The projections onto each axis are also shown.
Similar transverse energy cuts are used for the event class
definitions for both the 1995 and 1996 data sets. However, quite
different cuts are used for the magnet on and magnet off data sets due
to the change in the apparent transverse energy scale noted above.

%%fig5%%

We note that while the central event samples in the four data sets
should be very similar in terms of the impact parameter range
selection and particle multiplicity, the peripheral event samples of
the different data sets are likely to be more variable.  This is due
to the rapid variation of the overlap geometry in the peripheral
region and the variations in the minimum bias cross section and
background corrections for the four data sets noted above.

In the offline analysis the various trigger signals are checked to
remove events with inconsistent trigger information and to remove
events with another beam particle within 100~ns or another interaction
within about 300~ns of the triggered event.  These trigger cuts
discard about $10\%$ of the events.  For the 1996 data set the Plastic
Ball trigger was not required in the online trigger or used in the
offline analysis for the final data sample. However, events which did
not satisfy the Plastic Ball trigger requirement were used in the
offline analysis to investigate the downstream interaction
contributions in more detail.  Since the downstream interactions are
on light materials, such as air, mylar, and aluminum ($A<30$) and have
underestimated emission angles, they produce small measured transverse
energies with the result that their contamination is almost entirely
in the peripheral event sample (see Fig.~\ref{fig:events}).  The  
peripheral data sample for the 1996 data set was smaller than that
for the 1995 data set due to the factor of two larger prescale factor
used in the 1996 peripheral trigger.

In addition to a selection of the data sample based on trigger cleanup
cuts, the lead-glass shower data was analyzed in a preliminary scan of
the data and modules with questionable gain were eliminated for the
subsequent analysis. This selection was made by accumulating the
shower energy spectrum for each individual module where the shower
centroid was within that module.  The results were compared to the
average dependence across the detector surface. Modules whose spectrum
deviated from the average behavior, with rather strict criteria, were
flagged as bad. In the actual data analysis, showers with positions
within a module which was flagged as bad were eliminated.  As a result
of the rejected bad modules and the modules eliminated around the
edges of the detector and dead modules, the effective LEDA acceptance
was reduced by about 40\%.

\subsection{\label{sec:cpv} Charged Particle Veto}

The Charged Particle Veto detector provides essential information for
the photon analysis.  It allows charged showers in LEDA to be
identified and associated with charged hadrons or photon conversions
(see step G3 of Sec.~\ref{sec:pid}).  When the photon selection is
made without invoking the CPV directly in the shower identification
criteria (criteria S1 and S2 of Sec.~\ref{sec:pid}) then the CPV is
used to accumulate the transverse momentum spectrum of charged LEDA
showers. This spectrum is corrected for the $p_T$-independent CPV
efficiency and then subtracted from the total LEDA shower transverse
momentum spectrum to obtain the neutral shower transverse momentum
spectrum. The charge/neutral ratio is extracted and fitted as a
function of the transverse momentum and used to calculate a correction factor
applied to the total shower spectrum to obtain the neutral shower
spectrum. The neutral shower spectrum is then corrected for neutrons
and anti-neutrons, and for conversions to obtain the raw photon
spectrum.

%%fig6%%
\begin{figure}[hbt]
\begin{center}
   \includegraphics[scale=0.45]{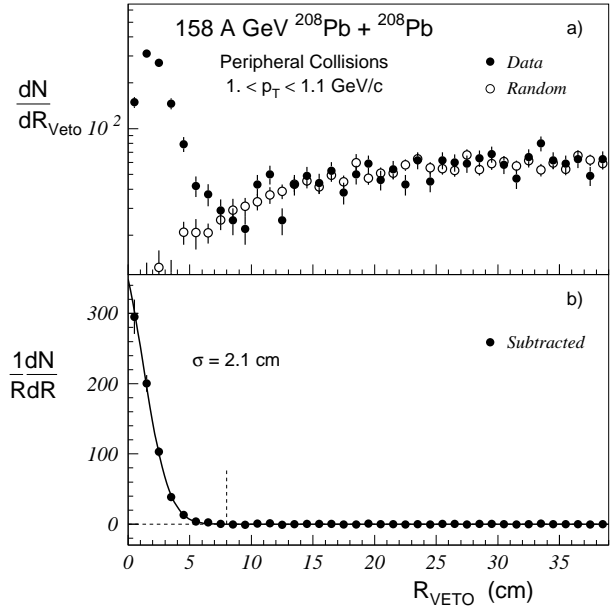}
\caption{The distribution of distances, $R_{Veto}$, to the
   nearest CPV hit from a shower particle in the lead-glass for
   peripheral events. In part a) the distribution for real events is
   shown by the solid circles. The distribution for randomly vetoed
   hits is shown by the open circles. The random veto distribution is
   obtained using test GEANT showers inserted into real events. The
   true veto radius distribution after subtraction of the random veto
   distribution is shown in part b).}
\label{fig:cpv_radius_per}
\end{center}
\end{figure}

Alternatively, the CPV can be used directly in the shower selection
criteria to choose non-charged photon candidates (criteria S3 and S4
of Sec.~\ref{sec:pid}). However, because the CPV detector was fully
installed and operated for the 1996 run period only, it was not
possible to perform the analysis with the CPV in this way for the 1995
data sample. Instead it was necessary to extract the charge/neutral
correction from the 1996 data sample and apply this correction to the
1995 data. Therefore it was important to verify the consistency of the
two methods in which the CPV information was used.

\begin{figure}[hbt]
\begin{center}
   \includegraphics[scale=0.5]{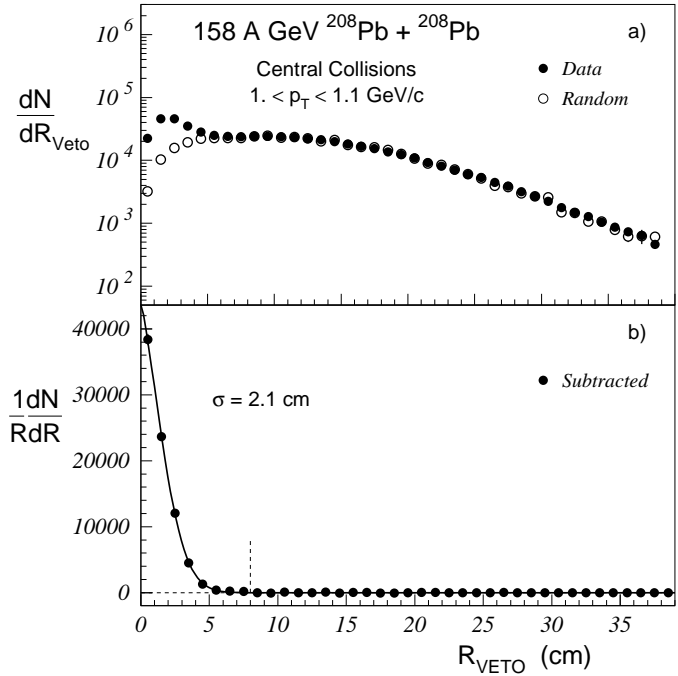}
\caption{The distribution of distances, $R_{Veto}$, to the
   nearest CPV hit from a shower particle in the lead-glass for central
   events. In part a) the distribution for real events is shown by the
   solid circles. The distribution for randomly vetoed hits is shown by
   the open circles. The random veto distribution is obtained using
   test GEANT showers inserted into real events. The true veto radius
   distribution after subtraction of the random veto distribution is
   shown in part b).}
\label{fig:cpv_radius_cen}
\end{center}
\end{figure}

In the shower analysis procedure described in Sec.~\ref{sec:pideff},
each shower in the list of individual localized showers in LEDA is
compared with the list of hits in the CPV and the distance between the
LEDA shower position and the nearest CPV hit position is recorded. An
example of this distance distribution is shown in
Fig.~\ref{fig:cpv_radius_per}a) for showers in peripheral collisions
with transverse momenta $1<p_T<1.1$ GeV/c.  The distribution shows a
clear peak at small veto distance with a long tail extending to large
veto radii.  The long tail results from random associations between
showers in LEDA and hits in the CPV. The veto radius distribution of
these random associations is also shown in
Fig.~\ref{fig:cpv_radius_per}a).  This distribution is extracted from
the veto distance distribution obtained for the GEANT test photons
introduced into the LEDA event, as described in Sec.~\ref{sec:pideff}.
Since these test showers have no correlated hit in the CPV their veto
distance distribution is strictly random. The random hit distribution
is normalized to the distribution for real LEDA showers at large veto
distance and subtracted from the distribution for real showers to
obtain the distance distribution of real charged showers shown in
Fig.~\ref{fig:cpv_radius_per}b). While the random veto contribution is
quite small for peripheral collisions, the detector occupancies are
much greater in central collisions with the result that there is a
much higher probability for a CPV hit to be randomly associated with a
shower in LEDA. This is shown in Fig.~\ref{fig:cpv_radius_cen} where
the same veto distance distributions are shown for central collisions.
In this case the correction for the random CPV hits is essential.

%%fig7%%

\begin{figure}[hbt]
\begin{center}
   \includegraphics[scale=0.5]{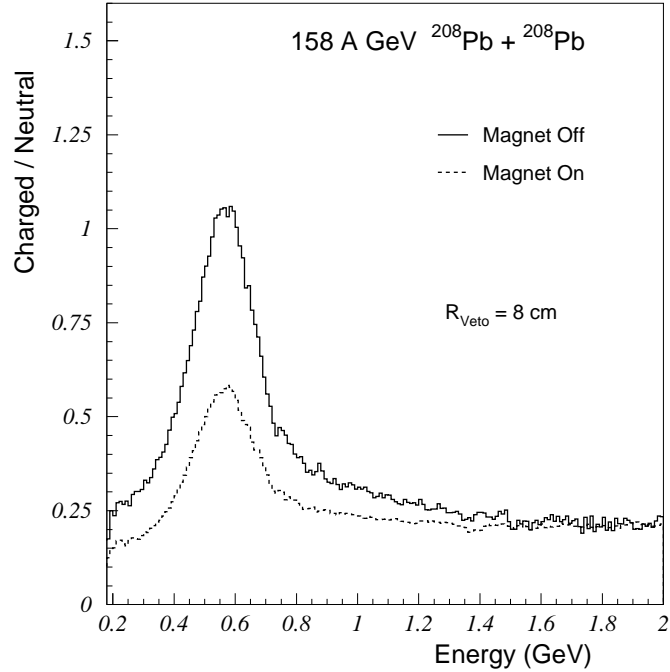} \vspace{0.3in}
\caption{The ratio of charged to neutral clusters identified in the lead-glass
   detector for peripheral collisions with magnet on or magnet off run
   conditions as a function of the total energy of the cluster in the
   low transverse momentum region.  The solid curve shows the ratio for
   clusters identified as charged hits by a coincident hit in the CPV
   within a distance of $R_{Veto} = 8$~cm of the cluster.  The peak at
   550 MeV from non-showering minimum ionizing particles is clearly
   seen in the distributions.  A minimum energy threshold of 750 MeV is
   used in the present analysis.}
\label{fig:cpv_select}
\end{center}
\end{figure}

Based on these distributions, showers with a CPV hit within a distance
cut of 8~cm are tagged as charged showers.  For the analysis methods
S1 and S2, in which the CPV is used to extract a charged/neutral
correction factor, the charged shower yield is extracted for each
$p_T$ bin as the random subtracted yield in the veto peak within the
8~cm distance cut.  The result is the charged shower $p_T$
distribution which is then used to obtain the charged/neutral shower
correction factor as a function of $p_T$. The charged/neutral shower
ratio as a function of the shower energy is shown in
Fig.~\ref{fig:cpv_select} for peripheral collisions using a 8~cm
distance cut. The peak in the spectrum at about 550~MeV is due to
non-showering hadrons which pass through the lead-glass and
deposit similar energy by $dE/dx$ only. A minimum
%%tp%% -->
%% Shouldn't we mention that it's the Cherenkov that we measure not
%% dE/dx, although Cherenkov is part of dE/dx?
%%tp%% <--
shower energy threshold of 750~MeV has been applied in the present
analysis to eliminate this minimum-ionizing particle, or MIP, peak
from the photon candidates.  It is seen that the charged hadron
background differs for the magnet on and magnet off run conditions.

%%fig8%%

For the analysis methods S3 and S4 charged hadrons are rejected
directly from photon candidate showers by an associated CPV hit within
the 8~cm distance cut. In this case there is no charged/neutral
correction necessary to the photon candidate spectrum. On the other
hand, the loss of photon showers due to random vetos and conversions
is treated as an efficiency loss which is then taken into account at a
later step in the efficiency correction (see Sec.~\ref{sec:gameff}).

\begin{figure}[hbt]
\begin{center}
   \includegraphics[scale=0.5]{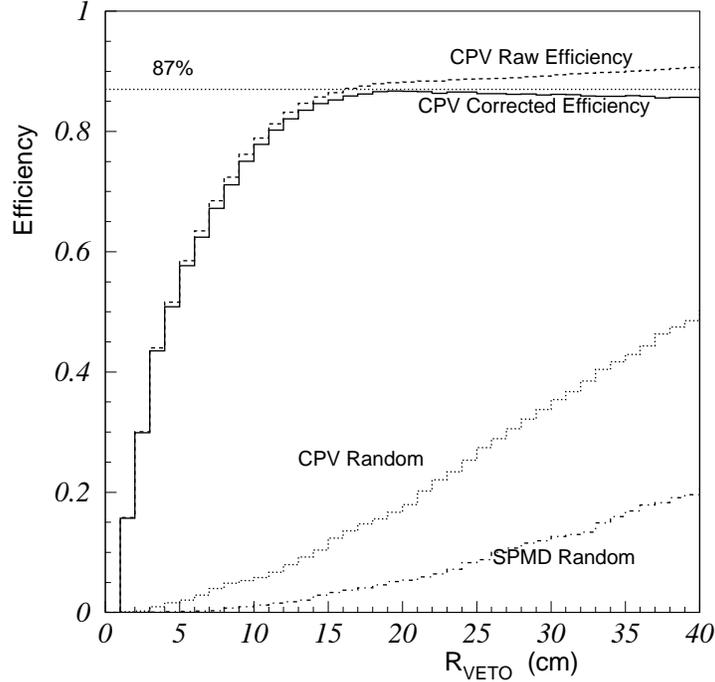}
\caption{The efficiency of the charged particle veto as a function of
   veto radius. The results are obtained from magnetic field off
   peripheral data.  The dashed curve shows the probability that a hit
   in the CPV is found within a distance $R_{Veto}$ of a charged hit in
   the lead-glass. The lead-glass hit has been tagged as charged by a
   coincident hit in the SPMD within distance $R_{Veto}$.  The distance
   $R_{Veto}$ is measured in the plane projected onto the lead-glass
   surface. The dashed curve includes random vetoes which artificially
   increase the apparent efficiency with increasing radius.  The dotted
   curve shows the random veto efficiency of the CPV. The dot-dashed
   curve shows the random veto efficiency of the SPMD.  The solid curve
   shows the final CPV efficiency result. }
\label{fig:cpv_efficiency}
\end{center}
\end{figure}

The charged shower identification must be corrected for the efficiency
of the Charged Particle Veto. As mentioned in Sec.~\ref{sec:ledacpv},
the CPV efficiency has been determined {\em in situ} using LEDA and
the Silicon Pad Multiplicity Detector. The analysis is performed using
magnet off data to allow straight line tracking between the SPMD and
LEDA. Peripheral collisions are used to keep the detector occupancies
low in order to minimize random hit associations. Hits in the SPMD are
projected to the LEDA detector surface and associated with a hit in
LEDA if they fall within the SPMD projected pad area.  If a hit is
found within LEDA the distance from the LEDA shower to the nearest hit
in the CPV is extracted. The raw CPV efficiency, $\epsilon^0_{CPV}$,
obtained as the ratio of the number of CPV hits found to the number of
SPMD-LEDA coincidence tracks, is shown in
Fig.~\ref{fig:cpv_efficiency} as a function of the distance between
the LEDA shower and the CPV hit. It is seen to increase rapidly up to
a veto distance of about 15~cm and then increase very slowly for
larger distances. This slow rise at large distances is due to random
coincidences with hits in the CPV which artificially increase the
apparent efficiency. The random CPV hit efficiency,
$\epsilon^R_{CPV}$, also shown in Fig.~\ref{fig:cpv_efficiency}, is
extracted by an event mixing technique in which the CPV data are taken
from a different peripheral event than the one used for the SPMD-LEDA
track. The CPV efficiency is corrected for such random CPV
associations as $\epsilon_{CPV} = (\epsilon^0_{CPV} -
\epsilon^R_{CPV})/ (1-\epsilon^R_{CPV})$. In addition, due to the SPMD
inefficiency or interactions, there can be random coincidences between
the SPMD and LEDA, for which a hit in the CPV would not be expected,
which decrease the apparent CPV efficiency.  The shape of the
SPMD-LEDA random coincidence distribution is also shown in
Fig.~\ref{fig:cpv_efficiency}. A small additional correction is
applied to $\epsilon_{CPV}$ to account for this effect.

%%fig9%%

As shown in Fig.~\ref{fig:cpv_efficiency}, an asymptotic corrected CPV
efficiency of $\epsilon_{CPV}=87\%$ is obtained for veto distance cuts
greater than about 17~cm. It will be noted that this distance is
considerably larger than indicated by the width of the veto distance
distributions shown in Figs.~\ref{fig:cpv_radius_per} and
\ref{fig:cpv_radius_cen}.  This difference is due to the fact that the
CPV efficiency analysis uses all charged hits in the SPMD. Therefore
it is dominated by charged hadrons which deposit very little energy in
LEDA either due to poorly developed hadronic showers, or due to 
energy deposit by $dE/dx$ only.  These low
energy showers have very poor position determination in LEDA.  This
effect is also evident by the observation that the width of the
LEDA-CPV correlation (see Figs.~\ref{fig:cpv_radius_per} and
\ref{fig:cpv_radius_cen}) increases at very low transverse momentum.
As a result the R$_{Veto}$ distance cut is increased for low
transverse momenta to insure that the asymptotic CPV efficiency is
attained. The CPV efficiency is assumed to be 87\% independent of
transverse momentum.

During datataking, CPV readout errors occurred with apparently random
frequency which resulted in loss of portions of the CPV data in the
event. These errors were not excluded in the present analysis and
account for most of the extracted CPV inefficiency. Since the readout
errors resulted in loss of trailer information in the CPV data packets
it was possible to identify such readout errors and determine the CPV
efficiency for a data sample without readout errors. An intrinsic
detector efficiency of better than 98\% was
obtained~\cite{the:rey99}. It was also
possible to verify that the readout error rate did not vary with
detector occupancy or with time during the run period. The readout
errors occurred relatively frequently but affected only a small
portion of the CPV data of an event when they did occur.  Therefore
rejection of all events in which a CPV readout error occurred would
have resulted in an unacceptable reduction of the data sample.

\subsection{\label{sec:photons} Photon Analysis}

The method to extract the inclusive photon transverse momentum spectra
are described in Sec.~\ref{sec:pid}. In summary, the procedure is to
remove the charged hadron contamination from the photon candidates or
candidate spectrum, remove the neutron and anti-neutron contamination,
correct for photon conversions, then correct for the photon
identification efficiency, and finally to correct for the detector
acceptance.  The details of this procedure are described in this
section.

%\subsubsection{Identification Criteria}

\subsubsection{\label{sec:gamchrg} Charged Particle
   Contamination and Conversions}

The CPV detector is used to select charged showers.  The charged
shower transverse momentum distribution is constructed from the random
corrected yield in the veto peak for each $p_T$ bin, as discussed in
Sec.~\ref{sec:cpv} (see Figs.~\ref{fig:cpv_radius_per} and
\ref{fig:cpv_radius_cen}).  The random corrected charged shower $p_T$
distribution is then corrected for the CPV efficiency to obtain the
total charged shower distribution. This corrected charged shower
distribution includes both charged hadrons and photon conversions.
Next, the corrected charged shower $p_T$ distribution is subtracted
from the total shower distribution to obtain the raw neutral shower
$p_T$ spectrum. The neutral shower spectrum is depleted uniformly as a
result of photon conversions and must be corrected by a factor
$1/(1-P_C)$ where $P_C$ is the photon conversion probability.  The
amount of conversion material between LEDA and the target was the same
for both the 1995 and 1996 run periods. The air plus vacuum exit
window contributed 5.4\% to the conversion probability and the CPV
material before the active volume of the CPV contributed an additional
2.0\% to the conversion probability. Finally, the 1995 and 1996 target
(half) thicknesses contributed an additional 3.0\% and 1.45\%,
respectively. Thus the total photon conversion probability for the
1996 run period during which the CPV was in operation was
$P_C=8.6\%$.

%%fig10%%
\vbox{
\begin{figure}[hbt]\vspace{-0.5in}
\begin{center}
   \includegraphics[scale=0.5]{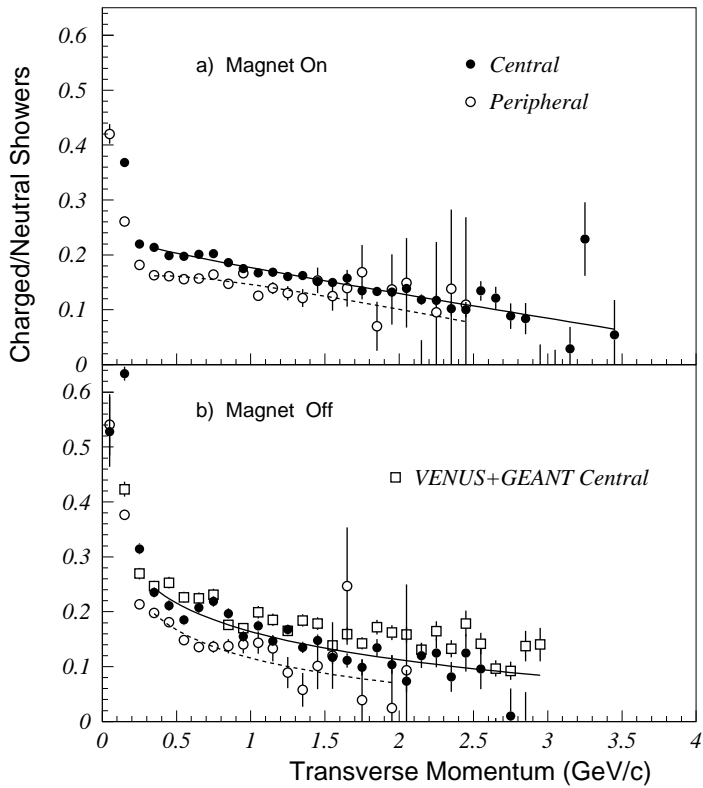}
\caption{The measured ratio of charged to neutral
   showers in the lead-glass detector for peripheral (open circles) and
   central (solid circles) 158~{\it A}~GeV $^{208}$Pb\/+\/$^{208}$Pb
   collisions as a function of the transverse momentum of the cluster.
   The results are shown for all clusters for a) magnet on and b)
   magnet off.  The data have been corrected for photon conversions and
   the CPV efficiency.  The lines are fitted curves.}
\label{fig:charge_neut}
\end{center}
\end{figure}
}

%%fig11%%
\begin{figure}[hbt] \vspace{-0.3in}
\begin{center}
   \includegraphics[scale=0.5]{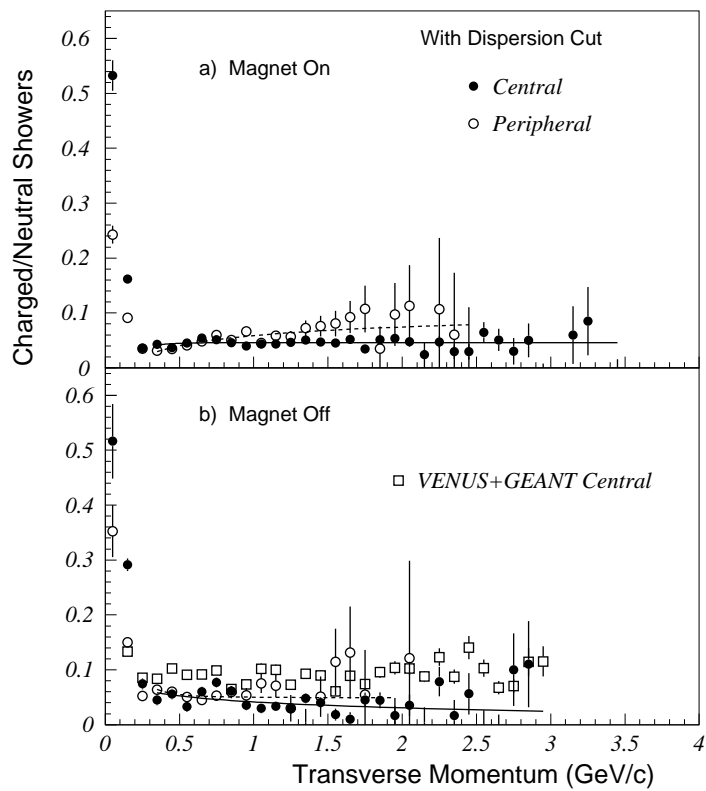}
\caption{The measured ratio of charged to neutral
   showers in the lead-glass detector for peripheral (open circles) and
   central (solid circles) 158~{\it A}~GeV $^{208}$Pb\/+\/$^{208}$Pb
   collisions as a function of the transverse momentum of the cluster.
   The results are shown for clusters which pass a photon-like
   dispersion cut for a) magnet on and b) magnet off.  The data have
   been corrected for photon conversions and the CPV efficiency.  The
   lines are fitted curves.}
\label{fig:charge_neut_disp}
\end{center}
\end{figure}

The neutral shower spectrum, after correction for conversions, 
is subtracted from
the total shower spectrum to obtain the final charged hadron shower
spectrum.  By this procedure the amount of contamination of the
selected showers due to charged hadrons is determined for each shower
identification criterion and for the magnet on and magnet off run
conditions. The extracted ratio of charged/neutral showers under the
condition to use all showers (S1) is shown in
Fig.~\ref{fig:charge_neut} for central and peripheral collisions and
for magnet on and magnet off.  The charged hadrons are seen to
constitute about 20\% of all showers at low $p_T$ decreasing to about
5\% at the highest $p_T$. When the shower dispersion cut is applied in
the shower selection criterion (S2) the hadron contamination is
reduced to about 5\% nearly independent of $p_T$ as seen in
Fig.~\ref{fig:charge_neut_disp}. For comparison, simulation results
are also shown for the case of magnet off. The simulation was
performed using full VENUS~4.12 \cite{pr:wer93} events calculated for
central $^{208}$Pb\/+\/$^{208}$Pb collisions.  All particles incident
on LEDA were tracked with GEANT~\cite{crn:geant} with full tracking of
the produced \v{C}erenkov photons, which is the dominant component of the
shower observed in the lead-glass.  The results shown were obtained
using the GCALOR hadronic shower package for
hadrons~\cite{ornl:gab77}.  Reasonably good
agreement with measurement is observed.  In addition, the GEANT
calculations were performed using the GHEISHA and FLUKA hadronic
shower packages. The GHEISHA results were in better agreement with
measurement while the FLUKA results overpredicted the observed
charged/neutral ratio with the GCALOR result intermediate between the
other two results.  The proper description of hadronic showers in the
lead-glass is an especially severe test of the hadronic shower
packages. Since only that component of the shower which produces
\v{C}erenkov light which reaches the photomultiplier contributes to
the observed signal, the lead-glass response to hadrons is sensitive
to details of the hadronic shower composition. For the charged/neutral
corrections the measured results have been used. In contrast, the
corrections for neutrons and anti-neutrons discussed below are, by
necessity, based solely on the contributions calculated using GEANT.

The measured charged/neutral ratio spectra, $(c/n)$, are fitted
(as seen in Figs.~\ref{fig:charge_neut} and
\ref{fig:charge_neut_disp}) to remove statistical fluctuations and to
extrapolate to high $p_T$. The lowest $p_T$ bins are not used in the
present analysis due to the sharply increasing rise in the hadron
contamination.\footnote{Note that while results will be shown to low
   $p_T$ for various intermediate analysis steps, the final results
   will be presented only for the region $p_T>0.5$ GeV/c due to various
   systematical error sources which increase at low $p_T$.} The fitted
result provides the neutral/total shower ratio $1/(1+c/n)$ which is
then used as a multiplicative factor to extract the neutral shower
spectrum from the total shower spectrum. This charged hadron
correction procedure was necessary for the analysis of the 1995 data
set since there was no CPV measurement to extract the charged hadron
contamination information.  It also allowed to use the full 1996 data
sample including periods when the CPV was not fully operational. It
should be noted that the 1995 and 1996 run conditions and analysis
procedures were the same, and so the amount of charged hadron
contamination extracted from the 1996 data sample should be the same
for the 1995 data sample. A minor difference between the two data
samples is the thicker target used for the 1995 run period which
resulted in an additional 1.55\% conversion probability. For the
magnet off data sample this is expected to have no effect since
essentially all converted photons are identified as single showers,
which means that the total shower spectrum is unchanged. Therefore the
charged/neutral shower correction factor is the same nearly
independent of the amount of conversions.  For the magnet on run
conditions the $e^+e^-$ pair produced at the target will separate in the
magnetic field and the total shower spectrum will depend on the amount
of photon conversions. Therefore a small correction ($\approx 1\%$)
for this effect has been applied to the charge/neutral ratio when used
for the 1995 magnet on data.  Based on the results shown in these
figures, a conservative 30\% uncertainty has been assumed for the
charged/neutral ratio for the photon analysis.

%%fig12%%
\begin{figure}[hbt]
\begin{center}
   \includegraphics[scale=0.5]{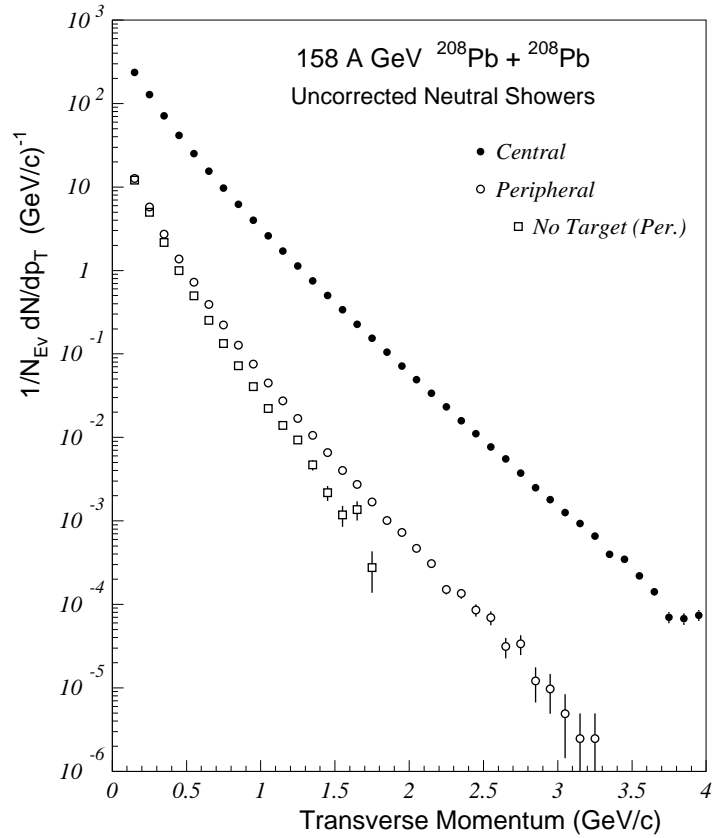}
\caption{The multiplicity  of all neutral showers as a function of
   transverse momentum for peripheral (open circles) and central (solid
   circles) 158~{\it A}~GeV $^{208}$Pb\/+\/$^{208}$Pb collisions. The
   results are the sum of 1995 and 1996 magnet on data sets. Results are also
   shown (open squares) for peripheral event selection for runs with no
   target.  The spectra have not been corrected for efficiency or
   acceptance. They are used as the first iteration in the iterative
   efficiency correction.}
\label{fig:raw_photons}
\end{center}
\end{figure}

The raw neutral shower multiplicity per event as a function of
transverse momentum, uncorrected for neutrons, efficiency, or
acceptance, are shown in Fig.~\ref{fig:raw_photons}. The results are
shown for central and peripheral collisions for the sum of 1995 and
1996 data samples. The measured distributions extend over more than 6
orders of magnitude and extend to about 3 and 4 GeV/c for peripheral
and central collisions, respectively. A flattening of the distribution
is observed in going from peripheral to central collisions.  Also
shown is the distribution obtained for events satisfying the
peripheral trigger condition for runs with no target. No events from
empty target runs satisfy the central event trigger condition (see
Fig.~\ref{fig:events}). The empty target data is seen to have a
neutral shower multiplicity which is similar to the peripheral
$^{208}$Pb\/+\/$^{208}$Pb case but to fall more steeply with
transverse momentum. The similarity is not surprising since the
trigger selection on the transverse energy for peripheral Pb
collisions will select events with a similar number of nucleon-nucleon
collisions as for an empty target interaction.  The no-target
interactions most likely occur on light materials of the target wheel
and vacuum system (aluminum or carbon). The steeper falling empty
target spectrum results from the fact that many of those interactions
occur on downstream materials and therefore have underestimated angles
and correspondingly smaller apparent transverse momenta.  Since the
spectral shapes differ, the peripheral spectrum should be corrected
for the effect of the empty target contribution.  The true peripheral
spectrum, $S^{True}_{Per}$, should be obtained from the raw peripheral
spectrum, $S_{Per}$, and empty target spectrum $S_{Empty}$ as
\begin{equation}
S^{True}_{Per} = S_{Per} + f\cdot(S_{Per}-S_{Empty}),
\label{eq:spect_corr}
\end{equation}
where f is the fraction of empty target events to true Pb target
events in the peripheral data sample. Based on the live-beam scalers,
downscale factors, and number of peripheral triggers a value of
$f=0.098$ is obtained for the results shown in
Fig.~\ref{fig:raw_photons} where the empty target results are obtained with
the same trigger cleanup cuts as used for the Pb data. On the other
hand, as will be discussed in regard to the $\pi^0$ result in
Sec.~\ref{sec:pi0s} below, there are reasons to believe that the
target out contribution to the peripheral data may be larger than
indicated by the properly normalized empty target data. With the
requirement of a coincident hit in the forward region of the Plastic
Ball detector (see Sec.~\ref{sec:selection}) as for the 1995 data
sample, the properly normalized fraction of empty target events in the
peripheral data sample is only $f=0.01$.

\subsubsection{\label{sec:nnbar} Neutron and Anti-neutron Corrections}

The neutral shower spectra must be corrected for neutron and
anti-neutron contributions to obtain the raw photon spectrum.  For
this correction it is necessary to rely entirely on results from
simulation. The incident neutron and anti-neutron flux into the LEDA
acceptance has been estimated using VENUS. In
Fig.~\ref{fig:neut_antineut}a) the VENUS predictions of the number of
particles per decay photon are shown as a function of their incident
transverse momentum for neutrons, anti-neutrons, and $\pi^+$. In the
region of greatest interest at high $p_T$, the neutron flux is seen to
exceed the $\pi^+$ flux by roughly an order of magnitude. Even the
incident anti-neutron flux is similar to the $\pi^+$ flux at large
$p_T$. Fortunately, as for the charged hadrons, the neutrons and
anti-neutrons deposit only a small portion of their incident energy in
the lead-glass detector which results in a much reduced apparent
transverse momentum.

The response of the lead-glass detector to the neutrons and
anti-neutrons, and to photons was simulated with GEANT
with full tracking of the produced \v{C}erenkov photons. 
The resulting
ratio of neutron+anti-neutron $(n+\overline{n})$ to total neutral
$(n+\overline{n}+\gamma)$ showers as a function of their apparent
transverse momentum is shown in Fig.~\ref{fig:neut_antineut}b). The
$(n+\overline{n})$ contamination is seen to be significantly
reduced in comparison to the incident flux due to the small energy
deposit. The contamination is dominated by the neutron contribution.
Results are shown with and without application of the shower
dispersion cut. It is seen that the requirement of a narrow
photon-like dispersion significantly reduces the $(n+\overline{n})$
contamination from about 5\% to 1-2\%.

The results shown have been calculated using the GCALOR hadronic
shower package~\cite{ornl:gab77} of GEANT. Calculations were also performed using the
GHEISHA and FLUKA hadronic shower packages.  GHEISHA predicted
$(n+\overline{n})$/neutral ratios which were nearly a factor of two
lower than the FLUKA predictions while the GCALOR result was about
1-2\% in value below the FLUKA result for the case of all showers. With the
narrow shower condition applied GCALOR and FLUKA gave consistent
results. While no hadronic shower package has been clearly
demonstrated to be superior~\cite{atl:fer96}, we have chosen to use
the GCALOR results, since GCALOR, as well as FLUKA, is considered to
be more reliable for neutron transport.

The proton spectrum predicted by VENUS for central
$^{208}$Pb\/+\/$^{208}$Pb is considerably flatter than the measured
spectrum reported by NA49~\cite{npc:afa96}. This would suggest that
VENUS also overpredicts the neutron and anti-neutron yield at high
transverse momentum, which would suggest a smaller contamination.  On
the other hand, VENUS also overpredicts the $\pi^0$ and hence
inclusive photon yield at high $p_T$~\cite{prl:agg98}. The open points
in Fig.~\ref{fig:neut_antineut}a) show an experimental estimate of the
neutron$/\gamma$ ratio based on the NA49 proton measurement and the
present WA98 photon measurement. The NA49 proton transverse mass
distribution (measured over the interval $0.< m_T - m_p < 0.3$
GeV/c$^2$ ~\cite{npc:afa96}) has been fitted to an exponential with
the integral yield normalized to the predicted VENUS neutron
multiplicity.  It is seen that the neutron$/\gamma$ ratio estimated
from the experimental results is quite similar to that predicted by
VENUS, even in the region extrapolated beyond the NA49 measurement.
Moreover, after GEANT response the final $(n+\overline{n})/$neutral
ratio is very similar to the VENUS result over the entire $p_T$
region. In view of the uncertainties inherent in comparing different
experimental results with different event selections, especially for
peripheral collisions, the VENUS predictions have been used for the
neutron and anti-neutron corrections. Due to these uncertainties, and
especially those uncertainties associated with the simulation of the
neutron response, a 50\% uncertainty has been assumed for the
$n+\overline{n}$ contribution to the photon result.

\vbox{
%%fig13%%
\begin{figure}[hbt]
\begin{center}
   \includegraphics[scale=0.5]{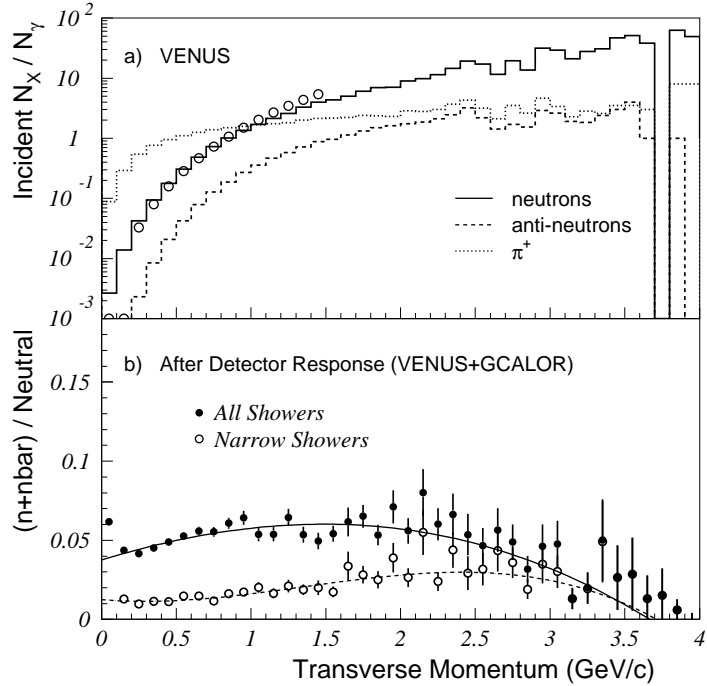}
\caption{In part a) the ratio of neutrons (solid line), anti-neutrons (dashed
   line), and $\pi^+$ (dotted line) to photons into the LEDA acceptance
   is shown as a function of the transverse momentum. The results are
   predictions of the VENUS 4.12 event generator for central
   158~{\it A}~GeV $^{208}$Pb\/+\/$^{208}$Pb collisions.  The open
   points are described in the text.  In part b) the ratio of neutrons
   plus anti-neutrons to neutral ($n + \overline{n}+\gamma$) is shown
   after the effects of the lead-glass detector response calculated
   with GEANT. Results are shown after application of shower
   identification criteria of using all showers (solid circles) or
   narrow showers (open circles). The curves are fit results.}
\label{fig:neut_antineut}
\end{center}
\end{figure}
}

The $(n+\overline{n})/$neutral ratio is fitted (as shown in
Fig.~\ref{fig:neut_antineut}b)) and the uncorrected neutral spectra
(see Fig.~\ref{fig:raw_photons}) are then corrected by the factor
$\gamma/$neutral$=1-(n+\overline{n})/(n+\overline{n}+\gamma))$ to
obtain the raw photon spectrum.
%tca \footnote{Actually, the
%tca    $(n+\overline{n})$ contribution is increased by the factor
%tca    $1/(1-P_C)$ to account for the conversion correction applied to the
%tca    measured neutral spectra as discussed in Sec.~\ref{sec:gamchrg}.}.
The anti-neutrons comprise about one third of the total neutral
correction.  The result is the raw inclusive photon spectra which must
then be corrected for the photon reconstruction efficiency and
acceptance.

\subsubsection{\label{sec:gameff} Reconstruction Efficiency}

As described in Sec.~\ref{sec:pideff}, the photon reconstruction
efficiency is extracted by the method of inserting GEANT photon test
showers into real events and determining how the test showers are
modified or lost. The efficiency is extracted as the ratio of the
transverse momentum spectrum of found photons divided by the
transverse spectrum of input photons. The input photon and its
associated found photon are weighted such that the input distribution
reproduces the measured transverse momentum distribution for each
event class. The input weights are taken according to the raw photon
distribution (see Fig.~\ref{fig:raw_photons}) for the initial
efficiency result and the weights and resulting efficiency are
iterated until the input distribution agrees with the final result.

%%fig14%%
\vbox{
\begin{figure}[hbt]
\begin{center}
   \includegraphics[scale=0.5]{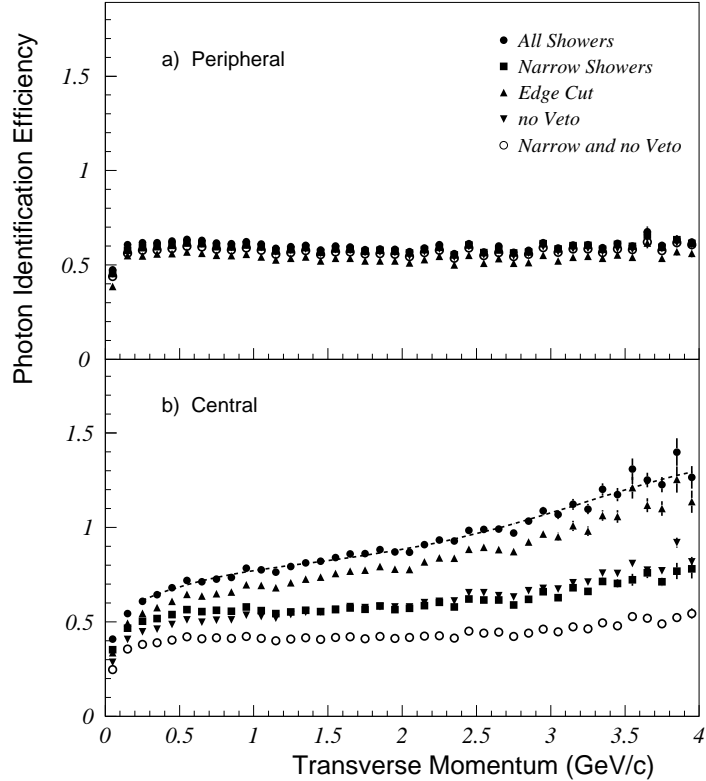}
\caption{The photon identification efficiency including effects
   of resolution, shower overlap, and excluded lead-glass modules
   as a function of transverse
   momentum for a) peripheral and b) central 158~{\it A}~GeV
   $^{208}$Pb\/+\/$^{208}$Pb collisions.  The efficiencies are shown
   for several shower identification methods using the 1996 magnetic
   field on data. The efficiency is shown for the following criteria:
   all showers (solid circles), showers satisfying a dispersion cut
   (solid squares), all showers within a further restricted acceptance
   (solid triangles), all showers with no associated hit in the CPV
   (inverted solid triangles), and all showers satisfying a dispersion
   cut with no associated hit in the CPV (open circles). An example of
   the fitted parameterization of the efficiency is shown by the dashed
   curve in part b) for the criterion of using all showers.}
\label{fig:photon_eff}
\end{center}
\end{figure}
}

The final photon identification efficiency results are shown in
Fig.~\ref{fig:photon_eff} for peripheral and central collisions in
parts a) and b), respectively. The results are shown for the various
photon idenification criteria (see Sec.~\ref{sec:pid}): All showers
(S1); Narrow showers surviving a shower dispersion cut (S2); All
showers within an increased edge cut (detector edge dead region
increased from 2 modules to 3 modules); All showers with no associated
CPV hit (S3); and Narrow showers with no associated CPV hit (S4). As
previously discussed, what is called identification efficiency should
more properly be called a response correction. It includes acceptance
losses resulting from the fiducial cuts to define the useable detector
region as well as dead or bad modules eliminated from the analysis.
The acceptance is calculated for the full geometrical area of the
lead-glass detector while the identification efficiency corrects for
modules removed from the acceptance in the analysis.  The efficiency
correction also includes resolution corrections which can increase the
apparent efficiency to values greater than one.

For peripheral reactions it is seen that the photon identification
efficiency is only about 60\% independent of transverse momentum. This
reflects the loss of acceptance due to such eliminated modules. The
efficiency is seen to be slightly smaller with the increased edge cut
reflecting the decreased fiducial region. Otherwise, the photon
efficiency is nearly independent of the identification method for
peripheral collisions.  This indicates that few photons are lost by
these identification criteria or by shower overlap effects.

In contrast, for central collisions a large difference in photon
identification efficiency is obtained for the different identification
criteria.  For the case of using all recovered showers the efficiency
is seen to rise strongly with transverse momentum. This is understood
as a result of the increase in the shower energy, and hence its
transverse momentum, as a result of absorbing the energy of underlying
showers which have been overlapped. Due to the steeply exponential
transverse momentum distribution of the photons (see
Fig.~\ref{fig:raw_photons}) there is a large feeddown of showers from
low $p_T$ to high $p_T$ when the shower energy is increased due to
overlap. When the shower dispersion cut is applied this rise in the
efficiency is dramatically reduced, indicating that many overlapping
showers are eliminated because they are found to be too broad to be
single photons (recall that the clustering algorithm separates obvious
shower overlap clusters with multiple maxima into multiple clusters
with single maxima). The photon identification efficiency is reduced
further when it is required that there be no associated hit in the CPV
detector. This reduction is due to the elimination of photon showers
which overlap with charged hadrons which would otherwise appear as a
single photon shower.  As discussed previously (see
Secs.~\ref{sec:pid} and \ref{sec:cpv}), all photon identification
methods use the CPV information to eliminate the charged hadron
contamination. The difference between methods S1, S2 and methods S3,
S4 is mostly a matter of procedure. In methods S3 and S4 in which the
CPV is used directly to reject charged showers no charged hadron
correction of the photon candidate spectrum is necessary. On the other
hand, as seen in Fig.~\ref{fig:photon_eff} the photon losses due to
random overlap with charged hits are greater, resulting in a reduced
photon identification efficiency.

The photon identification efficiencies are fitted (see
Fig.~\ref{fig:photon_eff}) to remove fluctuations\footnote{Note: The
   fluctuations for different identification methods are correlated
   since the efficiencies are extracted from the same simulated shower
   sample for all methods} and the fitted efficiency corrections are
applied to the raw photon spectra to obtain the final photon
transverse momentum distributions which need only be corrected for the
LEDA acceptance.  The identification efficiencies obtained by
iteration were compared to efficiencies obtained by the
two-dimensional unfolding method~\ref{sec:pideff} .
The systematical error on the photon identification
efficiency is estimated to be 2\%.

\subsubsection{\label{sec:gamerr} Systematical error}

The systematical errors relevant for the direct photon analysis which
contribute solely to the extraction of the inclusive photon yield are listed
in Table~\ref{table3} at two representative transverse momenta. The
errors are estimated for the case of photons identified with the
narrow shower criterion (S2).  The 30\% uncertainty in the
charged/neutral measurement discussed in regard to
Figs.~\ref{fig:charge_neut} and \ref{fig:charge_neut_disp} leads to
the listed $p_T$ dependent uncertainty in the photon yield due to the
charged particle background correction. Based on a comparison of the
magnetic field on and field off results, and a comparison with and
without the CPV requirement, a 0.5\% uncertainty on the photon
conversion probability has been assumed.  The assumed 50\% uncertainty
on the neutron and anti-neutron contribution discussed in regard to
Fig.~\ref{fig:neut_antineut} results in the listed $p_T$ dependent
uncertainty in the photon yield.  Finally a 2.0\% uncertainty in the
photon identification efficiency has been assumed. The systematical
errors are added in quadrature to give the total systematical error on
the photon yield measurement listed in Table~\ref{table3}. Other
sources of systematical error, such as the energy calibration and
non-target backgrounds, which also affect the $\pi^0$ yield extraction
will be discussed in Sec~\ref{sec:excess}.

\begin{figure}[hbt]
\begin{center}
   \includegraphics[scale=0.5]{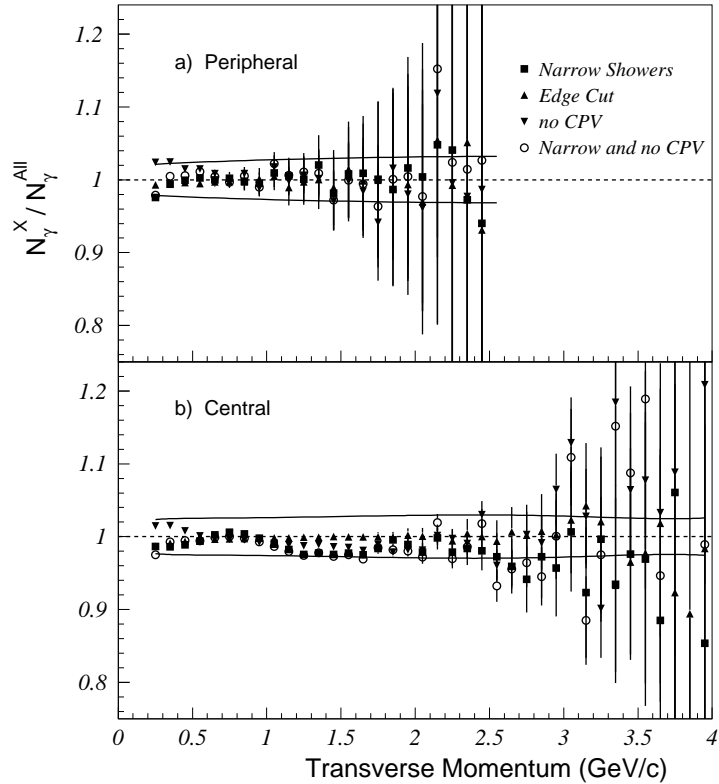}
\caption{The ratio of the efficiency-corrected inclusive photon yield for
   various methods compared to the method of using all showers
   ($N^{All}_\gamma$) is shown as a function of transverse momentum.
   The results are shown for a) peripheral and b) central
   158~{\it A}~GeV $^{208}$Pb\/+\/$^{208}$Pb collisions for the 1996
   magnet on data set.  The error bars indicate statistical errors
   only.  The deviation from unity, beyond statistical error, is
   indicative of the systematical error attributed to the photon
   identification method.  The horizontal lines indicate the estimated
   systematical error on the photon yield determination.}
\label{fig:photon_err}
\end{center}
\end{figure}

The total systematical error  on the photon yield measurement
can be investigated by comparison of the inclusive photon results
obtained with the different photon identification criteria. In
particular, it was shown above that the charged and neutron
backgrounds are reduced by about a factor of two when the shower
dispersion cut is applied. Also, the photon identification
efficiencies are observed to vary by a factor of 2-3 for central
collisions depending on the identification criterion (see
Fig.~\ref{fig:photon_eff}). The results of such a comparison are shown
in Fig.~\ref{fig:photon_err} for peripheral and central collisions
where the photon transverse momentum spectra obtained with the various
photon identification criteria are divided by the photon spectrum
obtained using the criterion of all showers.  Since the corrections
are largest for the condition of using all showers, it is the method
expected to have the largest systematical error, with errors which
should be larger than those which have been estimated for the narrow
shower condition.  The $p_T$ dependent upper and lower systematical
errors on the photon yield measurement are indicated by the horizontal
lines. In general, to the extent allowed by the statistical
uncertainties, one may conclude that the various results are consistent
within the systematical error estimates.
%tca The small deviation in the region of $p_T=0.5-1.$ is likely attributed
%tca to an overestimate of the neutron contamination, which would
%tca result in a larger underestimate of the photon yield for the
%tca case in which all photons are used (see Fig.~\ref{fig:neut_antineut}).

%%fig15%%

\subsection{\label{sec:pi0s} $\pi^{0}$ Analysis}

The method to extract the $\pi^{0}$ transverse momentum spectra is
described in Sec.~\ref{sec:pid}. In summary, the procedure is to
calculate the two-gamma invariant mass spectrum for each $p_T$
bin and extract the yield in the $\pi^0$ peak. This is done using the
various photon identification criteria and run conditions and the
final results are checked for consistency. While charged hadron and
neutron contamination in the selected photon showers will not directly
contribute to the $\pi^0$ peak and yield, they will contribute excess
photon pairs in the combinatorial background which must be subtracted
to obtain the yield in the peak. The large combinatorial background in
the $m_{\gamma\gamma}$ invariant mass distribution, especially for central
collisions, poses a special difficulty for the $\pi^0$ analysis. After
the raw $\pi^0$ yield is extracted it is corrected for the $\pi^0$
identification efficiency, and finally for the lead-glass detector
acceptance.  The details of this procedure are described in this
section.

\subsubsection{\label{sec:pi0yield} $\pi^{0}$  Yield Extraction}

\begin{figure}[hbt]
\begin{center}
   \includegraphics[scale=0.5]{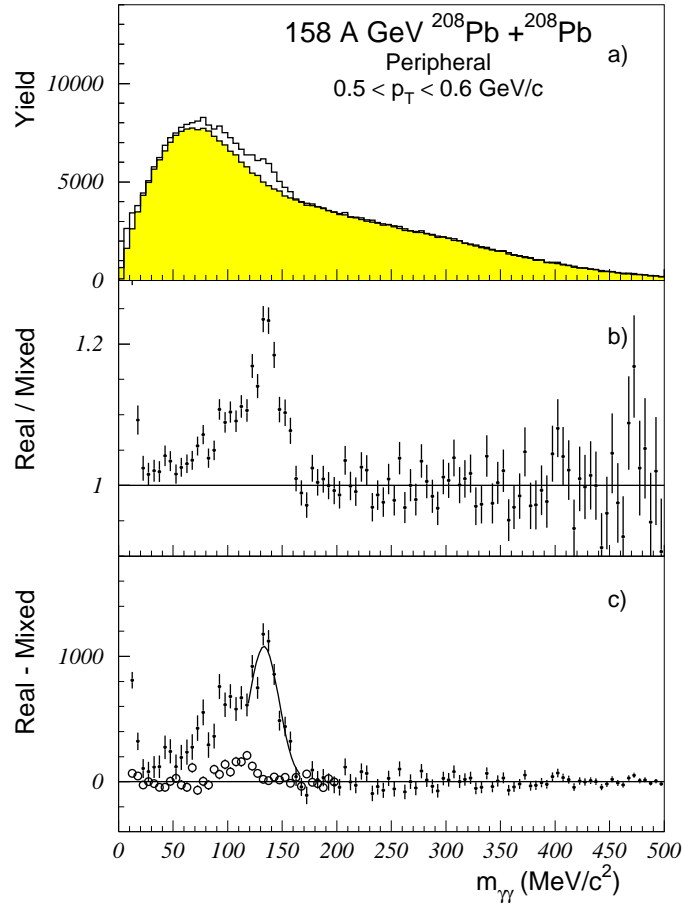}
\caption{The two-photon invariant mass distribution for peripheral
   events in the $\pi^{0}$ mass region with pair transverse momentum
   $0.5 < p_T < 0.6$ GeV/c. Results are shown for the 1996 magnet on
   data sample.  Part a) shows the invariant mass distribution for all
   shower pairs for real events (solid histogram) and mixed events
   (filled histogram). Part b) shows the ratio of real and mixed event
   mass distributions. Part c) shows the real invariant mass
   distribution after subtracting the normalized mixed event
   distribution. The normalized target out background contribution is
   shown by the open circles in part c).}
\label{fig:pi0_per_lo_pt}
\end{center}
\end{figure}

\begin{figure}[hbt]
\begin{center}
   \includegraphics[scale=0.5]{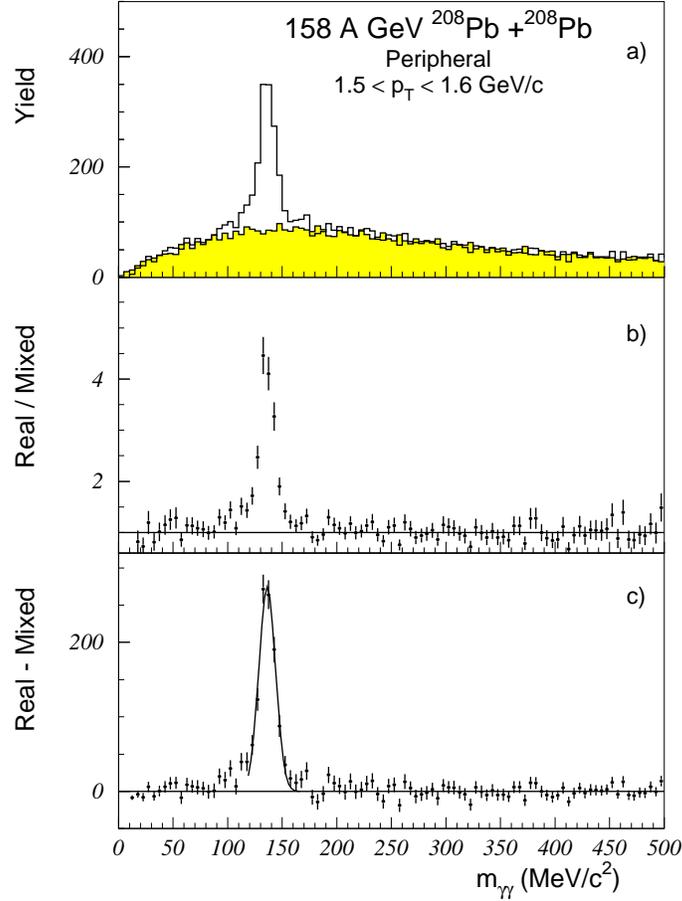}
\caption{The two-photon invariant mass distribution for peripheral
   events in the $\pi^{0}$ mass region with pair transverse momentum
   $1.5 < p_T < 1.6$ GeV/c. Results are shown for the 1995 magnet on
   data sample. Part a) shows the invariant mass distribution for all
   shower pairs for real events (solid histogram) and mixed events
   (filled histogram). Part b) shows the ratio of real and mixed event
   mass distributions. Part c) shows the real invariant mass
   distribution after subtracting the normalized mixed event
   distribution.}
\label{fig:pi0_per_hi_pt}
\end{center}
\end{figure}

The two-photon invariant mass distributions for peripheral
$^{208}$Pb\/+\/$^{208}$Pb collisions are shown in part a) of
Figs.~\ref{fig:pi0_per_lo_pt} and \ref{fig:pi0_per_hi_pt} for photon
pair transverse momenta in the range of $0.5 < p_T < 0.6$ GeV/c and
$1.5 < p_T < 1.6$ GeV/c, respectively. The distributions are obtained
using all showers which pass the 750 MeV energy threshold.  While it
is evident that the $\pi^0$ peak content can easily be extracted at
high $p_T$, it is seen that the peak sits on top of a rather large and
broad background with a complicated shape for the case of low pair
$p_T$. This combinatorial background arises as a result of ``random''
combinations of photons within the detector acceptance where the pair
of photons did not originate from the same radiative decay (e.g.
different $\pi^0$'s) and hence the two photons have little or no
correlation. Although correlations between the photon pairs could
exist, such as from a residual $\pi^0\pi^0$ Bose-Einstein correlation,
or from a multi-$\pi^0$ decay final state, or from collective flow, it
is expected that the shape of the photon pair combinatorial background
depends mainly on the photon spectrum and on the detector acceptance.
To accurately determine its shape without resort to a complicated fit
procedure, an event-mixing technique has been used in which all
photons of an event are paired with all photons of the next analyzed
event within the same centrality selection. This procedure removes all
resonance pair correlations but leaves some of the higher order
correlations which might also exist in the real-pair background
distribution.  No attempt was made to impose cuts on the mixed event
shower pairs to implement the effect of merging showers. This has the
result that the mixed events are not expected to accurately reproduce
the very low mass region of the real event mass spectra.

%%fig16%%

The mixed event $m_{\gamma\gamma}$ invariant mass distributions are
shown as the shaded histograms in part a) of
Figs.~\ref{fig:pi0_per_lo_pt} and \ref{fig:pi0_per_hi_pt}.  The ratios
of the real to mixed invariant mass distributions are shown in part
b). The ratios have been normalized to unity in the region outside of
the $\pi^0$ mass interval. For peripheral collisions the normalization
has been calculated as the ratio of integrated yields in the mass
intervals $50-60$ and $200-430$ MeV/c$^2$. The normalization extracted
in this way was then fit to a smooth function of $p_T$ to determine
the final mixed event background normalization. The normalized
background subtracted results are shown in part c). The mixed event
distribution is seen to provide a good description of the
combinatorial background outside the $\pi^0$ mass region. However, a
low-mass tail on the $\pi^0$ peak is observed at the lowest $p_T$ (see
Fig.~\ref{fig:pi0_per_lo_pt}c). Such a tail can result from $\pi^0$'s
produced downstream from the target, such as from $K^0_S$ decays (see
Sec.~\ref{sec:simulation}) or from background interactions on
downstream materials.  This will be further discussed below.

%%fig17%%

\begin{figure}[hbt]
\begin{center}
   \includegraphics[scale=0.5]{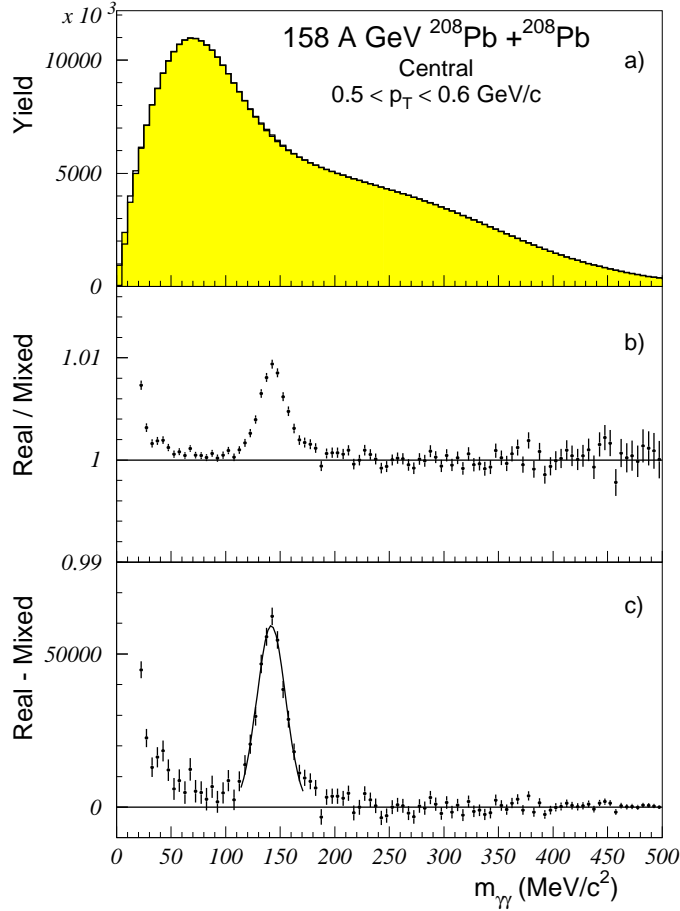}
\caption{The two-photon invariant mass distribution for central
   events in the $\pi^{0}$ mass region with pair transverse momentum
   $0.5 < p_T < 0.6$ GeV/c. Results are shown for the 1995 magnet on
   data sample. Part a) shows the invariant mass distribution for all
   shower pairs for real events (solid histogram) and mixed events
   (filled histogram). Part b) shows the ratio of real and mixed event
   mass distributions. Part c) shows the real invariant mass
   distribution after subtracting the normalized mixed event
   distribution.}
\label{fig:pi0_cen_lo_pt}
\end{center}
\end{figure}

\begin{figure}[hbt]
\begin{center}
   \includegraphics[scale=0.5]{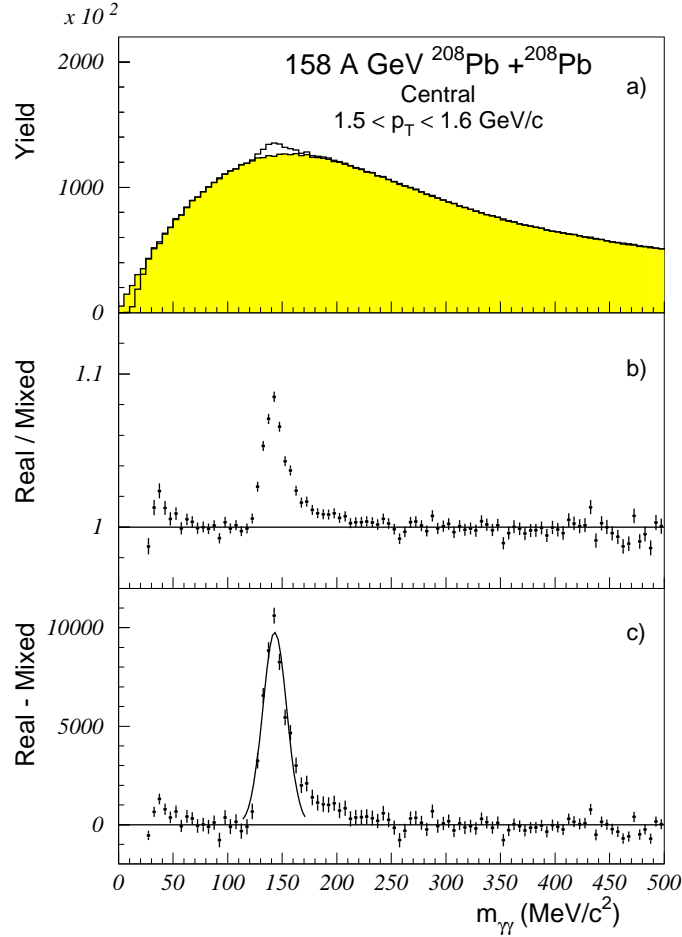}
\caption{The two-photon invariant mass distribution for central
   events in the $\pi^{0}$ mass region with pair transverse momentum
   $1.5 < p_T < 1.6$ GeV/c. Results are shown for the 1995 magnet on
   data sample.  Part a) shows the invariant mass distribution for all
   shower pairs for real events (solid histogram) and mixed events
   (filled histogram). Part b) shows the ratio of real and mixed event
   mass distributions. Part c) shows the real invariant mass
   distribution after subtracting the normalized mixed event
   distribution.}
\label{fig:pi0_cen_hi_pt}
\end{center}
\end{figure}

The $m_{\gamma\gamma}$ invariant mass distributions for central
$^{208}$Pb\/+\/$^{208}$Pb collisions are shown in part a) of
Figs.~\ref{fig:pi0_cen_lo_pt} and \ref{fig:pi0_cen_hi_pt} for photon
pair transverse momenta in the range of $0.5 < p_T < 0.6$ GeV/c and
$1.5 < p_T < 1.6$ GeV/c, respectively.  Due to the much higher photon
multiplicity in central events (see Fig.\ref{fig:events}) the
combinatorial backgrounds are much greater. As a result it is seen
that the $\pi^0$ peak is hardly visible at low $p_T$. In the same way
as for peripheral events, the combinatorial background has been
calculated using mixed events and normalized to the real event
invariant mass distribution using the mass intervals $70-95$ and
$220-430$ MeV/c$^2$ with the normalizations fitted to a smooth $p_T$
dependence. The real/mixed event invariant mass distribution ratios
are shown in part b) and the final normalized background subtracted
invariant mass distributions are shown in part c). As for the case of
peripheral collisions, the event mixed background is seen to provide a
good reproduction of the combinatorial background in the region around
the $\pi^0$ peak.

%%fig18%%

Several features of the results for central collisions are notable in
comparison to the results for peripheral collisions. First, it is
observed that the $\pi^0$ peak is broader and shifted to higher mass
as compared to the case for peripheral collisions. This is due to the
effects of shower overlap and the modification of the energy and
position of the original shower. Secondly, the $\pi^0$ peak is seen to
exhibit a long non-gaussian tail to high mass which is also attributed
to shower overlap effects. And finally, the low mass tail on the
$\pi^0$ peak at low $p_T$ is not present. The mixed event invariant
mass normalization regions were chosen differently for central
collisions due to these last two observations. In addition, the
invariant mass distribution is observed to show structure in the very
low mass region below about 50 MeV/c$^2$. This structure is attributed
to several effects, such as splitting of overlapping showers and the
above-mentioned details of the treatment of the mixed events for
nearby showers. It does not affect the analysis presented here and
will not be discussed further.

\begin{figure}[hbt]
\begin{center}
   \includegraphics[scale=0.5]{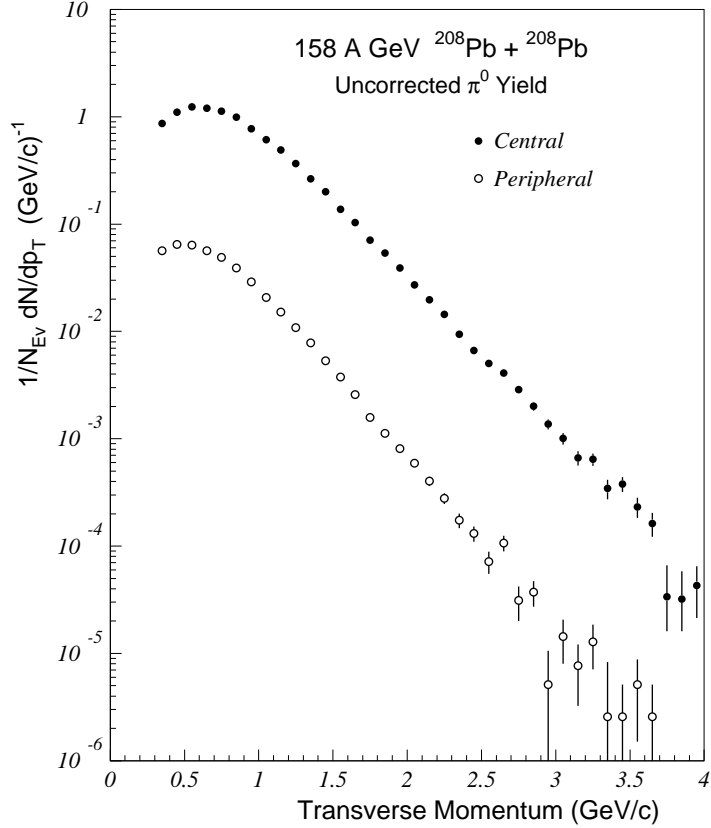}
\caption{The uncorrected $\pi^{0}$ multiplicity as a function of
   the transverse momentum for peripheral (open circles) and central
   (solid circles) 158~{\it A}~GeV $^{208}$Pb\/+\/$^{208}$Pb
   collisions. The results are for the full magnet on data sample.
   Narrow showers with energy above 750 MeV have been used in the
   invariant mass calculation. The data have not been corrected for
   efficiency or acceptance. These spectra are used as the first
   iteration in the iterative efficiency correction.}
\label{fig:raw_pi0}
\end{center}
\end{figure}

The $\pi^0$ transverse momentum spectra are obtained by integration of
the yield in the $\pi^0$ peak region of the combinatorial background
subtracted $m_{\gamma\gamma}$ invariant mass distributions for each $p_T$
bin. The integration regions used were $110 < m_{\gamma\gamma} < 170$
MeV/c$^2$ for peripheral collisions and $110 < m_{\gamma\gamma} < 200$
MeV/c$^2$ for central collisions.  The wider integration region used
for central collisions is due to the previously discussed high mass
tail on the $\pi^0$ peak which results from overlap effects.  The uncorrected
transverse momentum distributions are shown in Fig.~\ref{fig:raw_pi0}
for peripheral and central collisions. The results are obtained using
the full 1995 and 1996 magnet on data sample with the narrow shower
identification criterion (S2) (see Sec.~\ref{sec:pid}).  The
distributions are observed to extend over about four orders of
magnitude with similar $p_T$ coverage but lower statistical accuracy
as compared to the inclusive photon result shown in
Fig.~\ref{fig:raw_photons}. The distributions are observed to be
cut off at low $p_T$ at around 0.5 GeV/c due to the acceptance limit
(see Fig.~\ref{fig:acceptance}) imposed by the 750 MeV shower energy
threshold.

%%fig19%%

As noted in the discussion of the inclusive photon result, no empty
target events satisfy the central collision trigger requirement with
the result that no background corrections to the photon and $\pi^0$
spectra are necessary for central collisions. On the other hand, as
discussed with regard to Fig.~\ref{fig:pi0_per_lo_pt}, the
$m_{\gamma\gamma}$ invariant mass distributions for low $p_T$
peripheral events show a low mass tail which suggests a contribution
of $\pi^0$'s produced downstream of the target location. In principal,
this can be verified by comparison to the $m_{\gamma\gamma}$
distribution obtained for runs taken with no target. However, when 
the empty target events are analyzed with the trigger 
requirement of a hit in the forward region of
the Plastic Ball detector to eliminate 
downstream interactions, as used for analysis of the peripheral events, 
no significant peak is observed in
the empty target $m_{\gamma\gamma}$ distribution.
On the other hand, when the Plastic Ball hit condition is
removed, a small low mass peak is observed in the empty target
$m_{\gamma\gamma}$ distribution, as shown by the open symbols in
Fig.~\ref{fig:pi0_per_lo_pt}c. With the same Plastic Ball condition as
used for the Pb target the empty target contribution is reduced by a
factor of more than five (see the discussion of
Fig.~\ref{fig:raw_photons}). Even with the Plastic Ball condition
removed, the empty target yield,
appropriately normalized according to the relative number of live beam
triggers,  is about a factor of three lower than
the observed low mass yield with the Pb target.

%%fig20%%

The results shown in Fig.~\ref{fig:pi0_per_lo_pt} are for the 1996
magnet on data sample.  A slightly larger low mass contribution with a
more distinctive peak at around $m_{\gamma\gamma}\approx 100$
MeV/c$^2$ is observed for the 1995 data sample. As for the 1996 data,
there is no such contribution seen for central collisions or for high
$p_T$ peripheral collisions (the 1995 results are shown in
Figs.\ref{fig:pi0_per_hi_pt},\ref{fig:pi0_cen_lo_pt},\ref{fig:pi0_cen_hi_pt}).
The additional 1995 low mass contribution is attributed to downstream
interactions on an aluminum ring on the vacuum exit window which was
removed for the 1996 run period (see Sec.~\ref{sec:detectors}). For
the 1995 run period the Plastic Ball hit requirement was implemented
directly in the online trigger. With this condition, as for the 1996
run, no
significant low mass contribution was observed in the 1995 empty
target data. The explanation for these observations is that the
Plastic Ball condition was in fact effective to remove downstream
interactions when there was no target in place. However, when the
target was in place, $\delta$-electrons produced in the target could
be detected in the Plastic Ball and satisfy the Plastic Ball trigger
requirement. This meant that more downstream interactions could be
accepted with target in place than without target. Nevertheless, even
with the Plastic Ball trigger requirement removed, the empty target
result is about a factor of three below the observed low mass excess,
as noted above. This observation is not understood.

For the direct photon analysis the effect of the empty target
contribution has been studied under two extreme assumptions. In the
first case it has been ignored completely, while in the second case
the result without the Plastic Ball condition has been renormalized to
be as large as possible consistent with the result for the Pb target.
That is, the empty target result was increased to remove as much as
possible of the low mass excess in the $m_{\gamma\gamma}$ distribution
at low $p_T$.  While these two assumptions will give different photon
and $\pi^0$ $p_T$ distributions, it might be expected that the direct
photon result is not too dependent on this assumption as long as both
the photon and $\pi^0$ yields from all sources are taken into account
consistently.  Finally, we emphasize again that this non-target
background uncertainty is only relevant for the low $p_T$ 
(below $\sim 1$ GeV/c) peripheral data sample.

%%fig21%%
\begin{figure}[hbt]
\begin{center}
   \includegraphics[scale=0.5]{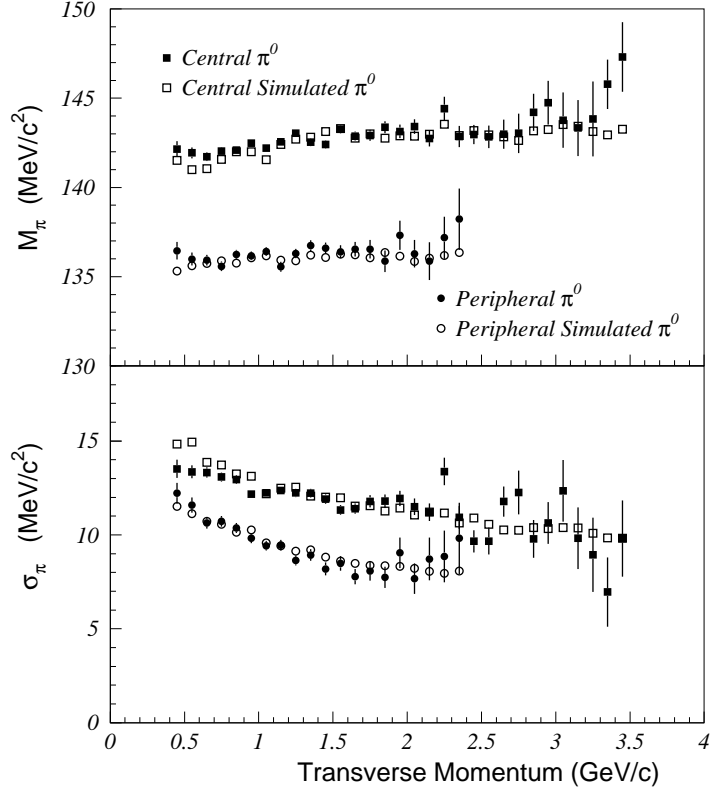}
\caption{The $\pi^{0}$ mass peak position and width, $\sigma$,
   resulting from a gaussian fit to the two-photon invariant mass
   distribution is shown as a function of the transverse momentum in
   the upper and lower portions of the figure, respectively.  The solid
   symbols show results for real $\pi^{0}$'s while the open points show
   the results for simulated $\pi^{0}$'s superimposed on real events.
   Results are shown for peripheral (circles) and central (squares)
   158~{\it A}~GeV $^{208}$Pb\/+\/$^{208}$Pb collisions using all
   shower candidates. The same fit region is used for the real and
   simulated data. }
\label{fig:pi_mass}
\end{center}
\end{figure}

\subsubsection{\label{sec:pi0eff} $\pi^0$ Reconstruction Efficiency}

The procedure to determine the $\pi^0$ reconstruction efficiency is
similar to the one for photons described in Sec.~\ref{sec:gameff}.  As
described in Sec.~\ref{sec:pideff}, the $\pi^0$ reconstruction
efficiency is extracted by inserting test $\pi^0$'s into real events
and determining how the test $\pi^0$'s are modified or lost. The test
$\pi^0$'s were initially generated uniformly in transverse momentum
and pseudo-rapidity. Only those $\pi^0$'s which decayed to give two
photons within the nominal LEDA acceptance were recorded and those
photons were tracked with GEANT.  The efficiency is extracted as the
ratio of the transverse momentum spectrum of found $\pi^0$'s divided
by the transverse momentum spectrum of input $\pi^0$'s. The input
$\pi^0$ and its associated found $\pi^0$ are weighted such that the
input distribution reproduces the measured transverse momentum
distribution for each event class. The initial input weights are taken
according to the raw $\pi^0$ distribution (see Fig.~\ref{fig:raw_pi0})
after application of the acceptance correction and an initial
estimated efficiency correction.  The weights and resulting efficiency
are iterated until the weighted input distribution agrees with the
final result.  The input $\pi^0$ distribution is weighted according to
a Gaussian rapidity distribution centered on mid-rapidty with an rms
width of $\sigma_y = 1.3$\cite{plb:agg99}, consistent with the final
result.

After being superimposed onto real events the simulated $\pi^0$'s are
considered to be found if both photons are recovered and the
reconstructed invariant mass of the photon pair falls within the same
$\pi^0$ mass window as applied in the analysis of the real events (see
Sec.~\ref{sec:pi0yield}).  For an accurate efficiency determination it
is essential that the simulated photon showers have the same
characteristics as the measured photon showers. This can be verified
by comparison of the extracted $\pi^0$ mass peak for real events with
that of the simulated $\pi^0$'s after superposition onto real events.
Such a comparison is shown in Fig.~\ref{fig:pi_mass}. The fitted mass
and width of the $\pi^0$ peak are shown as a function of transverse
momentum for central and peripheral event selections for the case of
using all shower candidates (S1) for the invariant mass calculation.
As noted in the discussion of the invariant mass spectra, the mass and
width of the $\pi^0$ peaks are observed to be significantly larger for
central collisions in comparison to peripheral collisions due to the
effects of shower overlap. With the narrow shower condition (S2), the
$\pi^0$ peak position is about 3 MeV/c$^2$ lower and the $\pi^0$ peak
width is about 1.5 MeV/c$^2$ smaller for central conditions
compared to the results shown in Fig.~\ref{fig:pi_mass}.  The
characteristics of the reconstructed simulated $\pi^0$'s are seen to
be in good agreement with those of the real $\pi^0$'s for both
peripheral and central event selections. This indicates that the
energy calibration and resolution, and their modification due to
detector occupancy effects, are accurately reproduced in the
simulation and therefore are properly taken into account in the
efficiency determination.

%%fig22%%
\begin{figure}[hbt]
\begin{center}
   \includegraphics[scale=0.5]{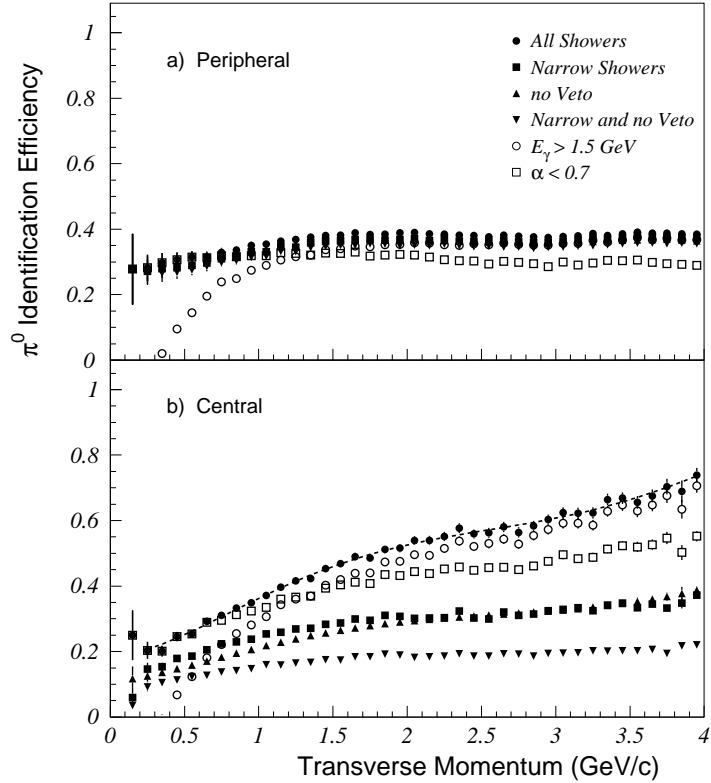}
\caption{The $\pi^{0}$ identification efficiency including effects
   of resolution, shower overlap, and excluded lead-glass modules.  The
   $\pi^{0}$ identification efficiency is shown as a function of
   transverse momentum for a) peripheral and b) central 158~{\it A}~GeV
   $^{208}$Pb\/+\/$^{208}$Pb collisions.  A 750 MeV shower energy
   threshold has been applied.  The efficiency is shown for the 1996
   magnet on data for the following criteria: all showers (solid
   circles), showers satisfying a dispersion cut (solid squares), all
   showers with no associated hit in the CPV (solid triangles), all
   showers satisfying a dispersion cut with no associated hit in the
   CPV (inverted solid triangles), all showers but with shower energy
   threshold increased to 1.5 GeV/c (open circles), and all showers
   with the requirement that the shower-pair energy asymmetry is less
   than 0.7 (open squares). An example of the fitted parameterization
   of the efficiency is shown by the dashed curve in part b) for the
   criterion of using all showers.  }
\label{fig:pi0_eff}
\end{center}
\end{figure}

The final $\pi^0$ identification efficiencies are shown in
Fig.~\ref{fig:pi0_eff} for peripheral and central collisions in parts
a) and b), respectively. The results are shown for the various photon
identification criteria (see Sec.~\ref{sec:pid}): all showers (S1);
narrow showers surviving a shower dispersion cut (S2); all showers
with no associated CPV hit (S3); and narrow showers with no associated
CPV hit (S4); all showers but with the photon energy threshold increased
from 750 MeV to 1.5 GeV; shower pairs which satisfy an energy
asymmetry cut $\alpha=|E_1-E_2|/|E_1+E_2| < 0.7$.  As previously
mentioned, what is called the identification efficiency should more
properly be called a response correction. It includes acceptance losses
resulting from the fiducial cuts to define the useable detector region
as well as dead or bad modules eliminated from the analysis. The
acceptance is calculated for the full geometrical area of the LEDA
while the identification efficiency corrects for modules removed from
the acceptance in the analysis.

For peripheral reactions it is seen that the $\pi^0$ identification
efficiency is only about 30\%, nearly independent of transverse
momentum, which reflects the loss of acceptance due to eliminated
modules. With the 1.5 GeV shower energy threshold the efficiency is
seen to drop sharply at low transverse momentum due to the loss of
acceptance which results from 
the higher energy threshold.  Similarly, the shower
energy asymmetry cut results in a loss of $\pi^0$ acceptance, and
hence a decreased efficiency, at large transverse momenta.  Otherwise,
the $\pi^0$ efficiency is nearly independent of the identification
method for peripheral collisions.  Again, this indicates that few
photons are lost by these identification criteria or by shower overlap
effects.

Similar to the photon identification efficiency, the $\pi^0$
efficiency dependence is quite different for the case of central
collisions.  A large difference in the $\pi^0$ identification
efficiency is obtained for the different identification criteria.  For
the case of using all recovered showers the efficiency is seen to rise
strongly with transverse momentum, exceeding the efficiency for
peripheral collisions. This is understood as a shower overlap effect which
results in an increase in the shower energy, and hence transverse
momentum, as a result of absorbing the energy of underlying showers
which get overlapped. Due to the steeply exponential transverse
momentum distribution of the $\pi^0$'s (see Fig.~\ref{fig:raw_pi0})
there is a large feeddown of the $\pi^0$ yield from low $p_T$ to high
$p_T$ when the shower energy is increased due to overlap.
When the shower dispersion cut is applied this rise in the efficiency
is dramatically reduced, and becomes more similar to the result for
peripheral collisions.  This indicates that many overlapping showers
are eliminated because they are found to be too broad to be single
photons.  The $\pi^0$ identification efficiency is reduced further
when it is required that there be no associated hit in the CPV
detector (methods S3 and S4). This reduction is due to conversions
and the elimination
of photon showers which overlap with charged hadrons which would
otherwise appear as a single photon shower.

The $\pi^0$ identification efficiencies are fitted (see
Fig.~\ref{fig:pi0_eff}) to remove fluctuations\footnote{Note: The
   fluctuations for different identification methods are correlated
   since the efficiencies are extracted from the same simulated shower
   sample for all methods.} and the fitted efficiency corrections are
applied to the raw $\pi^0$ spectra to obtain the final $\pi^0$
transverse momentum distributions which need only be corrected for the
LEDA acceptance.  The systematical error on the $\pi^0$ identification
efficiency is estimated to be 3\% for peripheral collisions and 4\%
for central collisions.

\subsubsection{\label{sec:pi0err} Systematical error}

\begin{figure}[hbt]
\begin{center}
   \includegraphics[scale=0.5]{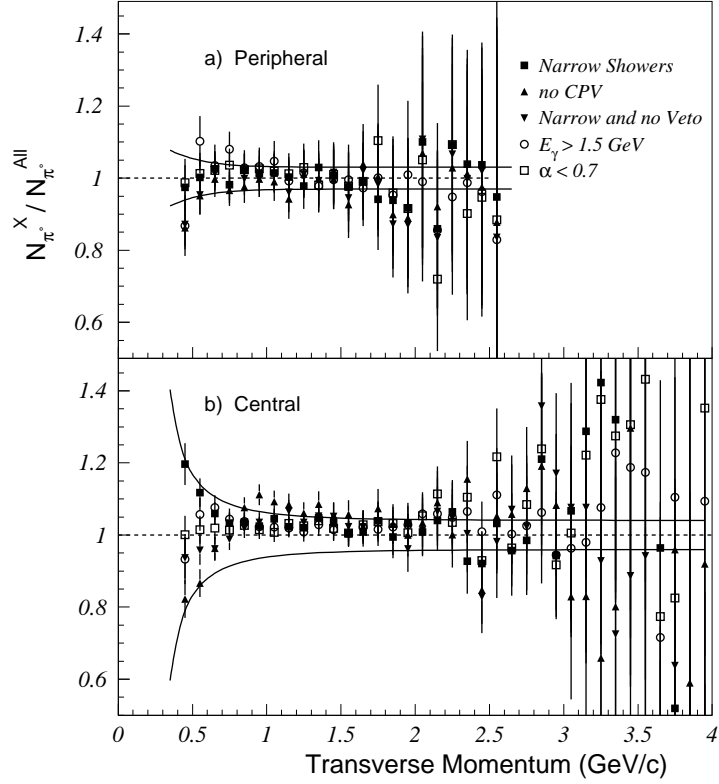}
\caption{The ratio of the efficiency-corrected $\pi^{0}$ yield for
   various shower identification methods compared to the method of
   using all showers is shown as a function of transverse momentum. The
   results are shown for a) peripheral and b) central 158~{\it A}~GeV
   $^{208}$Pb\/+\/$^{208}$Pb collisions for the 1996 data set.  The
   error bars indicate statistical errors only. The horizontal lines
   indicate the estimated systematical errors.  The deviation from unity,
   beyond statistical error, is indicative of the systematical error in
   the $\pi^{0}$ yield extraction. }
\label{fig:pi0_err}
\end{center}
\end{figure}

The systematical errors relevant for the direct photon analysis which
enter only for the inclusive $\pi^0$ yield extraction are listed in
Table~\ref{table3} at two representative transverse momenta. The
errors are estimated for the case of photons identified with the
narrow shower criterion (S2).  Based on a comparison of the magnetic
field on and field off results, and a comparison with and without the
CPV requirement a 0.5\% uncertainty in the $\pi^0$ yield due to photon
conversions has been assumed.  As mentioned above, the uncertainty in
the $\pi^0$ identification efficiency has been assumed to
be 3\% and 4\% for peripheral and central collisions, respectively.
An important source of systematical error in the $\pi^0$ yield
extraction at low $p_T$ for central collisions is the error associated
with the combinatorial background subtraction. The statistical error
of the combinatorial background subtraction is included in the
statistical error on the $\pi^0$ yield. An additional systematical error
of $10^{-3}$ of the background yield has been assumed. This error
contribution is estimated from the variability of the results at low
$p_T$ for central collisions for the different analysis methods (see
Fig.~\ref{fig:pi0_err}).  Besides the small peak/background ratios at
low $p_T$ for central collisions, the curvature of the 
background under the $\pi^0$
peak  may futher complicate the extraction of the
$\pi^0$ yield and increase the error beyond expectations from
the magnitude of the background.  The systematical errors are added in
quadrature to give the total systematical error on the $\pi^0$ yield
extraction listed in Table~\ref{table3}. Other sources of systematical
error, such as the energy calibration and non-target backgrounds,
which affect both the photon and $\pi^0$ yield extraction will be
discussed below.

As for the photon yield determination, the total systematical error
estimate on the $\pi^0$ yield extraction can be investigated by
comparison of the inclusive $\pi^0$ results obtained with the
different identification criteria.  The results of such a comparison
for peripheral and central collisions are shown in
Fig.~\ref{fig:pi0_err} where the $\pi^0$ transverse momentum spectra
obtained with the various conditions are divided by the $\pi^0$
spectrum obtained using the criterion of all showers (S1).  Since the
$\pi^0$ identification efficiencies vary by a factor of 2-3 for
central collisions depending on the identification criterion (see
Fig.~\ref{fig:pi0_eff}) this comparison provides a sensitive
indication of the accuracy of the efficiency determination.  Also, the
uncertainty in the $\pi^0$ peak yield extraction is probed since, for
example, the application of the shower dispersion cut reduces the
non-photon shower contamination and therefore improves the $\pi^0$
peak/background ratio by nearly a factor of two, which reduces the
necessary background correction.  Since the corrections are largest
for the condition of using all showers, it is the method expected to
have the largest systematical error, with errors which should be larger
than those which have been estimated for the narrow shower condition.
The $p_T$ dependent upper and lower systematical errors on the $\pi^0$ 
yield measurement are indicated by the horizontal lines.  In general,
to the extent allowed by the statistical uncertainties, one may
conclude that the various results are consistent within the given
systematical error estimates.  The increasing systematical deviations at
low transverse momentum for central collisions are due to the
above-mentioned systematical error of the combinatorial background
subtraction.

%%fig23%%

\subsection{\label{sec:etas} $\eta$ Analysis}

\begin{figure}[hbt]
\begin{center}
   \includegraphics[scale=0.5]{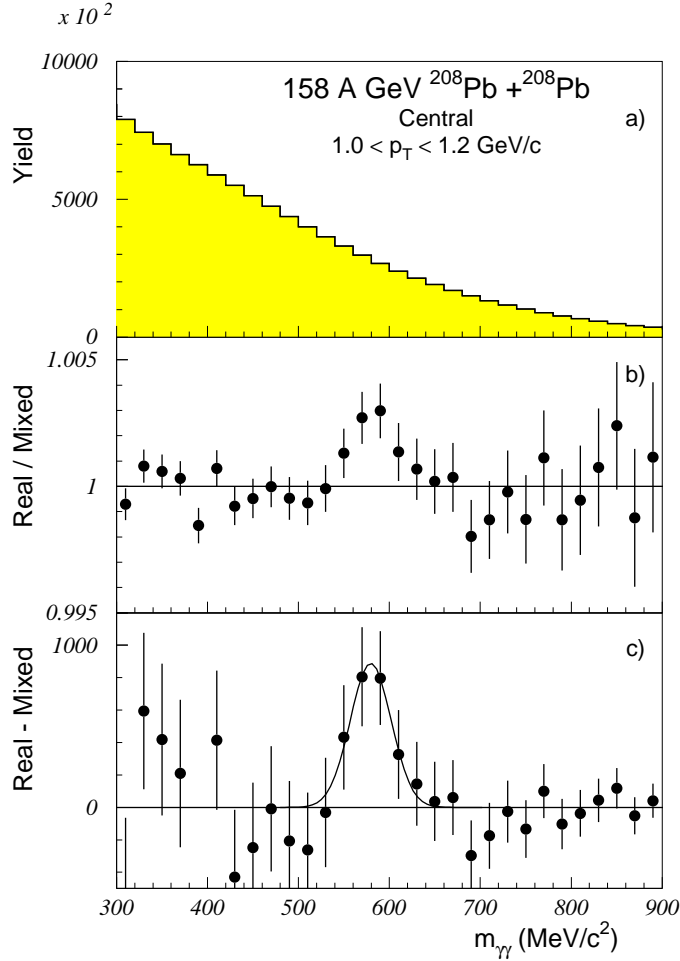}
\caption{The two-photon invariant mass distribution for central
   events in the $\eta$ mass region with photon pair transverse
   momentum $1.0 < p_T < 1.2$ GeV/c. Part a) shows the invariant mass
   distribution for all shower pairs for real events (solid histogram)
   and mixed events (filled histogram). Part b) shows the ratio of real
   and mixed event mass distributions. Part c) shows the real invariant
   mass distribution after subtracting the normalized mixed event
   distribution.}
\label{fig:eta_cent}
\end{center}
\end{figure}

The method to extract the $\eta$ transverse momentum spectrum is
similar to that used for the $\pi^0$ analysis described in
Sec.~\ref{sec:pid} and discussed in Sec.~\ref{sec:pi0s}.  As for the
$\pi^0$, the procedure is to calculate the  $m_{\gamma\gamma}$ invariant
mass spectrum for each $p_T$ bin and extract the yield in the $\eta$
peak.  The extraction of the $\eta$ yield is much more difficult due
to the lower $\eta$ production cross section and smaller branching
ratio to two photons together with the very large combinatorial
background in the $m_{\gamma\gamma}$ invariant mass distribution.  An
example of the two-photon invariant mass distributions for central
$^{208}$Pb\/+\/$^{208}$Pb collisions in the $\eta$ mass region is
shown in part a) of Fig.~\ref{fig:eta_cent} for photon pair transverse
momenta in the range of $1.0 < p_T < 1.2$ GeV/c. The distributions are
obtained using all narrow showers which pass the 750 MeV energy
threshold and include the full 1995 and 1996 magnet on data samples.
The invariant mass distribution for real events is shown by the open
histogram while the invariant mass distribution constructed with
photons from mixed events is shown by the shaded histogram. The
difference in the two distributions is scarcely visible, even in the
$\eta$ peak region.

%%fig24%%

The $\eta$ peak yield is extracted in the same way as the $\pi^0$
yield extraction but with one additional step. To insure that the high
mass tail of the $\pi^0$ peak does not affect the normalization of the
background the $\pi^0$ content is first removed from the invariant
mass distribution. This is done by extraction of the $\pi^0$ peak
yield, as described in Sec.~\ref{sec:pid}, and then subtraction of the
corresponding invariant mass distribution of simulated $\pi^0$'s after
event overlap, normalized to the same $\pi^0$ yield, from the
invariant mass distribution. This removes the $\pi^0$ peak and high
mass tail from the invariant mass distribution of real photon pairs.
The ratio of the $\pi^0$ subtracted real event to mixed event
invariant mass distributions is shown in part b) of
Fig.~\ref{fig:eta_cent}. The mixed event background normalization is
given by fixing to unity the ratio of integrated yields in the mass
interval $355-450$ and above $710$ MeV/c$^2$.  The normalized
background subtracted invariant mass distribution is shown in part c).
While the $\eta$ peak is clearly seen to contain several thousand
counts, the statistical significance of the peak is weak due the low
signal to background ratio of less than $0.5\%$.  The $\eta$ peak is
seen to be shifted to higher mass due to the effect of shower overlap
in central collisions, similar to observation for the $\pi^0$ (see
Fig.~\ref{fig:pi_mass}). The $\eta$ yield is obtained by integration
over the invariant mass region $500-640$ MeV/c$^2$.

\begin{figure}[hbt]
\begin{center}
   \includegraphics[scale=0.5]{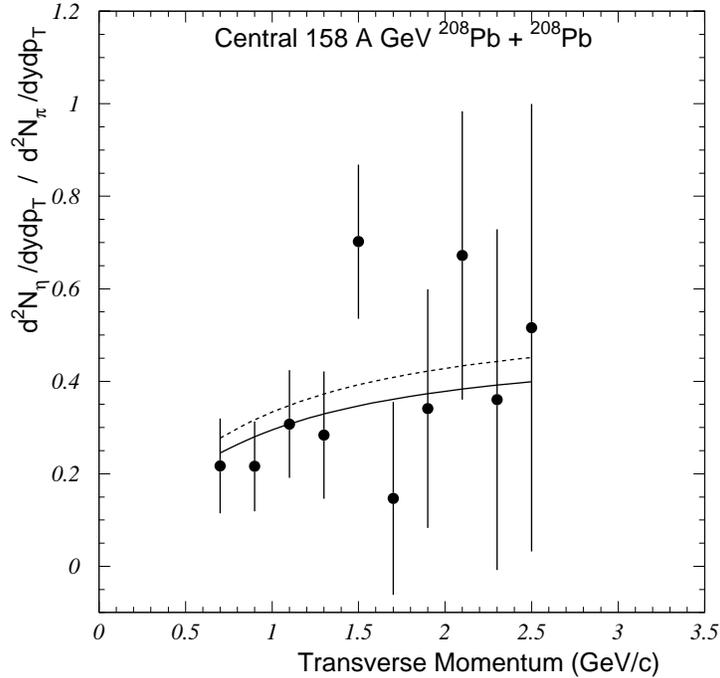}
\caption{\protect The $\eta/\pi^{0}$ ratio as a function of transverse
   momentum for central 158~{\it A}~GeV $^{208}$Pb\/+\/$^{208}$Pb
   collisions. The solid curve is the ratio expected for $m_T$-scaling
   of the fitted $\pi^0$ result of Fig.~\ref{fig:pi0_pt} with a fitted
   normalization $R_{\eta/\pi^0}=0.486$. The dashed curve is the
   expected result with fixed normalization$R_{\eta/\pi^0}=0.55$.}
\label{fig:eta_pi}
\end{center}
\end{figure}

The $\eta$ and $\pi^0$ yields are extracted from the invariant mass
distribution for each transverse momentum bin and are corrected for
identification efficiency, acceptance, and the two-photon decay
branching ratios.  The ratio of corrected yields, $\eta/\pi^0$, is
shown as a function of transverse momentum in Fig.~\ref{fig:eta_pi}.
While the statistical error at each point is large, due to the large
combinatorial background for extraction of the $\eta$ yield, there is
a general tendency for a rise in the ratio with increasing $p_T$.
This is the expected behavior if the $\eta$ and $\pi^0$ yields have
the same functional dependence on the transverse mass. The solid curve
in Fig.~\ref{fig:eta_pi} is the calculated ratio expected from the fit
to the $\pi^0$ spectrum (see Fig.~\ref{fig:pi0_pt} and discussion
below) assuming $m_T$-scaling. The absolute normalization is the
$m_T$-scaling parameter which is fitted to the results of
Fig.~\ref{fig:eta_pi} to be
$R_{\eta/\pi^0}=0.486\pm0.077$(stat.)$\pm0.097$(syst.).  A 20\%
systematical error has been assumed on the absolute normalization of the
$\eta$ yield relative to the $\pi^0$ yield. This systematical error
reflects the uncertainty due to the worse ratio of $\eta$ yield
to combinatorial background in
the $\eta$ region, and a less thorough investigation of the $\eta$
efficiency corrections due to the limited statistics.

%%fig25%%

Within errors, the fitted $R_{\eta/\pi^0}$ ratio is in agreement with
previous results. A value of $R_{\eta/\pi^0}=0.55\pm0.02$ has been
obtained from a compilation of previous
measurements~\cite{plb:ake86,plb:alb95} as discussed in
Sec.~\ref{sec:pideff} and listed in Table~\ref{table1}.  The present
$\eta$ measurement excludes any large enhancement of the $\eta$ yield
in central Pb+Pb collisions as has been suggested might occur as a
consequence of chiral symmetry restoration~\cite{prd:kap96,prd:hua96}
(see discussion of Sec.~\ref{sec:pideff}). The result also indicates
that the $m_T$-scaling assumption is valid, or at least not
significantly distorted by collective flow effects, for the transverse
momentum range of interest for the present direct photon analysis.

Finally, we note that due to the limited data sample and lower $\eta$
multiplicity it was not possible to obtain a significant $\eta$
measurement for the peripheral Pb+Pb event selection.  Instead,
$m_T$-scaling has been assumed as given by the measured peripheral
$\pi^0$ result with a $m_T$-scaling parameter of
$R_{\eta/\pi^0}=0.55$.

%\eject
\section{\label{sec:results} RESULTS}

In this section the final inclusive $\pi^0$ and inclusive photon
results are presented. The inclusive $\pi^0$ measurement is used,
together with the $\eta$ result of the previous section, as input to a
calculation of the expected inclusive photon distribution from
radiative decays. The difference between the measured and calculated
photon distributions is extracted as the direct photon excess. The
measured excess is presented and discussed in comparison to other
measurements and model calculations.

\subsection{$\pi^{0}$ and $\eta$ Production}

\begin{figure}[hbt]
\begin{center}
   \includegraphics[scale=0.5]{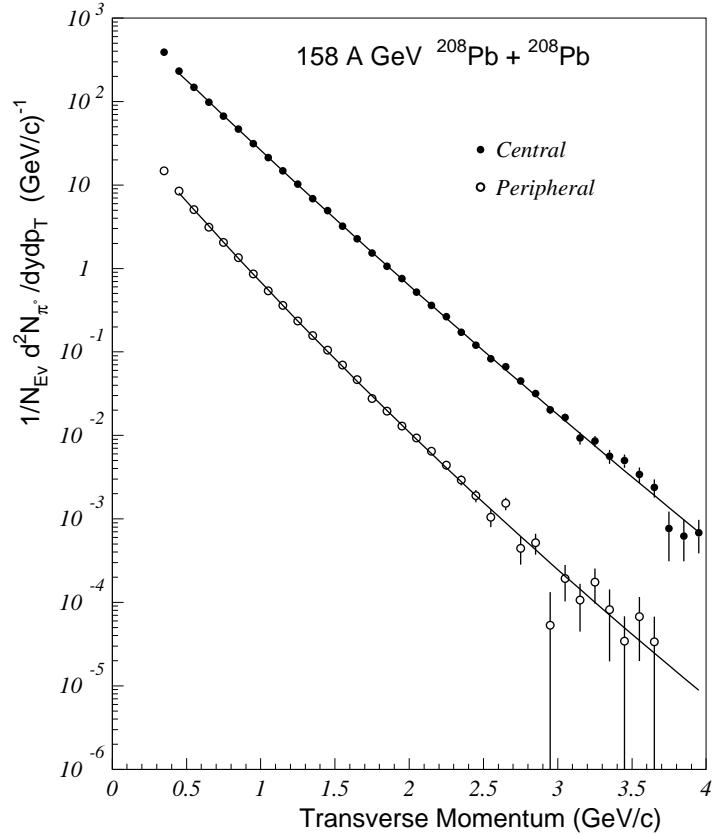}
\caption{The $\pi^{0}$ transverse momentum distribution
   for peripheral (open circles) and central (solid circles)
   158~{\it A}~GeV $^{208}$Pb\/+\/$^{208}$Pb collisions for the sum of
   1995 and 1996 magnet on data sets. The narrow shower photon
   identification criterion has been used. The data have been corrected
   for efficiency and acceptance. Only statistical errors are shown.
   The solid curves show the fit results described in the text.}
\label{fig:pi0_pt}
\end{center}
\end{figure}

The final inclusive $\pi^0$ transverse momentum spectra for central
and peripheral 158~{\it A}~GeV $^{208}$Pb\/+\/$^{208}$Pb collisions
are shown in Fig.~\ref{fig:pi0_pt}. The results presented are the sum
of the 1995 and 1996 data samples with magnet on.  The details of the
data selection have been discussed in Sec.~\ref{sec:selection} and the
characteristics of the peripheral and central data samples are
summarized in Table~\ref{table2}. The distributions have been
corrected for the $\pi^0$ identification efficiency and acceptance as
described in Secs.~\ref{sec:pid} and \ref{sec:pi0s}. The indicated
errors are the total statistical errors which include also 
the statistical errors
introduced by the combinatorial background subtraction.  The
systematical errors on the $\pi^0$ yield extraction have been discussed
in Sec.~\ref{sec:pi0err} and will be summarized again in
Sec.~\ref{sec:excess}. It should be noted that the $\pi^0$ yield per
event is given rather than the absolute $\pi^0$ cross section.  For
the direct photon analysis the measured photon multiplicity per event
will be compared to the background multiplicity calculated from the
measured $\pi^0$ yield per event for the same event sample.  Therefore
there are no systematical errors associated with the absolute cross
section determination. Also, it is recalled that the  definitions of
the peripheral data
samples were slightly different for the 1995 and 1996 runs
so that the result presented is for an averaged peripheral class. The
peripheral $\pi^0$ distribution has not been corrected for target-out
background (see Sec.~\ref{sec:pi0s}).

%%fig26%%

The $\pi^0$ transverse mass distributions are observed to be nearly
exponential over more than five orders of magnitude.  On the other
hand, the distributions exhibit a weak curvature which is best
described by a power-law. The spectra have therefore been fitted with
a QCD-inspired power-law functional form~\cite{cern:hag83},
\begin{equation}
\frac{1}{N_{Event}}\frac{d^2N}{dydp_T} =  C \cdot
   \left( \frac{p_0}{p_0+p_T} \right)^{n}.
\label{eq:dpt_fit}
\end{equation}
The fit region and fit results are shown by the solid curves in
Fig.~\ref{fig:pi0_pt}.  The extracted fit parameter values are
$C=66.5\pm2.3, p_0=9.6\pm1.3$ GeV/c, and $n=44.5\pm5.6$ with
$\chi^2/33=1.25$ for peripheral collisions and $C=1295\pm21,
p_0=19.98\pm0.24$ GeV/c, and $n=80.0\pm0.7$ with $\chi^2/36=1.06$ for
central collisions.

For the power-law functional form of Eq.~\ref{eq:dpt_fit} the local
inverse slope is calculated as
\begin{equation}
T = -\frac{f(p_T)}{\frac{\partial f(p_T)}{\partial p_T}} =
\frac{p_0}{n} + \frac{p_T}{n}.
\label{eq:slope}
\end{equation}
The ratio $p_0/n$ characterizes the slope of the distribution as $p_T
\rightarrow 0$, while $1/n$ characterizes the strength of the
curvature. The above fit results give $p_0/n=215.7$ and 249.8
MeV/c for peripheral and central collisions, respectively.  The
large fitted values of $n$ confirm the nearly exponential spectral
shapes.

More properly~\cite{cern:hag83,plb:boc96,epj:alb98}, the invariant
$\pi^{0}$ yields per event can be fitted to the same functional form
to provide a similarly good description.
\begin{equation}
\frac{1}{N_{Event}}E\frac{d^3 N}{dp^3} =  C' \cdot
  \left( \frac{p_0'}{p_0'+p_T} \right)^{n'}.
\label{eq:inv_fit   }
\end{equation}
Fitting the invariant transverse momentum distributions in the region
above 500 MeV/c gives fit parameters $C'=46.7\pm2.1, p_0'=3.64\pm0.06$ GeV/c,
and $n'=24.9\pm0.2$ with $\chi^2/32=1.41$ for peripheral collisions and
$C'=813.\pm21., p_0'=5.08\pm0.18$ GeV/c, and $n'=29.3\pm0.8$ with
$\chi^2/35=1.56$ for central collisions. This corresponds to
$p_0'/n'=146.0$ and 173.4 MeV/c for peripheral and central
collisions, respectively. The fitted slope and curvature parameters
are similar to those which have been extracted for the $\pi^{0}$
invariant cross sections for sulphur-induced
reactions~\cite{epj:alb98}.

The fit results of Eq.~\ref{eq:dpt_fit} shown in Fig.~\ref{fig:pi0_pt}
are the main experimental input to the calculation of the radiative
decay background.  In addition, the $\pi^0$ fit result for central
collisions has been used to fit the $\eta/\pi^0$ ratio shown in
Fig.~\ref{fig:eta_pi} assuming $m_T$-scaling and to extract the
$m_T$-scaling ratio $R_{\eta/\pi^0}=0.486$ for central collisions, as
discussed in Sec.~\ref{sec:etas}.  The calculated inclusive photon
distribution from radiative decays is compared to the measured
inclusive photon distribution to extract the excess which may be
attributed to direct photons.

%%fig27%%
\begin{figure}[hbt]
\begin{center}
   \includegraphics[scale=0.4]{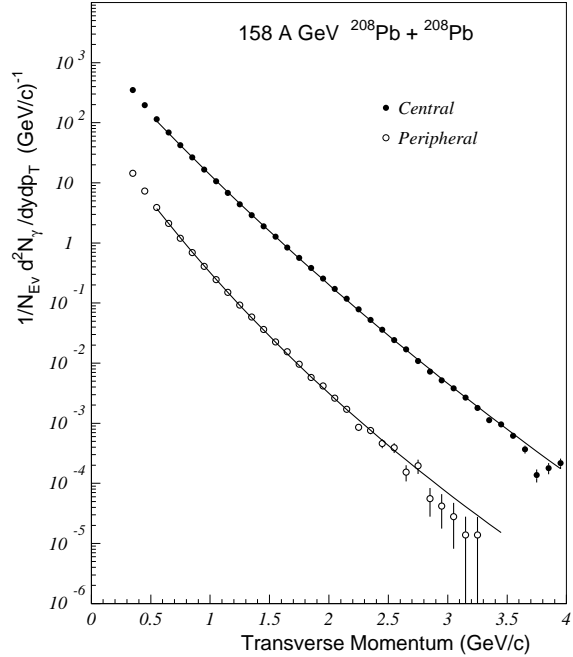}
\caption{The inclusive photon transverse momentum distribution
   for peripheral (open circles) and central (solid circles)
   158~{\it A}~GeV $^{208}$Pb\/+\/$^{208}$Pb collisions for the sum of
   1995 and 1996 magnet on data sets.  The narrow shower photon
   identification criterion has been used. The data have been corrected
   for efficiency and acceptance. Only statistical errors are shown.
   The solid curves show the fit results described in the text.}
\label{fig:photon_pt}
\end{center}
\end{figure}

\subsection{\label{sec:gam_result} Inclusive Photon Results}

The final inclusive photon transverse momentum spectra for central and
peripheral 158~{\it A}~GeV $^{208}$Pb\/+\/$^{208}$Pb collisions are
shown in Fig.~\ref{fig:photon_pt}. The results presented are the sum
of the 1995 and 1996 data samples with magnet on, which is exactly the
same data sample as used for the final $\pi^0$ and $\eta$ result.  The
details of the data selection have been discussed in
Sec.~\ref{sec:selection} and the characteristics of the peripheral and
central data samples are summarized in Table~\ref{table2}. The
distributions have been corrected for the photon identification
efficiency and acceptance as described in Secs.~\ref{sec:pid} and
\ref{sec:photons}. The indicated errors are statistical errors only.
The systematical errors on the photon yield extraction have been
discussed in Sec.~\ref{sec:gamerr} and will be summarized again in
Sec.~\ref{sec:excess}.  It is again noted that there are no systematical
errors associated with the absolute cross section determination for
the direct photon analysis.  Also, it is again noted that the
peripheral data sample definitions were slightly different for the
1995 and 1996 runs which means that the result represents an average
peripheral class. The peripheral photon distribution has not been
corrected for target-out background (see Sec.~\ref{sec:gamchrg}).

Similar to the $\pi^0$ transverse mass distributions, the inclusive
photon transverse momentum distributions are nearly exponential over
more than six orders of magnitude.  The inclusive photon distributions
have been fitted with the power-law functional form of
Eq.~\ref{eq:dpt_fit} with the fit region and results shown by the
solid curves in Fig.~\ref{fig:photon_pt}.  The extracted fit parameter
values are $C=115.9\pm3.3, p_0=3.2\pm0.4$ GeV/c, and $n=21.8\pm5.6$ for
peripheral collisions and $C=1570\pm18, p_0=7.35\pm0.19$ GeV/c, and
$n=37.2\pm0.8$ for central collisions.

The measured inclusive photon distributions are to be compared to the
background inclusive photon distributions calculated from radiative
decays of the $\pi^0$ and other hadrons to extract the direct photon
excess.

%%fig28%%
\begin{figure}[hbt]
\begin{center}
   \includegraphics[scale=0.5]{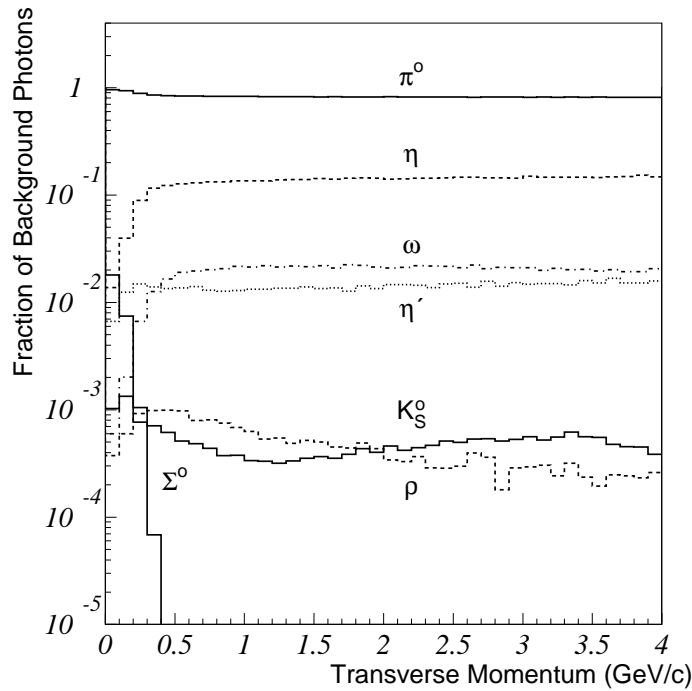}
\caption{The calculated fraction of all background inclusive photons
   from various radiative decay sources is shown as a function of the
   transverse momentum for central 158~{\it A}~GeV
   $^{208}$Pb\/+\/$^{208}$Pb collisions.}
\label{fig:photon_pt_bkg}
\end{center}
\end{figure}

\subsection{\label{sec:background} Background Photon Calculation}

For each data sample or event selection in which the inclusive yield
is extracted, the expected inclusive photon background from long-lived
radiative decays is calculated using a Monte Carlo program which uses
the JETSET 7.3 routines \cite{cern:jet89} to implement the hadron
decays with proper branching ratios and decay distributions.  As
described in Sec.~\ref{sec:simulation}, the most important input to
that calculation is the measured inclusive $\pi^0$ yield per event for
the same event sample.  In the present analysis, the direct photon
excess is to be extracted for central and for peripheral Pb\/+\/Pb 
collisions. The peripheral data sample is used to provide a
control measurement. Therefore, the fits to the final inclusive
$\pi^0$ transverse momentum distributions of Fig.~\ref{fig:pi0_pt} are
the primary input to the calculations.  The output of the simulation
is the calculated decay photon yield per $\pi^0$ into the LEDA
acceptance. This calculated $\gamma/\pi^0$ ratio, is to be compared to
the measured ratio. Since the simulated $\pi^0$ distribution must
agree with the measured $\pi^0$ distribution by construction, the
difference in the measured and calculated inclusive $\gamma$ yield per
event gives the direct photon excess.

While photons from $\pi^0$ decay constitute the dominant source of
background photons, other radiative decays are also included in the
background calculation. The calculated fraction of the total background
photons due to each of the various photon sources is shown in
Fig.~\ref{fig:photon_pt_bkg} for central Pb+Pb collisions. As
discussed in Sec.~\ref{sec:simulation}, the yields of the various
hadrons has been calculated with the assumption of $m_T$-scaling. This
implies that the transverse momentum distributions of the other
hadrons are determined by the measured $\pi^0$ transverse momentum
distributions. The $m_T$-scaling normalization factors $R_{X/\pi^0}$
are taken from the literature as discussed in
Sec.~\ref{sec:simulation} and given in Table~\ref{table1}. The quoted
$m_T$-scaling factors are used for both peripheral and central
collisions, with the exception that the measured $m_T$-scaling value
of $R_{\eta/\pi^0}=0.486$ extracted for the results shown in
Fig.~\ref{fig:eta_pi} and discussed in Sec.~\ref{sec:etas} is used for
central collisions. As seen from Fig.~\ref{fig:photon_pt_bkg}, the
$\eta$ radiative decay contribution is the most significant source of
background photons, beyond the $\pi^0$ contribution.\footnote{Note
   that decay photons from $\pi^0$'s which are themselves decay
   products of other hadrons are included in the measured $\pi^0$ decay
   contribution, and therefore not included as decay photons from the
   original hadron.} As will be discussed in the next section, the
uncertainty on the $\eta$ yield constitutes one of the largest sources
of systematical error on the background photon calculation.

For the background calculations, a Gaussian rapidity distribution of
particle production is assumed centered on mid-rapidity ($y=2.9$) with
an rms width of $\sigma_y=1.3$ according to measured results for
photon production in 158~{\it A}~GeV $^{208}$Pb\/+\/$^{208}$Pb
collisions~\cite{plb:agg99}. Since the LEDA is near to mid-rapidity
(see Fig.~\ref{fig:acceptance}) the rapidity distribution varies
little over the detector acceptance consistent with the $\sigma=1.3$
width. As an extreme, the background calculation has been performed
with the assumption of a flat rapidity distribution.  Under this
assumption the calculated background photon yield would increase by
about 2\% for all transverse momenta above $p_T=1$ GeV/c.  The
increase is greater at low $p_T$ with an increase in the photon yield
of about 5\% at $p_T=500$ MeV/c. This modest sensitivity to the
assumed rapidity distribution can be understood from the limited
acceptance for photons from $\pi^0$'s which are emitted away from
LEDA, as seen in part a) of Fig.~\ref{fig:acceptance}.  More important
however, is that the calculated $\gamma/\pi^0$ ratio, which is the
quantity relevant for the direct photon analysis, is much less
sensitive to the assumed rapidity distribution. With the assumption of
a flat rapidity distribution the $\gamma/\pi^0$ ratio is increased by
only 2\% at $p_T=500$ MeV/c, 0.5\% at $p_T=1$ GeV/c, with the size of
the increase rapidly decreasing at higher transverse momenta.
Therefore, one may conclude that the uncertainty in the calculated
$\gamma/\pi^0$ ratio due to extrapolation and uncertainty of the
$\pi^0$ rapidity distribution is negligible.

%\subsubsection{Acceptance}
%\subsubsection{$\pi^{0}$ Production}
%\subsubsection{m$_T$-scaling: $\eta, \eta', \omega$}
%\subsubsection{Other neutrals: $n,{\overline n},K_S,\Lambda$}

\subsection{\label{sec:excess} Direct Photon Excess}

In this section the final direct photon result is obtained and
compared to results from proton-induced reactions at similar $\sqrt{s}$
and to model calculations.  The direct photon yield is extracted as a
function of the photon transverse momentum for central and peripheral 
158~{\it A}~GeV $^{208}$Pb\/+\/$^{208}$Pb collisions.  The results
presented are the sum of the 1995 and 1996 magnet on data samples.
The details of the data selection have been discussed in
Sec.~\ref{sec:selection} and the characteristics of the peripheral and
central data samples are summarized in Table~\ref{table2}.  As
described in Sec.~\ref{sec:overview}, the direct photon excess is
extracted on a statistical basis as the difference between the
measured inclusive photon distributions discussed in
Sec.~\ref{sec:gam_result} and the background photon distributions
calculated from radiative decays of long-lived final state hadrons
discussed in Sec.~\ref{sec:background}.

%tca \begin{figure}
%tca \caption{The ratio $(\gamma/\pi^{0})^{X}/(\gamma/\pi^{0})^{Final}$
%tca as a function of transverse momentum
%tca for a) peripheral and b) central collisions of
%tca 158~{\it A}~GeV $^{208}$Pb\/+\/$^{208}$Pb.
%tca The final $(\gamma/\pi^{0})^{Final}$ result is compared to the
%tca result from the 1995 magnet on (solid circles) and magnet off (open
%tca circles) and the 1996 magnet on (solid squares) and magnet off (open
%tca squares) data sets.  The data have
%tca been corrected for efficiencies. Only statistical errors
%tca are shown.}
%tca \label{fig:gamma_pi0_err}
%tca \end{figure}

\subsubsection{\label{sec:final_results} Results}

Since $\pi^0$ radiative decays comprise the dominant source of
background photons (see Fig.~\ref{fig:photon_pt_bkg}), it is
instructive to first investigate the $\gamma/\pi^0$ ratio as a
function of the transverse momentum.  As a result of the steeply
falling $\pi^0$ transverse momentum spectra (see
Fig.~\ref{fig:pi0_pt}), photons of a given $p_T$ predominantly result
from asymmetric $\pi^0$ decays in which the photon carries most of the
momentum of the parent $\pi^0$.  (Since photons from symmetric decays
carry roughly half of the parent $\pi^0$ momentum they are suppressed
by the lower $\pi^0$ yield at the two times higher $p_T$, compared to
the asymmetric decay $\pi^0$'s.)  Therefore, the error on the
predicted photon yield at a given $p_T$ will be most strongly
correlated with the error in the measured $\pi^0$ yield at a similar,
slightly higher, $p_T$.  For the ratio $\gamma/\pi^0$ many sources of
systematical errors partially cancel, such as errors in the energy scale
calibration for the measurement, or errors in the assumed $\pi^0$
rapidity distribution for the background calculation (as discussed in
Sec.~\ref{sec:background}).  Therefore a comparison of the measured
and calculated $\gamma/\pi^0$ ratios provides a sensitive indication
whether a direct photon excess is observed.

%%fig29%%
\begin{figure}[hbt]
\begin{center}
   \includegraphics[scale=0.5]{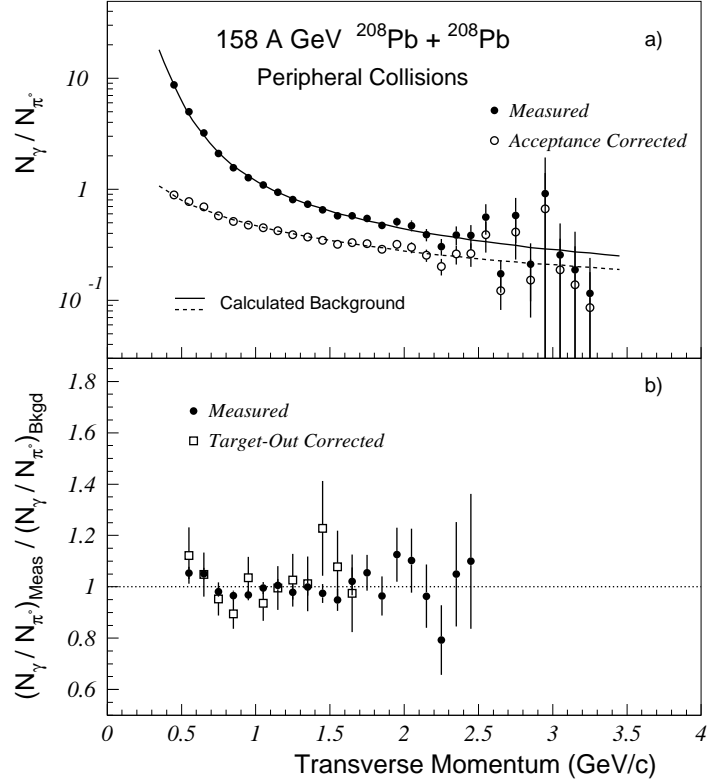}
\caption{The efficiency corrected (solid circles) and efficiency and
   acceptance corrected (open circles) $\gamma/\pi^{0}$ ratio is shown
   in part a) as a function of transverse momentum for peripheral
   158~{\it A}~GeV $^{208}$Pb\/+\/$^{208}$Pb collisions.  The
   calculated background $(\gamma/\pi^{0})_{\rm Bkgd}$ ratio in the
   WA98 acceptance is shown by the solid curve. The dashed curve shows
   the calculated background with acceptance corrections.  The ratio of
   measured $(\gamma/\pi^{0})_{\rm Meas}$ ratio to calculated
   $(\gamma/\pi^{0})_{\rm Bkgd}$ ratio for the WA98 acceptance is shown
   in part b).  The errors on the data points indicate the statistical
   errors only. }
\label{fig:gamma_pi0_per}
\end{center}
\end{figure}

The measured $\gamma/\pi^0$ ratio for peripheral
$^{208}$Pb\/+\/$^{208}$Pb collisions is shown as a function of
transverse momentum in part a) of Fig.~\ref{fig:gamma_pi0_per}. The
solid points show the measured $\gamma$ yield per event divided by the
measured $\pi^0$ yield per event in the LEDA acceptance, fully
corrected for efficiencies and backgrounds. The errors indicate the
total statistical error on the ratio from the $\gamma$ and $\pi^0$
yield extraction at each $p_T$.  The open points show the measured
results additionally corrected for the $\gamma$ and $\pi^0$
acceptances.  The results shown are obtained using the photon
identification criterion in which all showers are considered photon
candidates (condition S1 of Sec.~\ref{sec:pid}). This shower
identification criterion provides the greatest $\gamma$ and $\pi^0$
efficiency, but also requires the largest corrections and therefore
has the largest expected systematical error.  The measured result is
compared to the predicted $\gamma/\pi^0$ ratio shown by the solid
curve in part a). The predicted result is from the Monte Carlo
calculation of the background radiative decay photons discussed in
Sec.~\ref{sec:background} which is based on the measured peripheral
$\pi^0$ spectrum of Fig.~\ref{fig:pi0_pt}. The simulation provides the
yield per event of photons and $\pi^0$'s into the LEDA acceptance. The
calculated result is also shown with the $\gamma$ and $\pi^0$
acceptance corrections.

The predicted $(\gamma/\pi^0)_{\rm Bkgd}$ ratio from background decays
is seen to be in good agreement with the measured results. This is
shown more clearly in part b) of Fig.~\ref{fig:gamma_pi0_per} where
the solid points show the ratio of the measured $(\gamma/\pi^0)_{\rm
   Meas}$ ratio to predicted background $(\gamma/\pi^0)_{\rm Bkgd}$
ratio. Even without consideration of possible systematical errors, the
rather good agreement between the measured and calculated ratio,
within statistical errors, suggests that no significant direct photon
excess is observed for peripheral collisions over the transverse
momentum region of measurement.

The results in part a) and solid points in part b) of
Fig.~\ref{fig:gamma_pi0_per} have not been corrected for target out
backgrounds. As discussed in Sec.~\ref{sec:gamchrg} the measured
target out background rate implied less than 10\% of the events of the
peripheral data sample were due to non-target background.  The photon
multiplicities per non-target background event were measured to be
similar to peripheral events, but with a steeper falling $p_T$
distribution (see Fig.~\ref{fig:raw_photons}).  From the low $p_T$
$m_{\gamma\gamma}$ invariant mass distributions for peripheral events (see
Fig.~\ref{fig:pi0_per_hi_pt}) the non-target background events were
deduced to be due to downstream interactions, as indicated by a
$\pi^0$ peak shifted to lower mass, with indications for a non-target
contamination larger than determined by the target out measurements,
as discussed in Sec.~\ref{sec:pi0yield}.  The peripheral result has
therefore been checked with the assumption of the maximum possible
non-target correction consistent with the low mass structure observed
in the low $p_T$ invariant mass spectra. It was only possible to do
this for the 1996 data sample since no low mass peak was observed in
the target out data for the 1995 run due to the Plastic Ball
multiplicity condition requirement in the online trigger (see
Sec.~\ref{sec:pi0yield}).  The result is shown by the open squares in
part b) of Fig.~\ref{fig:gamma_pi0_per}. Due to the smaller 1996
peripheral data sample and the maximally increased target out
correction of roughly 50\% of the 1996 peripheral events, the
statistical errors on the corrected results are very large.
Nevertheless, the corrected $\gamma/\pi^0$ is found to be consistent
with the uncorrected ratio and confirms the lack of a significant
photon excess for the case of peripheral collisions.

%%fig30%%
\begin{figure}[hbt]
\begin{center}
   \includegraphics[scale=0.5]{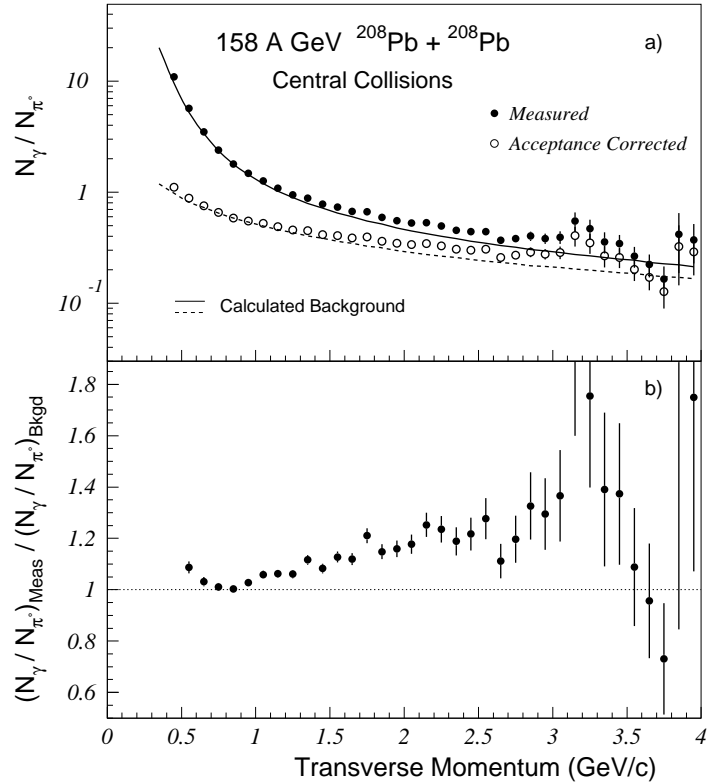}
\caption{The efficiency corrected (solid circles) and efficiency and
   acceptance corrected (open circles) $\gamma/\pi^{0}$ ratio is shown
   in part a) as a function of transverse momentum for central
   158~{\it A}~GeV $^{208}$Pb\/+\/$^{208}$Pb collisions.  The
   calculated background $(\gamma/\pi^{0})_{\rm Bkgd}$ ratio in the
   WA98 acceptance is shown by the solid curve. The dashed curve shows
   the calculated background with acceptance corrections.  The ratio of
   measured $(\gamma/\pi^{0})_{\rm Meas}$ ratio to calculated
   $(\gamma/\pi^{0})_{\rm Bkgd}$ ratio for the WA98 acceptance is shown
   in part b).  The errors on the data points indicate the statistical
   errors only. }
\label{fig:gamma_pi0_cen}
\end{center}
\end{figure}

The corresponding $\gamma/\pi^0$ ratio results for central
$^{208}$Pb\/+\/$^{208}$Pb collisions are shown in
Fig.~\ref{fig:gamma_pi0_cen}.  The results are obtained using the
photon identification criterion in which all showers are considered
photon candidates, as was used for the results of
Fig.~\ref{fig:gamma_pi0_per}.  The background photon calculation is
based on the measured central $\pi^0$ spectrum of
Fig.~\ref{fig:pi0_pt} and uses the measured $R_{\eta/\pi^0}$ ratio for
central collisions deduced from the results of Fig.~\ref{fig:eta_pi}.
Compared to the result for peripheral collisions, the statistical
uncertainty on the measured $\gamma/\pi^0$ ratio for the case of
central collisions is greatly reduced due to the slightly larger event
sample, and more importantly, the much higher particle multiplicities
per event.

In contrast to the case for peripheral collisions, the background
calculation does not account well for the measured $\gamma/\pi^0$
ratio for central collisions. Instead, an excess in the measured
$\gamma/\pi^0$ ratio compared to the background calculation is clearly
seen in part b) of Fig.~\ref{fig:gamma_pi0_cen}. The observed excess
increases with $p_T$ up to about a 20\% excess at high transverse
momentum.

To determine whether the observed excess is significant, a detailed
consideration of the various sources of systematical error is necessary.
The final direct photon excess is to be extracted as
\begin{equation}
\gamma_{\rm Excess} = \gamma_{\rm Meas}-\gamma_{\rm Bkgd}
= \left(1-\frac{\gamma_{\rm Bkgd}}{\gamma_{\rm Meas}}\right)
\cdot \gamma_{\rm Meas}.
\label{eq:excess}
\end{equation}
If the ratio $\gamma_{\rm Meas}/\gamma_{\rm Bkgd}$ is equal to one
within errors, then no significant photon excess is observed. The
ratio $\gamma_{\rm Meas}/\gamma_{\rm Bkgd}$ and its error are
determined from the measured $(\gamma/\pi^{0})_{\rm Meas}$ and
calculated $(\gamma/\pi^{0})_{\rm Bkgd}$ ratios.  Since, by
construction, $(\pi^{0})_{\rm Bkgd} \equiv (\pi^0)_{\rm Meas}$ within
uncertainties we have
\begin{equation}
\frac{\gamma_{\rm Meas}}{\gamma_{\rm Bkgd}}
= \frac{(\gamma/\pi^{0})_{\rm Meas}}{(\gamma/\pi^{0})_{\rm Bkgd}}.
\label{eq:gamgam}
\end{equation}
Additional systematical errors on $\gamma_{\rm Meas}$ which partially
cancel in Eq.\ref{eq:gamgam} must also be included in
Eq.~\ref{eq:excess}.  Furthermore, when the absolute excess photon
cross section is extracted the absolute cross section normalization
errors on $\gamma_{\rm Meas}$ must be introduced.

The total systematical error on $\gamma_{\rm Meas}/\gamma_{\rm Bkgd}$ is
obtained from the separate error contributions to
$(\gamma/\pi^{0})_{\rm Meas}$, $(\gamma/\pi^{0})_{\rm Bkgd}$, and
$(\pi^{0})_{\rm Meas}/(\pi^0)_{\rm Bkgd}$.  These various systematical
error contributions to the direct photon result are summarized in
Table~\ref{table3}.  
The errors quoted correspond to the shower identification criterion of
using narrow showers (S2), which minimizes the background corrections
and therefore is expected to have the smallest systematical error.
This shower identification criteria will be used to extract the
final WA98 result. A comparison of the final result with the
results of Figs.~\ref{fig:gamma_pi0_per} and ~\ref{fig:gamma_pi0_cen}
where all showers (S1) have been used provides an additional check
of the overall systematical errors.
In general, many of the systematical errors are
dependent on the transverse momentum and so must be estimated for each
$p_T$. The systematical errors are listed at $p_T \approx$ 1 GeV/c and 2.5
GeV/c for peripheral and central collisions to give an indication of
the $p_T$ and centrality dependence of the systematical error
contributions.

\vbox{
\begin{table}
\caption{
Various sources of systematical error in the
WA98 158~{\it A}~GeV $^{208}$Pb\/+\/$^{208}$Pb direct photon analysis
specified as a percentage of
$(\gamma/\pi^{0})_{\rm Meas}$ (items a) ), $(\gamma/\pi^0)_{\rm Bkgd}$
(items b) ), or
$(\pi^{0})_{\rm Meas}/(\pi^0)_{\rm Bkgd}$ (item c) ).
The systematical errors are quoted at two $p_T$ values to give an
indication of the dependence on transverse momentum. The errors
are estimated for the narrow shower identification criterion (S2).
The total estimated systematical error on
$\gamma_{\rm Meas}/\gamma_{\rm Bkgd}$ is given as the quadratic sum
of the various contributions.
}
\label{table3}
\begin{tabular}{lcccc}
Source of Error& \multispan2
\hfill Peripheral Collisions $(20\%\ \sigma_{\rm mb})$ \hfill &
\multispan2 \hfill Central\ Collisions\ $(10\%\ \sigma_{\rm mb})$ \hfill \\
& $p_T \approx 1.0$ GeV/c & $ p_T \approx 2.5$ GeV/c
& $p_T \approx 1.0$ GeV/c & $ p_T \approx 2.5$ GeV/c \\
\hline
\ \ \ Charged Particle background$^+$ & 1.7 & 2.2 & 1.3 & 1.3 \\
\ \ \ $\gamma$ conversion correction$^+$ & 0.5 & 0.5 & 0.5 & 0.5 \\
\ \ \ Neutrons$^+$ & 0.6 & 1.0 & 0.9 & 1.9  \\
\ \ \ $\gamma$ reconstruction efficiency$^+$ & 2.0 & 2.0 & 2.0 & 2.0 \\
\hline
a) $\gamma$ yield measurement & 2.7 & 3.2 & 2.6 & 3.1 \\
\hline
\hline
\ \ \ $\gamma$ conversion correction$^*$ & 0.5 & 0.5 & 0.5 & 0.5 \\
\ \ \ $\pi^0$ yield extraction$^*$ & 0.3 & $<$0.1 & 5.1 & 1.0 \\
\ \ \ $\pi^0$ reconstruction efficiency$^*$ & 3.0 & 3.0 & 4.0 & 4.0 \\
\hline
a) $\pi^0$ yield measurement  & 3.1 & 3.0 & 6.5 & 4.2 \\
\hline
\hline
a) Non-target background & 1.5 & $<$0.1 & $<$0.1 & $<$0.1 \\
a) Energy scale calibration  & 0.9 & 1.7 & 0.8 & 1.7 \\
b) Detector acceptance & 0.5 & 0.5 & 0.5 & 0.5 \\
b) $\eta / \pi$ ratio, $m_T$-scaling & 2.9 & 3.2 & +3.4 (-4.8) &
+3.7 (-5.2) \\
b) Other radiative decays & 1.0 & 1.0 & 1.0 & 1.0 \\
c) $\pi^0$ fit  & 1.6 & 6.8 & 2.9 & 0.4 \\
\hline
Total: (quadratic sum)  & 5.7 & 8.9 & +8.3 (-9.1) & +6.7 (-7.6) \\
\end{tabular}
$^+$ Included in $\gamma$ yield measurement error.
$^*$ Included in $\pi^0$ yield measurement error.
\end{table}
\vskip -0.3cm }

The separate systematical errors on the photon and $\pi^0$ measurement
as they contribute to $(\gamma/\pi^{0})_{\rm Meas}$ have been
discussed previously in Secs.~\ref{sec:gamerr} and \ref{sec:pi0err}.
As discussed in Sec.~\ref{sec:gamerr}, the photon error includes an
assumed 30\% uncertainty in the charged particle correction related to
the results shown in Fig.~\ref{fig:charge_neut_disp}, and an assumed
50\% uncertainty in the neutron+anti-neutron correction related to the
results shown in Fig.~\ref{fig:neut_antineut}.  A 2\% uncertainty is
assumed for the photon identification efficiency. The
overall systematical error on the determination of the photon yield is
$\sim 3\%$ which was shown to be consistent with observed systematic
variations by a comparison of the $\gamma$ yield determination using
different shower identification methods (see
Fig.~\ref{fig:photon_err}).

The systematical errors on the $\pi^0$ yield determination listed in
Table~\ref{table3} were discussed in Sec.~\ref{sec:pi0err}. The
systematical error associated with the determination of the $\pi^0$
identification efficiency was estimated to be 3\% and 4\% for
peripheral and central collisions, respectively.  An additional source
of systematical error on the $\pi^0$ yield extraction is attributed to
the subtraction of the combinatorial background underlying the $\pi^0$
peak in the $\gamma\gamma$ invariant mass distributions which becomes
important for the case of central collisions at low $p_T$.  This error
was estimated to contribute at the level of $\sim 10^{-3}$ of the
background.  The overall systematical error estimate on the
determination of the $\pi^0$ yield was shown by a comparison of the
$\pi^0$ yield results using different shower identification methods
(see Fig.~\ref{fig:pi0_err}) to be consistent with observed systematical
variations.

Other systematical error contributions to the measured
$(\gamma/\pi^0)_{\rm Meas}$ ratio listed in Table~\ref{table3} are the
target out background contribution and the calibration of the energy
scale. Although the magnitude of the non-target correction was found
to have large uncertainties, as discussed in Secs.~\ref{sec:gamchrg}
and \ref{sec:pi0yield}, the peripheral $(\gamma/\pi^0)_{\rm Meas}$
ratio showed little systematic dependence on the non-target background
correction, as discussed in regard to Fig.~\ref{fig:gamma_pi0_per}. No
target out background correction was necessary for central collisions,
as previously discussed.

Based on the observed agreement between the measured and simulated
$\pi^0$ peak positions shown in Fig.~\ref{fig:pi_mass}, the
calibration of the energy scale is estimated to be accurate to 0.5\%.
Using the measured $\gamma$ and $\pi^0$ transverse momentum
distributions the error on the $(\gamma/\pi^0)_{\rm Meas}$ ratio is
calculated as a function of $p_T$ assuming a 0.5\% uncertainty in the
$p_T$ scale.

The uncertainty in the detector acceptance listed in
Table~\ref{table3} is relevant for the background photon calculation.
As discussed in Sec.~\ref{sec:background}, the calculated
$(\gamma/\pi^0)_{\rm Bkgd}$ ratio is not very sensitive to the
uncertainty in the detector acceptance, or to extreme assumptions on
the extrapolation of the hadron distributions outside of the detector
acceptance. The quoted error estimate of 0.5\% includes both effects.

One of the larger contributions to the systematical error in this
analysis is attributed to the uncertainty in the $\eta$ yield. While
the $\eta$'s constitute a relatively large source of background
photons (see Fig.~\ref{fig:eta_pi}), their yield has not been measured
very precisely in the present analysis (see discussion of
Sec.~\ref{sec:etas}).  According to the discussion of
Sec.~\ref{sec:background}, $m_T$-scaling has been assumed in order to
allow to relate the $p_T$ spectrum of the $\eta$ to the measured
$\pi^0$ spectrum with only a single overall normalization factor
$R_{\eta/\pi^0}$.  For peripheral collisions the $\eta$ yield has not
been measured. In this case, it should be a good approximation to use
the $m_T$-scaling factor of $R_{\eta/\pi^0}=0.55\pm0.02$
\cite{plb:ake86,plb:alb95} obtained from a compilation of proton and
$\pi$-induced results, where the $m_T$-scaling assumption has also
been shown to be valid.  Nevertheless, we assume a larger systematical
error of 20\% on $R_{\eta/\pi^0}$ since it is an unmeasured quantity
and also to accommodate deviations from the $m_T$-scaling assumption.
This results in a systematical error contribution to
$(\gamma/\pi^0)_{\rm Bkgd}$ of about 3\% for peripheral collisions, as
shown in Table~\ref{table3}.

For central collisions, the transverse momentum dependence of the
$\eta$ yield was extracted with modest precision, as discussed in
Sec.~\ref{sec:etas}. The measured $\eta/\pi^0$ ratio was found to be
consistent with the $m_T$-scaling assumption with a best fit overall
normalization of
$R_{\eta/\pi^0}=0.486\pm0.077$(stat.)$\pm0.097$(syst.). For the direct
photon analysis, we have added these statistical and systematical errors
in quadrature to obtain a low estimate on the $\eta/\pi^0$ ratio of
$R_{\eta/\pi^0}=0.36$ which is used for the lower error estimate on
the $\gamma/\pi^0$ ratio due to the $\eta$ background contribution.
For the upper error estimate on the $\gamma/\pi^0$ ratio due to the
$\eta$ we have assumed a high estimate of $R_{\eta/\pi^0}=0.66$, which
is the same upper estimate used for the case of peripheral collisions.
This results in a larger upper error than lower error on the
$(\gamma/\pi^0)_{\rm Bkgd}$ ratio with the associated asymmetric
errors listed in Table~\ref{table3}\footnote{Note that an upper error
   estimate on the background photons contribute to the lower error
   estimate on the photon measurement.}.

The expected total background contribution from radiative decays other
than those of the $\pi^0$ and $\eta$ (predominantly the $\omega$ and
$\eta$') is of the order of a few percent only (see
Fig.~\ref{fig:photon_pt_bkg}).  Since this contribution is unmeasured,
a systematical error of $\sim 30\%$ in their yield, or 1\% on the
background photon contribution, has been assumed.

 From the results shown in Fig.~\ref{fig:pi0_pt} it is apparent that
the statistical error on the $\pi^0$ measurement limits the
significance of the $\gamma/\pi^0$ results shown in
Figs.~\ref{fig:gamma_pi0_per} and ~\ref{fig:gamma_pi0_cen}, especially
at high transverse momenta. This statistical error may be removed by
fitting the $\pi^0$ spectra of Fig.~\ref{fig:pi0_pt}, but at the cost
of an additional source of systematical error. The last error listed in
Table~\ref{table3} is the estimated $\pi^0$ fit error.  This error was
estimated by fitting separately the $\pi^0$ spectra for the 1995,
1996, and sum 1995+1996 data samples and separately for the shower
condition of all showers (S1) or narrow showers (S2) ({\it i.e.} 6
different spectra each for peripheral and central) in the high $p_T$
region only ($p_T > 1.5$ GeV/c) and using the half-width of the maximum
variation of the fit results over the full $p_T$ region to define the
fit error as a function of $p_T$.  This error is observed to be
strongly $p_T$ dependent and increases rapidly outside of the range of
measurement.

Since $(\pi^{0})_{\rm Meas}/(\pi^0)_{\rm Bkgd} \equiv 1$ within
uncertainties by construction, the individual systematical errors on
$(\gamma/\pi^{0})_{\rm Meas}$, $(\gamma/\pi^0)_{\rm Bkgd}$, and
$(\pi^{0})_{\rm Meas}/(\pi^0)_{\rm Bkgd}$, can be combined to give the
total systematical error on $\gamma_{\rm Meas}/\gamma_{\rm Bkgd}$.  The
various systematical errors listed in Table~\ref{table3} are added in
quadrature to obtain the total systematical error on the $\gamma_{\rm
   Meas}/\gamma_{\rm Bkgd}$ ratio. While some of the errors are
correlated, the correlations are such that they tend to cancel in the
final ratio. Also, the $\pi^0$ fit error includes some of the errors
of the $\pi^0$ yield extraction.  Therefore, the assumption of
independent errors is considered a conservative assumption.

\begin{figure}[hbt]
\begin{center}
   \includegraphics[scale=0.5]{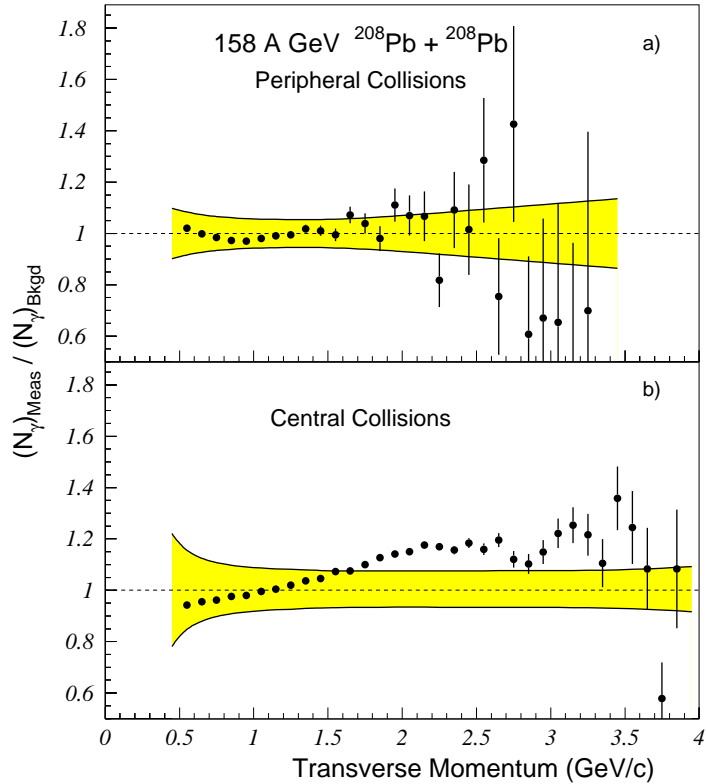}
\caption{The $\gamma_{\rm Meas}/\gamma_{\rm Bkgd}$ ratio
   as a function of transverse momentum for peripheral (part a)) and
   central (part b)) 158~{\it A}~GeV $^{208}$Pb\/+\/$^{208}$Pb
   collisions for the sum of 1995 and 1996 magnet on data sets.  The
   narrow shower photon identification criterion has been used. The
   errors on the data points indicate the statistical errors only. The
   $p_T$-dependent systematical errors are indicated by the shaded bands
   ($\equiv 1^{+\sigma_L}_{-\sigma_U}$).  (Note: Upper and lower errors
   are inverted for display purposes.) }
\label{fig:gamma_excess}
\end{center}
\end{figure}

The $\gamma_{\rm Meas}/\gamma_{\rm Bkgd}$ ratio as a function of
transverse momentum is shown in Fig.~\ref{fig:gamma_excess} for
peripheral and central 158~{\it A}~GeV $^{208}$Pb\/+\/$^{208}$Pb
collisions.  The results are shown for photons identified using the
narrow shower condition (S2).  As expected according to the discussion
of Eq.~\ref{eq:gamgam}, the $\gamma_{\rm Meas}/\gamma_{\rm Bkgd}$
ratios of Fig.~\ref{fig:gamma_excess} show the same behavior as the
$(\gamma/\pi^{0})_{\rm Meas}/(\gamma/\pi^0)_{\rm Bkgd}$ ratios of
Figs.~\ref{fig:gamma_pi0_per} and \ref{fig:gamma_pi0_cen}.  This
comparison provides an additional systematic check
since the two sets
of results were obtained with different shower identification
criteria.  As discussed previously, in addition to the larger
systematical errors expected when using all showers as photon
candidates, the results of Figs.~\ref{fig:gamma_pi0_per} and
\ref{fig:gamma_pi0_cen} have larger statistical error because the
$\pi^0$ measurement at each $p_T$ is used and contributes to the
statistical error, while on the other hand, the fitted $\pi^0$ results
of Fig.~\ref{fig:pi0_pt} are used for the results of
Fig.~\ref{fig:gamma_excess}, which reduces the statistical error but
increases the systematical error.  The general agreement of the two
sets of results again suggests that the systematical errors are not
large.

%%fig31%%

The total $p_T$-dependent systematical errors listed in
Table~\ref{table3} are shown by the shaded regions in
Fig.~\ref{fig:gamma_excess}. The systematical errors increase strongly
at low transverse momenta due the combinatorial background
uncertainties in the $\pi^0$ yield extraction. They increase also at
large transverse momenta due to the uncertainties in the fits to the
$\pi^0$ distributions. For peripheral collisions, all measured results
fall within $\sigma_{\rm stat.}+\sigma_{\rm syst.}$ of $\gamma_{\rm
   Meas}/\gamma_{\rm Bkgd} = 1$, which indicates that no significant
photon excess is observed. On the other hand, the results for central
collisions do indicate a significant photon excess in the region above
about 1.5 GeV/c.

%%fig32%%
\begin{figure}[hbt]
\begin{center}
   \includegraphics[scale=0.45]{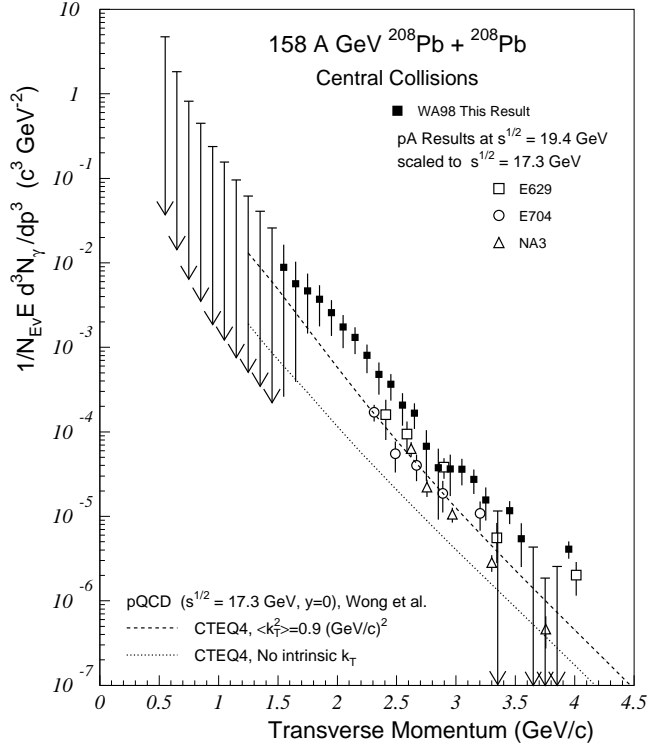}
\caption{The invariant direct photon multiplicity
   for central 158~{\it A}~GeV $^{208}$Pb\/+\/$^{208}$Pb collisions.
   The error bars indicate the combined statistical and systematical
   errors.  Data points with downward arrows indicate unbounded $90\%$
   CL upper limits.  Results of several direct photon measurements for
   proton-induced reactions have been scaled to central
   158~{\it A}~GeV $^{208}$Pb\/+\/$^{208}$Pb collisions for comparison.
The calculation is described in the text.}
\label{fig:gamma_excess_cs}
\end{center}
\end{figure}

The direct photon excess for central $^{208}$Pb\/+\/$^{208}$Pb
collisions is shown in Fig.~\ref{fig:gamma_excess_cs}. The results are
presented as the invariant direct photon yield per central collision.
The direct photon excess is obtained according to Eq.~\ref{eq:excess}
with the measured results of Fig.~\ref{fig:gamma_excess} and
Fig.~\ref{fig:photon_pt}. The statistical and asymmetric systematical
errors of Fig.~\ref{fig:gamma_excess} are added in quadrature to
obtain the total upper and lower errors shown in
Fig.~\ref{fig:gamma_excess_cs}. An additional $p_T$-dependent error is
included to account for that portion of the uncertainty in the energy scale
which cancels in the $(\gamma/\pi^0)_{Meas}$ ratio, but enters again
with $\gamma_{Meas}$ in Eq.~\ref{eq:excess}.  In the case that the
lower error is less than zero a downward arrow is shown with the tail
of the arrow indicating the 90\% confidence level upper limit
($\gamma_{Excess}+1.28\/\sigma_{Upper}$).  The results are also
tabulated in Table~\ref{table4}.

For comparion, other published fixed
target prompt photon measurements for proton-induced reactions
at 200 GeV are also shown in Fig.~\ref{fig:gamma_excess_cs}.
Results are shown from FNAL experiment E704~\cite{plb:ada95}
($-0.15 < x_F < 0.15$)
for proton-proton reactions, and from FNAL experiment
E629~\cite{prl:mcl83} ($-0.75 < y_{cm} < 0.2$) and CERN SPS experiment
NA3~\cite{zpc:bad86} ($-0.4 < y_{cm} < 1.2$) for proton-carbon reactions.
These results have been divided by the total pp inelastic cross
section ($\sigma_{int}=30$~mb) and by the mass number of the
target to obtain the invariant direct photon yield per nucleon-nucleon
collision. They have then been multiplied by the calculated number
of nucleon-nucleon collisions for the central Pb+Pb event selection
for comparison with the present measurements. Based on a Glauber model
calculation, the number of nucleon-nucleon collisions is calculated
to be 660 for the 635~mb most central event selection used for the
present analysis. This scaling of the proton-induced results is
estimated to have an uncertainty of less than 10\% due to
uncertainties in the assumed nuclear density distribution and
nucleon interaction cross section. 
The proton-induced results have also been scaled from $\sqrt{s}=19.4$ GeV to
the lower $\sqrt{s}=17.3$ GeV of the present measurement.
The proton-induced results have also been scaled 
from $\sqrt{s}=19.4$ GeV to the lower $\sqrt{s}=17.3$ GeV of the 
present measurement under the assumption that $E
d^3\sigma_{\gamma}/dp^3 = f(x_T)/s^2$, where  
$x_T=2p_T/\sqrt{s}$\cite{rmp:owe87}. The
$\sqrt{s}$-scaling effectively reduces the $19.4$ GeV 
proton-induced results by about a factor of two. 
This comparison 
suggests an excess direct photon
production in central $^{208}$Pb\/+\/$^{208}$Pb collisions
as compared to proton-induced reactions at the same  $\sqrt{s}$ of
17.3 GeV. The interpretation of the present result is
further discussed in the next section.

\vbox{
\begin{table}
\caption{
The direct photon invariant yield per event
$1/N_{events} \cdot Ed^3 N_{\gamma}/dp^3$ for
central 158~{\it A}~GeV $^{208}$Pb\/+\/$^{208}$Pb collisions.
The central event selection is defined as those events with the
largest measured transverse energy with a total cross section of
635~mb for the selected central sample.
The lower and upper region of uncertainty ($\pm\sigma$, including
statistical and systematical errors added in quadrature) are
indicated. In the case that the lower limit is less than or equal to
zero, only the 90\% confidence level ($+1.28\sigma$) upper limit is
listed.}
\label{table4}
\begin{tabular}{ddddd}
$p_T$ & Lower & $1/N_{events} \cdot Ed^3 N_{\gamma}/dp^3$ &  Upper  &
90\% C.L. Upper Limit\\
(GeV/c) & (c$^3$/GeV$^2$) & (c$^3$/GeV$^2$) & (c$^3$/GeV$^2$) & 
(c$^3$/GeV$^2$)  \\
\tableline
   0.55 &  - & - & - & 4.72  \\
   0.65 &  - & - & - & 1.83  \\
   0.75 &  - & - & - & 0.815 \\
   0.85 &  - & - & - & 0.450 \\
   0.95 &  - & - & - & 0.237 \\
   1.05 &  - & - & - & 0.156 \\
   1.15 &  - & - & - & 0.955E-01 \\
   1.25 &  - & - & - & 0.617E-01 \\
   1.35 &  - & - & - & 0.409E-01 \\
   1.45 &  - & - & - & 0.258E-01 \\
   1.55 &  0.259E-03 & 0.884E-02 & 0.164E-01 & -  \\
   1.65 &  0.392E-03 & 0.565E-02 & 0.103E-01 & -  \\
   1.75 &  0.145E-02 & 0.465E-02 & 0.748E-02 & -  \\
   1.85 &  0.177E-02 & 0.371E-02 & 0.543E-02 & -  \\
   1.95 &  0.136E-02 & 0.257E-02 & 0.363E-02 & -  \\
   2.05 &  0.981E-03 & 0.174E-02 & 0.241E-02 & -  \\
   2.15 &  0.830E-03 & 0.131E-02 & 0.173E-02 & -  \\
   2.25 &  0.496E-03 & 0.806E-03 & 0.107E-02 & -  \\
   2.35 &  0.274E-03 & 0.480E-03 & 0.662E-03 & -  \\
   2.45 &  0.234E-03 & 0.365E-03 & 0.481E-03 & -  \\
   2.55 &  0.118E-03 & 0.207E-03 & 0.286E-03 & -  \\
   2.65 &  0.108E-03 & 0.166E-03 & 0.217E-03 & -  \\
   2.75 &  0.262E-04 & 0.675E-04 & 0.104E-03 & -  \\
   2.85 &  0.915E-05 & 0.376E-04 & 0.632E-04 & -  \\
   2.95 &  0.173E-04 & 0.364E-04 & 0.537E-04 & -  \\
   3.05 &  0.232E-04 & 0.360E-04 & 0.478E-04 & -  \\
   3.15 &  0.183E-04 & 0.273E-04 & 0.357E-04 & -  \\
   3.25 &  0.893E-05 & 0.156E-04 & 0.220E-04 & -  \\
   3.35 &  - & - & - & 0.116E-04 \\
   3.45 &  0.812E-05 & 0.116E-04 & 0.151E-04 & -  \\
   3.55 &  0.251E-05 & 0.543E-05 & 0.826E-05 & -  \\
   3.65 &  - & - & - & 0.432E-05 \\
   3.75 &  - & - & - & 0.186E-05 \\
   3.85 &  - & - & - & 0.255E-05 \\
   3.95 &  0.314E-05 & 0.409E-05 & 0.503E-05 & -  \\
\end{tabular}
\end{table}
\vskip -0.3cm }

\subsubsection{\label{sec:gam_calc} Comparison to Calculations}

Considerable progress has been made  in the theoretical description
of prompt photon production in hadron-induced reactions.
It is now possible to perform a complete and fully consistent
next-to-leading order (NLO) QCD calculation of the prompt photon
cross section and obtain a quite good description of the data
at incident energies from $\sqrt{s}=23$ GeV up to Tevatron
energies ($\sqrt{s}=1.8$ TeV)\cite{jpg:vog97}. On the other hand,
these same calculations which provide a good description at high
incident energy, underpredict prompt photon production at
$\sqrt{s}=19.4$ GeV. For the results shown in
Fig.~\ref{fig:gamma_excess_cs} the E704 and NA3 results are
underpredicted by about a factor of two, while the E629 result
is underpredicted by about a factor of five~\cite{jpg:vog97}.

It has been proposed~\cite{rmp:owe87} that the intrinsic transverse
momentum of the partons, as a consequence of confinement within the
hadron, or of soft gluon radiation,
may significantly increase the theoretical prompt photon
predictions at low incident energies, or
low transverse momenta. The effect of intrinsic $k_T$
is normally neglected in state of the art perturbative QCD calculations
due to the formidable technical difficulty to perform the integration
over transverse degrees of freedom in the NLO calculations.
Nevertheless, there has been a renewed interest to investigate
intrinsic $k_T$ effects in prompt photon
production~\cite{prc:won98,prc:pap99,prd:apa99} largely motivated
by recent high precision prompt photon measurements of FNAL experiment
E706~\cite{prd:alv93,prl:apa98}, 
although the necessity for such intrinsic $k_T$ effects
remains a topic of debate~\cite{epj:aur99}.

Recently,
Wong and Wang~\cite{prc:won98} have investigated the effects of
parton intrinsic $k_T$ on photon production under the assumption
that the NLO corrections are independent of intrinsic
$k_T$. Correction factors, or K-factors, were determined
as a function of photon transverse momentum
as the ratio of the NLO+LO calculation result to the leading-order
(LO) result,
without intrinsic $k_T$. The LO calculations were reevaluated with
the inclusion of
parton intrinsic $k_T$ assumed to be characterized by a Gaussian
distribution with $\left<k_T^2\right> = 4\left<k_T\right>/\pi $, where
$\left<k_T\right>$ is the average intrinsic $k_T$ of a parton. The K-factors
determined without intrinsic $k_T$ were then applied to the LO result
with intrinsic $k_T$ included to obtain the final prompt photon
prediction.

%%fig33%%
\begin{figure}[hbt]
\begin{center}
   \includegraphics[scale=0.5]{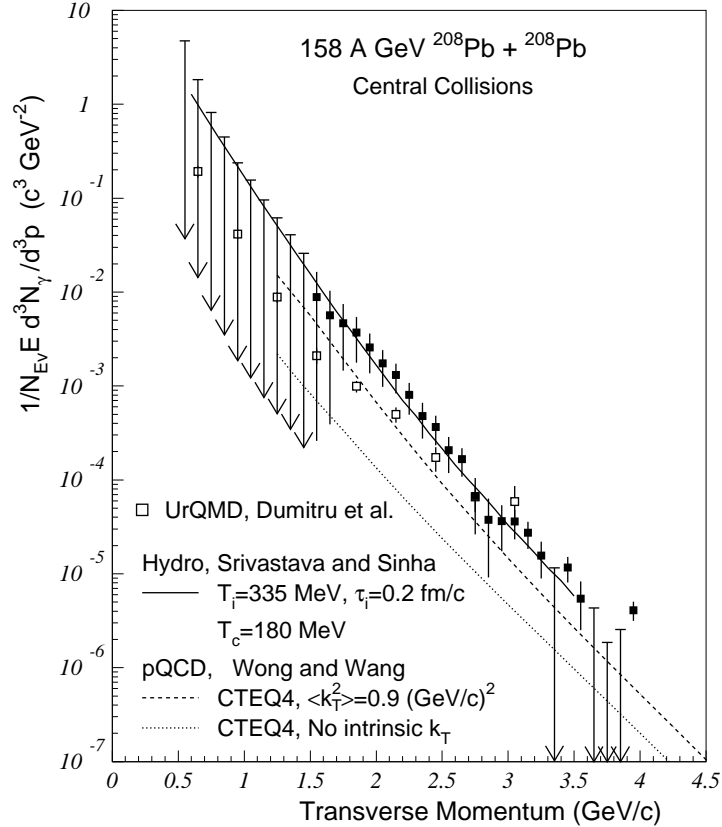}
\caption{The invariant direct photon multiplicity
   for central collisions of 158~{\it A}~GeV $^{208}$Pb\/+\/$^{208}$Pb
   compared to various predictions discussed in the text. The error
   bars indicate the combined statistical and systematical errors.
   Data points with downward arrows indicate unbounded $90\%$ CL upper
   limits. }
\label{fig:gamma_excess_cs_compare}
\end{center}
\end{figure}

With this prescription the E704 and NA3 results shown in
Fig.~\ref{fig:gamma_excess_cs} were well described
with a parton intrinsic $k_T$ of  $\left<k_T^2\right> = 0.9$
(GeV/c)$^2$~\cite{prc:won98}. The importance of the intrinsic $k_T$
effect decreases with increasing $\sqrt{s}$ and increasing photon
$p_T$ such that the prompt photon data at higher $\sqrt{s}$ are
equally well described with or without intrinsic $k_T$.

Predictions of the direct photon production
for $\sqrt{s}=17.3$ GeV central $^{208}$Pb\/+\/$^{208}$Pb collisions
calculated in this manner~\cite{prc:won98} are compared to the measured
results in Fig.~\ref{fig:gamma_excess_cs} and also in 
Fig.~\ref{fig:gamma_excess_cs_compare}. The calculation
has been scaled to central $^{208}$Pb\/+\/$^{208}$Pb 
collisions in the same way as described
above for the proton-induced data. Results are shown with and without
the effects of parton intrinsic  $k_T$ by the long-dashed and
short-dashed curves, respectively. At this low incident energy, the
parton intrinsic $k_T$ is seen to increase the predicted photon
yield by a factor which increases with decreasing $p_T$ from about
4 to 8. The predicted direct photon yield with intrinsic
$k_T$ effects included is in good agreement with the $\sqrt{s}=19.4$
GeV proton-induced results scaled to $\sqrt{s}=17.3$ GeV.
It is also in general agreement with the 
shape of the observed photon spectrum in central
$^{208}$Pb\/+\/$^{208}$Pb collisions, but underpredicts the
observed yield by about a factor of 2.5.
This discrepancy could be a result of further deficiencies in the
prompt photon calculations when applied at low incident energy,
or it may indicate new effects attributable
to nuclear collisions. A possible explanation might be
additional $p_T$ broadening of the incoming partons due to
soft scatterings prior to the hard scattering which produces
the photon~\cite{prc:pap99}. Alternatively, it may be expected
that the photon production is enhanced by the additional
scatterings which occur as a result of rescattering in nucleus-nucleus
collisions.

It is anticipated that if a quark gluon plasma is formed
in relativistic heavy ion collisions, it will evolve from the
initial hard-scattering partonic stage, through a preequilibrium
stage, to a finally thermalized partonic QGP phase. Therefore
the Parton Cascade Model (PCM)~\cite{npb:gei92} has been widely regarded
as a promising tool to investigate the evolution of the initial
partonic state. The VNI~\cite{cpc:gei97} implementation
of the PCM has recently been used to predict direct photon production
in central Pb+Pb collisions at the SPS~\cite{prc:sri98}. Those
predictions are in reasonable agreement with the present result,
although they slightly overpredict the yield above $p_T= 3$ GeV/c
while underpredicting the yield below 2.5 GeV/c. However, it has
recently been pointed out that the method which was used
to implement higher order pQCD corrections in those calculations,
by rescaling $Q^2$ in the running coupling constant $\alpha_s$,
is highly questionable at SPS energies~\cite{plb:bas99}.

Non-equilibrium direct photon emission has also been investigated 
within the context of the Ultrarelativistic Quantum Molecular Dynamics 
model (UrQMD)~\cite{ppnp:bas98}. The UrQMD predictions of the total 
direct photon production in central
Pb+Pb collisions are shown by the open squares in 
Fig.~\ref{fig:gamma_excess_cs_compare}~\cite{prc:dum98}. It is 
interesting that the UrQMD predictions are very similar to the
the VNI parton cascade 
predictions~\cite{prc:sri98}.  In the UrQMD calculations,
the direct photon production results strictly from meson-meson 
scattering, predominantly $\pi\pi \rightarrow \rho\gamma$ and
$\pi\rho \rightarrow \pi\gamma$. In these calculations it was
observed that the photon production in the transverse 
momentum region with $p_T>1.5$ GeV/c is dominated by 
pre-equilibrium emission in which the scattering mesons have
locally non-isotropic momentum distributions. The thermal photon
emission was found to be important only in the low transverse momentum 
region below  $p_T<1.5$ GeV/c. 

It is of particular interest to compare the observed direct
photon yield to predictions of the expected yield in the event
of quark gluon plasma formation. Hydrodynamic model calculations
of the predicted direct photon yield are shown by the solid curve
in Fig.~\ref{fig:gamma_excess_cs_compare}~\cite{epj:sri99,nth:sri00}.
The hydrodynamic calculations include transverse
expansion~\cite{prc:cle97}  with a rich hadronic equation of
state including all hadrons and resonances with mass up to 2.5 GeV. 
The photon emission rates used for the hadronic phase are those which
have been obtained from a two-loop 
approximation of the photon self energy using a model where the
$\pi\rho$ interactions have been included~\cite{Kap91}, and also
include the important contribution from the $A_1$
resonance~\cite{prd:xio92}.  The photon emission rates from the 
quark matter phase include the lowest order contributions from 
the Compton ($q(\overline{q})g \rightarrow q(\overline{q})\gamma $)
and annihilation ($q\overline{q} \rightarrow g\gamma $) processes,
as well as the new contributions from two-loop diagrams which have
recently been shown~\cite{prd:aur98} to give a large bremsstrahlung
($qq(g) \rightarrow qq(g)\gamma $) contribution and a previously
neglected process of $q\overline{q}$ annihilation accompanied by
q(g) rescattering which is found to dominate the photon emission
rate.

As shown in Fig.~\ref{fig:gamma_excess_cs_compare},
very good agreement with the experimental result is obtained for the
case of an equation of state which includes a QGP phase
transition which occurs at a critical temperature of
$T_C=180$ MeV~\cite{nth:sri00} with a hadron thermal 
freeze-out temperature of $T_F=100$ MeV.
The calculation has been performed for the 10\% most central
Pb+Pb collisions for a particle density $dN/dy = 750$ with the
assumption of a fast equilibration time of $\tau_0= \frac{1}{3} T_0$
given by the uncertainty relation. This results in 
an initial temperature of $T_0=335$ MeV with a thermalization time of
$\tau_0=0.2$ fm/c. The contribution from the quark matter in the
QGP and mixed phase dominates the calculated high $p_T$ photon
yield~\cite{nth:sri00}. This is attributed to the
``annihilation with rescattering'' process which is a special
feature of dense quark matter.  This calculation does not include
prompt or preequilibrium contributions.

Further study will be necessary to improve the theoretical predictions
of the prompt photon contribution as well as the non-equilibrium
contributions. On the other hand, the hydrodynamical model
calculations which provide a time integration of the
emission rate with a very short initial formation time and
high initial temperature may provide a reasonable approximation 
of the contributions from the early non-equilibrium phase of 
the collision~\cite{nth:sri00}. Further studies will also be
necessary to determine how the present direct photon results
might further constrain the QGP and non-QGP scenarios. The WA80 direct
photon upper limit for central S+Au collisions~\cite{prl:alb96} 
was able to rule out thermal emission from purely 
hadronic matter in which the
hadronic matter consisted of a simple pion gas with few degrees
of freedom~\cite{prl:sri94,prc:neu95,prc:dum95,plb:arb95} due
to the constraints which the result implied on the initial
temperature~\cite{prc:sol97}. The S+Au direct photon limit
has been shown to be consistent with the QGP transition
using the new rates for quark matter, although with somewhat
different parameters than presented
here~\cite{epj:sri00,nth:sri00}. 
It will be important to determine whether the present result, 
combined with other measurements including the S+Au direct
photon upper limit, can provide further constraints and provide
compelling evidence for the QGP phase transition scenario.

\section{\label{sec:conclusion} SUMMARY and CONCLUSIONS}

A search for direct photons in $^{208}$Pb\/+\/$^{208}$Pb collisions
at 158~{\it A}~GeV has been performed using the large-area
finely-segmented lead-glass calorimeter of the WA98 experiment.
The analysis has been performed on nearly equal-sized event samples of
the 10\% most central and 20\% most peripheral fractions of
the total minimum bias cross section. The inclusive photon and
$\pi^0$ transverse momentum distributions were measured for each event
sample. Constraints were also obtained
on the $\eta$ transverse momentum distribution for central
collisions with a modest accuracy which was limited by statistics.
The systematical error sources were discussed at length with
particular attention to demonstrate the
accuracy of the systematical error estimates. The total
systematical error for the direct photon analysis varied from about
6\% to 10\% over the $p_T$ region of interest, and varied with
the centrality selection.
The direct photon excess was extracted on a statistical basis
as the difference between
the measured inclusive spectrum of photons within the detector
acceptance and the calculated spectrum of photons which result
from all hadrons with significant radiative decay contributions.
The dominant decay contributions are those from the measured $\pi^0$'s
and $\eta$'s.

No significant direct photon excess was observed for the peripheral
event sample for transverse momenta up to about 2.5 GeV/c. This
upper $p_T$ limit of the measurement was imposed by the increasing
statistical error at large $p_T$
due to the low particle multiplicity for peripheral collisions.

In contrast, a significant direct photon excess was observed
over the $p_T$ region of about 1.5 GeV/c to 3.5 GeV/c for central
$^{208}$Pb\/+\/$^{208}$Pb collisions. At transverse momenta
where the observed excess was not significant, a 90\% confidence level
upper limit on the direct photon excess was given. This extended
the result to the full region of measurement from 0.5 GeV/c to 4 GeV/c.

The observed direct photon yield was compared to a next-to-leading order
perturbative QCD calculation at the same $\sqrt{s}$ (=17.3 GeV)
and scaled according to the number of nucleon-nucleon collisions
calculated for the selected central event class. The measured
direct photon yield was about a factor of 2-3 above the pQCD
calculation which included the effects of partonic intrinsic $k_T$
with a parameter which gave a good description of proton-induced
prompt photon results at the nearest incident energy
($\sqrt{s}$=19.4 GeV). A NLO pQCD calculation without intrinsic
$k_T$ was about an order of magnitude below the measured result.

The results suggest that the additional direct photon yield
in central $^{208}$Pb\/+\/$^{208}$Pb collisions is the result
of additional rescatterings in the dense matter. The measured
results can be well-described by hydrodynamical model calculations
which include a quark gluon plasma phase transition with an 
initial temperature of $T_0=335$ MeV and a critical temperature
of $T_C=180$ MeV~\cite{nth:sri00}. The photon emission from the
quark matter is found to dominate over the contribution from the
hadronic matter due to the recently identified process of annihilation
with rescattering~\cite{prd:aur98} which is only expected
to occur in dense quark matter. It will be important
to determine the uniqueness of this interpretation.

\acknowledgments{
We wish to express our gratitude to the CERN accelerator division for the
excellent performance of the SPS accelerator complex. We acknowledge with
appreciation the effort of all engineers, technicians and support staff who
have participated in the construction of this experiment. We
acknowledge helpful discussions with D.K.~Srivastava and C.-Y.~Wong.

This work was supported jointly by
the German BMBF and DFG,
the U.S. DOE,
the Swedish NFR and FRN,
the Dutch Stichting FOM,
the Stiftung f{\"u}r Deutsch-Polnische Zusammenarbeit,
the Grant Agency of the Czech Republic under contract No. 202/95/0217,
the Department of Atomic Energy,
the Department of Science and Technology,
the Council of Scientific and Industrial Research and
the University Grants
Commission of the Government of India,
the Indo-FRG Exchange Program,
the PPE division of CERN,
the Swiss National Fund,
% the International Science Foundation under Contract N8Y000,
the INTAS under Contract INTAS-97-0158,
ORISE,
Grant-in-Aid for Scientific Research
(Specially Promoted Research \& International Scientific Research)
of the Ministry of Education, Science and Culture,
the University of Tsukuba Special Research Projects, and
the JSPS Research Fellowships for Young Scientists.
ORNL is managed by UT-Battelle, LLC, for the U.S. Department of Energy
under contract DE-AC05-00OR22725.
The MIT group has been supported by the US Dept. of Energy under the
cooperative agreement DE-FC02-94ER40818.

}

% now the references.

% \begin{table}
% \caption{}
% \label{}
% \end{tabular}
% \end{table}


\begin{references}
%
\bibitem{QM99}
See for example, Proceedings of {\it Quark Matter '99}, 
Nucl. Phys. A {\bf 661}  (1999).                       
%
% J/psi suppression
\bibitem{Abr97}
M.C. Abreu et al., Phys. Lett. B {\bf 410}, 337 (1997).                       
%
% strangeness enhancement
\bibitem{And99}
E. Andersen et al., Phys. Lett. B {\bf 449}, 401 (1999).
%
% Direct production of photons and dileptons in thermodynamic models
% of multiple hadron production.
\bibitem{Fei76}
E.L.Feinberg, Nuovo Cimento {\bf 34} A, 391 (1976).
%
% Quark-gluon plasma and hadronic production of leptons, photons, and psions.
\bibitem{Shu78}
E.Shuryak, Phys. Lett. B {\bf 78}, 150 (1978).
%
\bibitem{Bjo76}
J.D.Bjorken and H.Weisberg, Phys. Rev. D {\bf 13}, 1405 (1976).
%
% Transition to hot quark matter in relativistic heavy-ion collisions.
\bibitem{Chi78}
S.A.Chin, Phys. Lett. B  {\bf 78}, 552 (1978).
%
% Central collisions between heavy nuclei at extremely high energies:
% the fragmentation region.
\bibitem{Ani80}
R.Anishetty, P.Koehler, and L.McLerran, Phys. Rev. D {\bf 22}, 2793 (1980).
%
% Experimental signatures of phase transition to quark matter in high-energy
% Quark gluon plasma in ultrarelativisitic nucleus-nucleus collisions.
\bibitem{Kaj83}
K.Kajantie and R.Raitio, Phys. Lett. B  {\bf 121}, 415 (1983).
%
% Quark-matter diagnostics.
\bibitem{Dom81}
G.Domokos and J.I.Goldman, Phys. Rev. D {\bf 23}, 203 (1981).
%
% Temperature measurement of quark-gluon plasma formed in high energy
% nucleus-nucleus collisions.
\bibitem{Kaj81}
K.Kajantie and H.I.Miettinen, Z. Phys. C  {\bf 9}, 341 (1981).
%
% Experimental signatures of phase transition to quark matter in high-energy
% collisions of nuclei.
\bibitem{Hal82}
F.Halzen and H.C.Liu, Phys. Rev. D {\bf 25}, 1842 (1982).
%
% Dilepton production from hot quark matter in an ultra-relativistic
% heavy ion collision.
\bibitem{Chi82}
S.A.Chin, Phys. Lett. B  {\bf 119}, 51 (1982).
%
% Universal signals of a quark-gluon plasma.
\bibitem{Sin83}
B.Sinha, Phys. Lett. B  {\bf 128}, 91 (1983).
%
% Photon and dilepton emission from the quark-gluon plasma: Some
% general considerations.
\bibitem{McL85}
L.D.McLerran and T.Toimela, Phys. Rev. D {\bf 31}, 545 (1985).
%
% Experimental signatures of phase transition to quark matter in high-energy
% Scatttering amplitudes in hot gauge theories.
\bibitem{Pis89}
R.D.Pisarski, Phys. Rev. Lett. {\bf 63}, 1129 (1989).
%
% Production of soft dileptons in the quark-gluon plasma.
\bibitem{Bra90}
E.Braaten, R.D.Pisarski, and T.C.Yuan, Phys. Rev. Lett.  {\bf 64}, 2242 (1990).
%
% High-energy photons from quark-gluon plasma versus hot hadronic gas.
\bibitem{Kap91}
J.Kapusta,P.Lichard, and D.Seibert, Phys. Rev. D {\bf 44}, 2774
(1991); Erratum, {\it ibid} D {\bf 47}, 4171 (1993).
%
\bibitem{prd:aur98} P.~Aurenche, et~al., Phys. Rev. D {\bf 58},
   085003 (1998).
\bibitem{epj:sri99} D.~K.~Srivastava,  Eur. Phys. J. C {\bf 10}, 487 (1999).
%
% Diagnosing quark matter by measuring the total entropy and the
% photon or dilepton emission rates.
% \bibitem{Hwa85}
% R.C.Hwa and K.Kajantie,  Phys. Rev. D {\bf 32}, 1109 (1985).
%
\bibitem{prl:mcl83} E629 Collaboration, M.~McLaughlin et~al.,
Phys. Rev. Lett. {\bf 51}, 971 (1983).
\bibitem{zpc:bad86} NA3 Collaboration, J.~Badier et al., Z. Phys. C
{\bf 31}, 341 (1986).
\bibitem{prd:dem87}
NA24 collaboration, C.~De Marzo et al., Phys. Rev. D  {\bf 36}, 8 (1987).
\bibitem{npb:ang86}
R108 collaboration, A.L.S.~Angelis et al., Nucl. Phys. B {\bf 263}, 228 (1986).
\bibitem{plb:ada95} E704 Collaboration, D.L.~Adams  et al.,
   Phys. Lett. B {\bf 345}, 569 (1995).
\bibitem{zpc:ann82}
R806 collaboration,  E.~Annassontzis et al., Z. Phys. C {\bf 13}, 277 (1982).
\bibitem{zpc:bon88}
WA70 collaboration, M.~Bonesini et al.,  Z. Phys. C {\bf 37}, 535 (1988).
\bibitem{zpc:bon88pi}
WA70 collaboration, M.~Bonesini et al.,  Z. Phys. C {\bf 38}, 371(1988).
\bibitem{npb:ang89}
R110 collaboration, A.L.S.~Angelis et. al., Nucl. Phys. B {\bf 327},
541 (1989).
\bibitem{sjn:aak90}
AFS/R807 collaboration,  T.~\AA kesson et al., Sov. J. Nuc. 
Phys. {\bf 51}, 836 (1990).
\bibitem{prd:alv93}
E706 collaboration, G.~Alverson et al., Phys. Rev. D {\bf
48}, 5 (1993).
\bibitem{prl:apa98}
E706 collaboration, L.~Apanasevich et al., Phys. Rev. Lett. {\bf 81}, 2642 (1998).
\bibitem{prd:apa99} L.~Apanasevich, et~al., Phys. Rev. D {\bf 59},
   074007 (1999).
\bibitem{plb:bal98}
UA6 collaboration, G.~Ballocchi et al., Phys. Lett. B  {\bf
436}, 222 (1998).
\bibitem{prl:abe94} CDF collaboration, F.~Abe et al.,
Phys. Rev. Lett. {\bf 73}, 2662 (1994).
%               CERN-EP-98-111
\bibitem{rmp:owe87} J.F.~Owens, Rev. Mod. Phys. {\bf 59}, 465 (1987).
\bibitem{jpg:vog97} See for example, W.~Vogelsang and M.R.~Whalley, J. Phys. G:
   Nucl. Part. Phys.  {\bf 23}, A1 (1997).
\bibitem{prd:hus95} J.~Huston et al.,  Phys. Rev. D {\bf 51},
   6139 (1995).
\bibitem{epj:aur99} P.~Aurenche et~al., Eur. Phys. J. C {\bf 9}, 107 (1999).

%
\bibitem{Ake90}
HELIOS Collaboration, T.{\AA}kesson et al., Z. Phys. C  {\bf 46}, 369 (1990).
%
\bibitem{Alb91}
WA80 Collaboration, R.~Albrecht et al., Z. Phys. C  {\bf 51}, 1 (1991).
%
% CERES photons
\bibitem{Bau96}
CERES Collaboration, R.Baur et al., Z. Phys. C {\bf 71}, 571 (1996).
%
% WA80 direct photon upper limits
\bibitem{prl:alb96} R.~Albrecht et~al., Phys. Rev. Lett. 
{\bf 76}, 3506 (1996).
\bibitem{prl:sri94} D.~K.~Srivastava and B.~Sinha, Phys. Rev. 
Lett.\ {\bf 73}, 2421 (1994).
\bibitem{prc:neu95} J.~J.~Neumann, D.~Siebert, and G.~Fai, Phys. Rev. 
C {\bf 51}, 1460 (1995).
\bibitem{prc:dum95} A.~Dumitru et~al., Phys. Rev. C {\bf 51}, 2166 (1995).
\bibitem{plb:arb95} N.~Arbex et~al., Phys. Lett. B {\bf 354}, 307 (1995).
\bibitem{prc:sol97} J.~Sollfrank et~al., Phys. Rev. C {\bf 55}, 392 (1997).
%

\bibitem{nim:bad82} A.~Baden et al.,
Nucl. Instrum. Methods Phys. Res. Sect. A {\bf 203}, 189 (1982).
\bibitem{nim:lin97} W.~T.~Lin  et al.,
Nucl. Instrum. Methods Phys. Res. Sect. A {\bf 389}, 415 (1997).
\bibitem{nim:rub95} J.~M.~Rubio  et al.,
Nucl. Instrum. Methods Phys. Res. Sect. A {\bf 367}, 358 (1995).
\bibitem{nim:car99} L.~Carl\'{e}n  et al.,
Nucl. Instrum. Methods Phys. Res. Sect. A {\bf 431}, 123 (1999).
\bibitem{nim:agg96} M.~M.~Aggarwal  et al.,
Nucl. Instrum. Methods Phys. Res. Sect. A {\bf 424}, 395 (1999).
\bibitem{nim:awe89} T.~C.~Awes et al., Nucl. Instrum. Methods Phys. Res.
Sect. A {\bf 279}, 497 (1989).
\bibitem{nim:chu96} T.Chujo, et al., Nucl. Instr. and Meth. A {\bf
   383}, 409 (1996).
\bibitem{nim:pei96} T.~Peitzmann et al., Nucl. Instrum. Methods  Phys.
Res. Sect. A {\bf 376}, 368 (1996).
\bibitem{nim:neu95} S.~Neumaier  et al., Nucl. Instrum. Methods Phys. Res.
Sect. A {\bf 360}, 593 (1995).
\bibitem{nim:win94} A.~L.~Wintenberg et al., Proceedings of
   Electronics
for Future Colliders Conference, May 1994, LeCroy Corp.
\bibitem{the:rey99} K.~Reygers, Doctoral thesis, University of
  M{\"u}nster, Germany (1999).
\bibitem{nim:ber92} F.~Berger  et al., Nucl. Instrum. Methods Phys. Res.
Sect. A {\bf 321}, 152 (1992).
\bibitem{nim:awe92} T.~C.~Awes et al., Nucl. Instrum. Methods Phys. Res.
Sect. A {\bf 311}, 130 (1992).
% \bibitem{crn:geant} R.~Brun et al., CERN report DD/EE/84-1 (1987)
% GEANT 3.x, revised edition.
\bibitem{crn:geant} R.~Brun and F.~Carminati, {\it GEANT Detector
Description and Simulation Tool},
   CERN Program Library, Long Writeup W5013, March, 1994.
\bibitem{the:buc99} D.~Bucher, Doctoral thesis, University of
  M{\"u}nster, Germany (1999).

%\bibitem{prc:fei72} E.L.~Feinberg, Phys. Rep. 5C, 237 (1972).
\bibitem{npb:bou76} M.~Bourquin and J.-M.~Gaillard, Nucl. Phys. B
{\bf 114} 334, (1976).
\bibitem{prd:ban74} R.~Blankenbecler and S.J.~Brodsky, Phys. Rev. D {\bf 10},
2973  (1974).
\bibitem{plb:shu80} E.~V.~Shuryak and O.~Zirhov, Phys. Lett. B {\bf 89},
253 (1980).
\bibitem{rnc:hag83} R.~Hagedorn, Riv. Nuovo Cimento {\bf 6}, 1 (1983).

\bibitem{npb:bus76} F.~B{\"u}sser et~al., Nucl. Phys. B {\bf 106}, 1
(1976).
\bibitem{npb:bad77} J.~Bartke et~al., Nucl. Phys. B {\bf 120}, 14 (1977).
\bibitem{prl:don78} G.~Donaldson et~al., Phys. Rev. Lett. {\bf 40}, 684
(1978).
\bibitem{plb:kou79} C.~Kourkoumelis et~al., Phys. Lett. B {\bf 84}, 277
(1979).
\bibitem{prl:pov83} J.~Povlis et~al., Phys. Rev. Lett. {\bf 51}, 967 (1983).
\bibitem{zpc:agu91} NA27 Collaboration, M.~Aguilar-Benitez et~al.,
Z. Phys. C {\bf 50}, 405 (1991).
\bibitem{zpc:aps92} Omega Photon Collaboration, R.~Apsimon et~al.,
Z. Phys. C {\bf 54}, 185 (1992).
\bibitem{epj:aga98} CERES/TAPS Collaboration, G.~Agakichiev et~al.,
Eur. Phys. J. C {\bf 4}, 231 (1998).
\bibitem{plb:ake86} AFS Collaboration, T.~{\AA}kesson et~al., 
Phys. Lett. B {\bf 178}, 447 (1986).
\bibitem{plb:alb95} WA80 Collaboration, R.~Albrecht  et~al., 
Phys.\ Lett.\ B {\bf 361}, 14 (1995).

\bibitem{prl:bea97} NA44 Collaboration, I.G.~Bearden et~al., 
Phys. Rev. Lett. {\bf 78}, 2080 (1997).
\bibitem{epj:app98} NA49 Collaboration, H.~Appelsh\"{a}usser et~al., 
Eur. Phys. J. C {\bf 2}, 661 (1998).
\bibitem{prl:agg99} WA98 Collaboration, M.M.~Aggarwal et~al., 
Phys. Rev. Lett. {\bf 83}, 926 (1999).
\bibitem{prc:nix98} J.R.~Nix, Phys. Rev. C {\bf 58},
2303 (1998).

\bibitem{prd:kap96} J.~Kapusta et~al., Phys. Rev. D {\bf 53},
5028 (1996).
\bibitem{prd:hua96} Z.~Huang and X.-N.~Wang, Phys. Rev. D {\bf 53},
5034 (1996).
\bibitem{prd:jal98} J.~Jalilian-Marian and B.~Tekin, Phys. Rev. D {\bf 57},
5593 (1998).
\bibitem{epj:pdb98} {\em Review of Particle Properties}, 
Eur. Phys. J. C {\bf 3}, 1 (1998).
\bibitem{plb:dia80} M.~Diakonou et~al., Phys. Lett. B {\bf 89}, 432 (1980).
\bibitem{prl:dao73} F.T.~Dao et~al., Phys. Rev. Lett. {\bf 30}, 1151 (1973).
\bibitem{zpc:der91} I.~Derado et~al., Z. Phys. C {\bf 50}, 31 (1991).
\bibitem{zpc:alb94} NA35 Collaboration, T.~Alber et~al., 
Z. Phys. C {\bf 64}, 195 (1994).
\bibitem{plb:aba96} WA85 Collaboration, S.~Abatzis et~al., 
Phys. Lett. B {\bf 376}, 251 (1996).
\bibitem{cern:jet89} T.~Sj\"{o}strand and M.~Bengtsson, 
{\it The Lund Monte Carlo Programs - JETSET},
  CERN Program Library, W5035, March, 1994.
\bibitem{plb:agg99} WA98 Collaboration, M.M.~Aggarwal et~al., 
Phys. Lett. B {\bf 458}, 422 (1999).
\bibitem{pr:wer93} K.~Werner, Phys. Rep. {\bf 232}, 87 (1993). 
\bibitem{ornl:gab77} T.A.~Gabriel et al., {\em CALOR: A Monte Carlo
    Program Package for the Design and Analysis of Calorimeters}, 
ORNL/TM-5619, (1977).
\bibitem{atl:fer96} A.~Ferrari and P.R.~Sala, {\em GEANT Hadronic
    Event Generators: a comparison at the single interaction level},
ATLAS Internal Note, PHYS-No-086, June 1996.
\bibitem{npc:afa96} NA49 Collaboration, S.V.~Afanasiev et~al., 
Nucl.\ Phys.\ A {\bf 610}, 188c (1996).
\bibitem{prl:agg98} WA98 Collaboration, M.M.~Aggarwal et~al., 
Phys. Rev. Lett. {\bf 81}, 4087 (1998).
\bibitem{cern:hag83} R.~Hagedorn, CERN-TH. 3684 (1983).
\bibitem{plb:boc96} UA1 Collaboration, G.~Bocquet  et~al.,
  Phys. Lett. B {\bf 366}, 434 (1996).
\bibitem{epj:alb98} R.~Albrecht et~al., Eur. Phys. J. C {\bf 5},
  255 (1998).

\bibitem{prc:won98} C.-Y.~Wong and H.~Wang, Phys. Rev. C {\bf 58}, 376 (1998).
\bibitem{prc:pap99} G.~Papp, P.~Levai, and G.~Fai, Phys. Rev. C {\bf
     61}, 021902 (1999).

\bibitem{npb:gei92} K.~Geiger and B.~M\"{u}ller,
Nucl.\ Phys.\ B {\bf 369}, 600 (1992); K.~Geiger,
Phys. Rep. {\bf 258}, 237 (1995).
\bibitem{cpc:gei97} K.~Geiger, Comp. Phys. Comm. {\bf 104}, 70 (1997).
\bibitem{prc:sri98} D.K.~Srivastava and K.~Geiger, Phys. Rev. C {\bf
     58}, 1734 (1998).
\bibitem{plb:bas99} S.A.~Bass and B.~M\"{u}ller,
   Phys. Lett. B {\bf 471}, 108 (1999).
\bibitem{ppnp:bas98} S.A.~Bass et~al., 
   Prog. Part. Nucl. Phys. {\bf 41}, 225 (1998).
\bibitem{prc:dum98} A.~Dumitru et~al.,
   Phys. Rev. C {\bf 57}, 3271 (1998).
\bibitem{nth:sri00} D.~K.~Srivastava and B.~Sinha, nucl-th/0000000.
\bibitem{prc:cle97} J.~Cleymans, K.~Redlich, and D.K.~Srivastav, 
Phys. Rev. C {\bf 55}, 1431 (1997).
\bibitem{prd:xio92} Li~Xiong, E.~Shuryak, and G.E.~Brown, Phys. Rev. D {\bf
     46}, 3798 (1992).
\bibitem{epj:sri00} D.~K.~Srivastava and B.C.~Sinha, Eur. Phys. J. C 
        {\bf 12}, 109 (2000).


%\bibitem{ref:alb89} R.~Albrecht {\em et al.}, Nucl. Instrum. Methods 
%Phys. Res.
%Sect. A {\bf 276}, 131 (1989).
%\bibitem{ref:ru92} See for example, P.~V.~Ruuskanen, Nucl.\ Phys.\
%A {\bf 544}, 169c (1992) and
%J.~Kapusta, Nucl.\ Phys.\ A {\bf 566}, 45c (1994), and references
%therein.
%\bibitem{ref:al91} R.~Albrecht {\em et al.}, Z.\ Phys.\ C {\bf 51}, 1 (1991).
%\bibitem{ref:sa94} R.~Santo {\em et al.}, Nucl.\ Phys.\ A {\bf 566}, 
%61c (1994).
%\bibitem{ref:sh94} E.~V.~Shuryak and L.~Xiong, Phys.\ Lett.\ B {\bf 333}, 316
%(1994).
%\bibitem{ref:sr94} D.~K.~Srivastava and B.~Sinha, Phys.\ Rev.\ 
%Lett.\ {\bf 73},
%2421 (1994).
%\bibitem{ref:ne95} J.~J.~Neumann, D.~Siebert, and G.~Fai, Phys. Rev. 
%C {\bf 51},
%1460 (1995).
%\bibitem{ref:du95} A.~Dumitru, U.~Katscher, J.~A.~Maruhn, H.~St\"ocker,
%W.~Greiner, and D.~H.~Rischke, Phys. Rev. C {\bf 51}, 2166 (1995).
%\bibitem{ref:ar95} N.~Arbex, U.~Ornik, M.~Pl\"umer, A.~Timmermann, 
%and R.~M.~Weiner,
%Phys. Lett. B {\bf 354}, 307 (1995).
%\bibitem{ref:al94} R.~Albrecht {\em et al.}, Phys.\ Rev.\ C {\bf 
%50}, 1048 (1994).
%\bibitem{ref:al95} R.~Albrecht {\em et al.}, Phys.\ Lett.\ B {\bf 
%361} 14 (1995).
%\bibitem{ref:ba96} R.~Baur {\em et al.}, Z.\ Phys.\ C {\bf 71}, 571 (1996).

\end{references}
\end{document}